\documentclass[fleqn,usenatbib]{mnras}
\usepackage{newtxtext,newtxmath}
\usepackage{longtable} 
\usepackage{multicol}
\usepackage[T1]{fontenc}
\usepackage{multirow}
\usepackage[table,dvipsnames]{xcolor}
\usepackage{rotating}
\usepackage{pdflscape}
\DeclareRobustCommand{\VAN}[3]{#2}
\let\VANthebibliography\thebibliography
\def\thebibliography{\DeclareRobustCommand{\VAN}[3]{##3}\VANthebibliography}

\usepackage{graphicx}	
\usepackage{amsmath}	
\usepackage{comment}
\newcommand\numberthis{\addtocounter{equation}{1}\tag{\theequation}}

\title[A Bayesian BBH
Model Selection combining Population Synthesis and
Galaxy Formation Models]{A Trifecta of Modelling Tools: A Bayesian Binary Black Hole
Model Selection combining Population Synthesis and
Galaxy Formation Models}
\author[Rauf \textit{et al.}]{
Liana Rauf$^{1,2,3}$\thanks{E-mail: liana.rauf@anu.edu.au},
Cullan Howlett$^{1,2}$,
Simon Stevenson$^{2,4}$,
Jeff Riley$^{2,5}$,
Reinhold Willcox$^{6}$
\\
$^{1}$School of Mathematics and Physics, The University of Queensland, Brisbane, QLD 4072, Australia\\
$^{2}$OzGrav: The ARC Centre of Excellence for Gravitational Wave Discovery, Australia\\
$^{3}$The Research School of Astronomy and Astrophysics, Australian National University, Stromlo, ACT 2601, Australia\\
$^{4}$Centre for Astrophysics and Supercomputing, Swinburne University of Technology, Mail H39, PO Box 218, VIC 3122, Australia\\
$^{5}$School of Physics and Astronomy, Monash University, Clayton VIC 3800, Australia\\
$^{6}$Institute of Astronomy, KU Leuven, Celestijnlaan 200D, 3001 Leuven, Belgium}
\date{Accepted XXX. Received YYY; in original form ZZZ}

\begin{document}
\label{firstpage}
\pagerange{\pageref{firstpage}--\pageref{lastpage}}
\maketitle

\begin{abstract}
Gravitational waves (GWs) have revealed surprising properties of binary black hole (BBH) populations, but there is still mystery surrounding how these compact objects evolve. We apply Bayesian inference and an efficient method to calculate the BBH merger rates in the \textsc{Shark} host galaxies, to determine the combination of \texttt{COMPAS} parameters that outputs a population most like the GW sources from the LVK transient catalogue. For our \texttt{COMPAS} models, we calculate the likelihood with and without the dependence on the predicted number of BBH merger events. We find strong correlations between hyper-parameters governing the specific angular momentum (AM) of mass lost during mass transfer, the mass-loss rates of Wolf--Rayet stars via winds and the chemically homogeneous evolution (CHE) formation channel. 
We conclude that analysing the marginalised \textit{and} unmarginalised likelihood is a good indicator of whether the population parameters distribution and number of observed events reflect the LVK data. In doing so, we see that the majority of the models preferred in terms of the population-level parameters of the BBHs greatly overpredict the number of events we should have observed to date. Looking at the smaller number of models which perform well with both likelihoods, we find that those with no CHE, AM loss occurring closer to the donor during the first mass-transfer event, and/or higher rates of mass-loss from Wolf--Rayet winds are generally preferred by current data. We find these conclusions to be robust to our choice of selection criteria. 
\end{abstract}




\section{Introduction}
The detection of gravitational waves (GWs) have prompted the astrophysical community to revise the models for binary black hole (BBH) formation and evolution. There are currently two main formation channels proposed: 
\begin{enumerate}
    \item Isolated: The formation of merging BBHs through the evolution of isolated binaries. These stars are gravitationally bound since birth and are not perturbed by other stars or compact objects \citep{2018MNRAS.479.4391M,2021A&A...647A.153B,2021MNRAS.502.4877S,2021MNRAS.508.5028B,2022ApJ...938...45B,2022MNRAS.516.2252O}.
    \item Dynamical: BBH mergers occurring in dense environments via three-body encounters followed by energy exchange and hardening of the binary. Some examples of dense environments are young star, nuclear or globular clusters, or in accretion disks of the central supermassive BHs in galaxies \citep{PhysRevD.100.043027,2019ApJ...876..122Y,2020ApJ...894..133A,2021ApJ...921L..43Z}. 
\end{enumerate}
GWs can be used to gain insight into the poorly constrained aspects of massive stellar binary evolution and help in distinguishing between binary formation channels. The detailed origins and formation channels for binary systems are still unknown, particularly what fraction of these systems come from isolated, which is the focus of this paper, and dynamical channels. 
 
There are also aspects of stellar evolution that are poorly understood, such as the common envelope phase, mass transfer efficiency and AM loss. 
This information is critical for understanding  the population of BBHs observed by LIGO, Virgo and KAGRA (LVK), including their mass and redshift distributions, and to inform us on what is required for future detectors \citep{2019IAUS..346..397M, 2020FrASS...7...38M, 2022LRR....25....1M, 2023MNRAS.520.5259A}. 
Combining the data from GWTC-1 \citep{2019PhRvX...9c1040A}, GWTC-2 \citep{2021PhRvX..11b1053A} and GWTC-3 \citep{2023PhRvX..13d1039A}, there is a total of 90 confident ($p_{\rm astro}>0.5$; see Section~\ref{Selection_criteria} and Eq.~\ref{eq:pastro}) GW event candidates observed to date, including 85 BBHs, 2 binary neutron stars (BNSs), and 3 black-hole-neutron-stars (BHNSs). This growing population of GWs, including those to be released from the ongoing fourth LVK observing run (O4), provides key information on the source masses, spins and local merger rates \citep{2023PhRvX..13a1048A} of binary compact objects, which can be used to answer some of these questions. 

However, there are many formation pathways that could lead to the same individual source properties. Simulations of ensembles of events, produced by varying the physical assumptions, can provide better insight into how these binary systems evolve and merge over time. When combined with the full set of data observed to date, they can potentially break the degeneracy between uncertainties in stellar evolution parameters and BBH populations. They also enable us to produce merger rates that can be compared to observations. \citet{mandel2021rates} provide a comprehensive review of the local merger rates for the isolated and dynamical formation channels. The studies in the review combine population synthesis codes with different assumptions for the metallicity dependent star formation rate (SFR) density, which plays a crucial role in the evolution of massive binaries \citep{2019MNRAS.490.3740N, 2024AnP...53600170C}. \citet{2022MNRAS.516.5737B} utilises \texttt{COMPAS} \citep{2017NatCo...814906S, 2018MNRAS.481.4009V, 2022ApJS..258...34R} to explore the impact of stellar evolution parameters and metallicity-specific star formation, $\mathcal{S}(z, Z)$, on the merger rate, chirp mass and coalescence time distributions. They find that uncertainties in the binary evolution models only led to variations in the BBH merger rate of up to a factor 2, with the BBH merger rates most sensitive to $\mathcal{S}(z, Z)$. \citet{2021MNRAS.502.4877S} and \citet{2019MNRAS.490.3740N} also come to similar conclusions. \citet{2014A&A...564A.134M,2021MNRAS.502.4877S,2022MNRAS.509.1557C} find variations in the merger rate over several orders of magnitude. Many of these studies choose to explore the variations in binary evolution models over a grid of metallicities, and generate populations for each grid point. Some studies have shown that population III stars can also contribute to the total BBH merger rate density \citep{2024MNRAS.528..954S, 2024arXiv240304389T}. These studies emphasise the need to further study the assumptions in binary evolution models to understand the impact on the merger rate, specifically the \textit{combined} impact of model variations \citep{2022MNRAS.517.4034S}. 

There have also been various studies proposing methods for constraining different binary evolution models. \citet{2023ApJ...950..181W} applies ``backward" modelling of each GW event to the progenitors to avoid fixed astrophysical assumptions, by allowing variations in the hyper-parameters. However, they note that their methods become less efficient as the dimensionality of the parameter space increases. \citet{2018MNRAS.477.4685B} uses Fisher information matrices to quantitatively analyse how the observed BBH population will constrain the \texttt{COMPAS} models, given the chirp mass distribution and accounting for the metallicity-specific cosmic SFR, selection effects and measurement uncertainty. \citet{2023ApJ...950...80R} uses artificial neural networks to generate an interpolant to calculate the merger detection rates for double compact objects for varying prescriptions of the metallicity-specific SFR. This is used to create GW populations for each model, where the chirp mass and redshift distribution are compared using Bayesian analysis to the LVK data. \citet{2023PhRvD.108d3023D} calculates a joint likelihood for binary evolution models. They fit normal distribution fits to the likelihood grids and implement Gaussian process regression to interpolate between models, the detected number of events and the likelihood. This is also used to quantify the systematic uncertainty in the parameters. However, they emphasise that the inference on a limited subspace of models could be biased. These studies highlight the high computational costs for multi-dimensional parameter studies. However, as more observations are made, the uncertain astrophysics will produce correlated impacts in the binary population, so there is a crucial need for more parameters covering a large range in model inference. Realistically, all formation channels (within the isolated and dynamical channels) must be considered to obtain a parameter space representative of the true BBH population \citep{2021ApJ...910..152Z, 2023ApJ...955..127C}.

To build on this body of work, we develop here a new method of combining a more accurate, non-parametric star-formation and metallicity history of the Universe with a wider range of different massive binary evolution models. We then use these models alongside data from GWTC-1 \citep{2019PhRvX...9c1040A}, GWTC-2 \citep{2021PhRvX..11b1053A} and GWTC-3 \citep{2023PhRvX..13d1039A} to identify the most likely avenues for generating binary compact objects, and to rule out those that are unable to produce the distributions of objects seen to date. We do this by improving on \citet{10.1093/mnras/stad1757}, where we combined runs from the \texttt{COMPAS} rapid population synthesis code, for formation and evolution of BBH formed through isolated binary evolution, with a SFR and metallicity history for the observed host galaxies predicted by the semi-analytic galaxy evolution code \textsc{Shark} \citep{2018MNRAS.481.3573L, 2019MNRAS.489.4196L}. In our original work, we predicted (among other things) the BBH merger rate as a function of redshift for four \texttt{COMPAS} runs, where we varied the remnant mass prescription. We found the merger rate ranged over several magnitudes and we concluded that the \citet{mandel2020simple} remnant mass prescription was the best fit out of the four models to the local merger rate currently predicted using the BBH mergers from GWTC-3. However, we also found that cross-matching between the \textsc{Shark} galaxies and the \texttt{COMPAS} population was prohibitively time-consuming for exploring larger portions of our parameter space. Since we explored only rates, we were also not able to confidently rule out any models that do not align with the GW population-level parameters, and did not test the impact of varying sub-parameters of these models such as those covering other aspects of binary evolution rather than just the mass of the remnant compact objects.

So, in this paper, we substantially extend on our work in \citet{10.1093/mnras/stad1757} and explore an alternative and efficient method of predicting the BBH merger rate and population level parameters such as the binary masses and redshift distributions. We do so by rescaling a high-fidelity single run of our original simulation, varying some of the most relevant \texttt{COMPAS} input parameters. This allows us to test in this work $>300$ different models for population synthesis, varying ingredients/model assumptions such as the common envelope efficiency, mass transfer accretion efficiency, mass-loss rates of Wolf--Rayet stars, chemically homogeneous evolution model and specific AM loss. For each of these we evaluate the ability of the model to reproduce the observed distribution of black hole masses and redshifts, using a Bayesian framework to account for selection effects. We also explore the impact on our findings with and without including the merger rate as a fitting parameter (i.e., including a comparison of the model-predicted and observed merger rates in evaluating our model posteriors). Compared to previous studies this work also has the improvement of incorporating a basic estimate of sampling error in the simulated model merger rates and posteriors. This allows us to set a threshold for what we consider to be tangible differences in model performance, and also demonstrates more generally the importance of further investigation into the uncertainties in the typical `simulation-based inference' used in GW population studies.

This paper is structured as follows. In Section \ref{Section 2}, we summarise recent progress in our understanding of various massive binary evolution models and parameters, motivating the range of models we explore in this work. In Section~\ref{Section 3}, we show that starting with a fiducial model, where the merger rate in a set of simulated galaxies has been evaluated by carefully cross-matching  between \texttt{COMPAS} and \textsc{Shark}, we can rescale the simulation to produce the time and metallicity dependent GW merger rate for another model where the \texttt{COMPAS} parameters have been altered. 
In Section~\ref{Section 4}, we implement Bayesian inference to determine which of these \texttt{COMPAS} runs output a BBH population that best matches the current observed population. We do so by comparing the chirp mass, mass ratio and redshift distributions from each model with the LVK posterior sample distributions. We then validate the total number of theoretical and detected GW events expected for the models and the models posteriors by comparing to the values obtained using the four `brute-force' evaluations from \citet{10.1093/mnras/stad1757}. We repeat the method for a wide range of new \texttt{COMPAS} runs in Section \ref{Section 5}, identifying the best and worst fit models to the data and studying the impact of different selection effects and analysis choices on the conclusions we can draw about stellar evolution. We conclude in Section \ref{Section 6}. In this work, we use the \textsc{Shark} cosmology $H_0, \Omega_m = \{67.8, 0.308\}$ in flat $\Lambda$CDM where necessary.


\section{\texttt{COMPAS}: Overview of Models} \label{Section 2}
In the following section, we detail the physics and the implementation in \texttt{COMPAS} of the common envelope (CE), mass transfer, mass loss via winds, angular momentum (AM) loss and formation channels such as the isolated and chemically homogeneous evolution. We review the literature on the assumptions and parameters associated with these processes that can impact the evolution of the binary stars and, hence, the number and source properties of GWs for each \texttt{COMPAS} model. We conclude with a summary of the big questions surrounding these prescriptions and their limitations in the context of BBH population modelling. Note that in this section, we refer to the more massive star as the donor and the lighter star as the accretor star.


\subsection{Isolated formation channels}

In this paper, we focus on BBHs in the isolated formation channel. This is the most commonly studied channel, but see Fig 1 in \citet{2020RAA....20..161H} for possible binary formation channels. It involves massive stars formed on the zero-age-main-sequence (ZAMS). We describe, one of the formation channels, the classical isolated binary evolution via a CE phase as follows: 
\begin{enumerate}
    \item The primary star in the binary reaches the end of main sequence (MS) evolution. After hydrogen burning in the core ends,  the core contracts and causes the outer stellar envelope to expand past the Roche lobe (due to the negative heat capacity of the star \citep{1968MNRAS.138..495L}) and transfer mass to the secondary star. 
    \item Mass transfer from the primary stops and leaves a Wolf-Rayet star.\footnote{This formation scenario for Wolf--Rayet stars is disputed by \citet{2020A&A...634A..79S} due to lack of observations of massive stars at low metallicity.} This then reaches the end of its life, undergoes a supernova and forms a BH. The system is now an X-ray binary. 
    \item The secondary evolves and transfers mass back to the primary BH. This process is dynamically unstable and forms a CE, which encompasses the compact object.
    \item Orbital energy released due to the compact object rapidly in-spiralling can remove the CE.
    \item The secondary star evolves and eventually also forms a BH. 
    \item The BBH system slowly in-spirals due to orbital energy loss via GWs. 
\end{enumerate}
However, there are many stages within this formation channel where the binary may not survive, or the events can subtly differ from the above prescription leading to dramatically different outcomes. The system may not survive the first supernova explosion or the CE may be not be ejected resulting in the compact object merging with the secondary core before a BBH is formed \citep{2020ApJ...892...13S,2023MNRAS.523..221G}. Alternatively, the second mass transfer could be stable and avoid the CE \citep[e.g.,][]{1976IAUS...73...35V,vandenHeuvel:2017MNRAS,2019MNRAS.490.3740N,2022ApJ...931...17V,2021ApJ...922..110G,2021A&A...650A.107M}. 
The wind mass-loss rates and remnant prescription can also impact the orbital separation, as well as the final masses. There are various circumstances and factors that can affect the separation of BBH systems, which are strongly correlated with the delay time and the merger rate. On one hand, this makes GW sources unique probes of cosmic history \citep{2019MNRAS.490.3740N, 2021hgwa.bookE..16M, 2022PhR...955....1M, 2023arXiv231115778C}, but on the other, leads to difficulties in interpreting or identifying the most important/critical points in the evolution of the system.

As an example, \citet{2017NatCo...814906S} used \texttt{COMPAS} to show that the first three GW events detected by LIGO: GW150914, GW151226 and GW151012, can be explained by the classical isolated channel with mass transfer and CE phase, and characterised their progenitor metallicity and masses. However, \citet{2022arXiv220708837D} find that the stable channel, which consists of two stable mass transfer phases, dominates the GW sources. The formation of GW sources via a radiative CE channel is not possible for low mass transfer efficiency ($f_{\rm MT} \leq 0.3$, see Eq.~\ref{eq:Fmt} in Section~\ref{CE+MT}) due to the mass ratio being below the critical mass ratio ($q_{\rm crit} = M_d/M_a$, where $M_a$ is the accretor mass and $M_d$ is the donor mass) for instability. \citet{2023arXiv230905736P} similarly find that stable mass transfer, where mass and AM is lost through the L2 point, is a possible formation channel for BBHs. 

Comparing the two channels, \citet{2022ApJ...931...17V} find that the CE channel tends to dominate BBH mergers with primary masses $< 30M_\odot$ in binaries with short delay times ($<1$Gyr). They also find that the CE phase is rare for massive stars due to heavy mass loss preventing them from filling their Roche lobe. As such, the stable mass transfer channel, which widens the binary orbit as mass is lost from the system during secondary mass transfer, forms primary BH masses $> 30M_\odot$ and systems with long delay times ($>1$Gyr). They also find that the CE or stable mass transfer channel dominates the mass distribution depending on redshift, which is in line with \citet{2019MNRAS.490.3740N} finding redshift dependence in the mass distribution. \citet{2022ApJ...940..184V} studies the stable mass transfer channel only and the impact of $f_{\rm MT}$ on the local BBH merger rate and the primary BH mass distribution. For $f_{\rm MT} = 0.75$ and $f_{\rm MT} = 1$, there are no BBHs formed through the stable mass transfer channel. The merger rate drops from $141~\rm Gpc^{-3} yr^{-1}$ with $f_{\rm MT} = 0.25$ to $16~\rm Gpc^{-3} yr^{-1}$ with $f_{\rm MT} = 0.5$, which implies the stable mass transfer contributes strongly to the merger rate for $f_{\rm MT} < 0.5$. The low mass end of the primary BH mass distribution is justified by the stable mass transfer channel --- as $f_{\rm MT}$ increases, the minimum primary BH mass increases. Overall, the relative contribution of both channels is uncertain. The CE channel in particular may be over-represented in population synthesis models (see Section~\ref{CE+MT} for relevant references). 

\subsubsection{Common envelope evolution ($\alpha_{\rm CE}$) and Mass transfer efficiency ($f_{\rm MT}$)} \label{CE+MT}
Given the above, in the formation of isolated binary stars, uncertainties remain in the  CE phase and mass transfer process, which play a key role in the final component and orbital properties. 

The mass transfer efficiency is the fraction of mass lost by the donor star and accreted by the companion star. \citet{2022ApJS..258...34R} uses a simplified approach for the mass transfer. They parameterise the fraction of mass lost by the donor to the accretor with $f_{\rm MT}$,
\begin{align}
    \dot{M}_a = -f_{\rm MT}\dot{M}_d, \label{eq:Fmt}
\end{align}
with $0 \leq f_{\rm MT} \leq 1$ and $\dot{M}_a$ and $\dot{M}_d$ are the mass loss rates for the accretor and donor respectively. This fraction depends on the properties of the binary prior to mass transfer and varies over time. We use a fixed fraction prescription in \texttt{COMPAS}, so the fraction of mass accreted is assumed to be constant over the mass transfer phase. $f_{\rm MT} = 1$ corresponds to fully conservative mass transfer, so all the mass lost from the donor is accreted. $f_{\rm MT} < 1$ is non-conservative and $f_{\rm MT} = 0$ is fully non-conservative. We choose to explore the full range of possibilities in \texttt{COMPAS}. We also explore the default setting, where the mass transfer efficiency is determined by the ratio of the Kelvin–Helmholtz thermal timescale of the two stars, 
\begin{align}
    \tau_{\rm KH} = \frac{E_{\rm int}}{L}.
\end{align}
This is the time taken for the star's internal energy, $E_{\rm int}$, to be radiated at its current luminosity, $L$. This ratio translates to a fixed mass ratio, $0 \leq f_{\rm MT} \leq 1$, but depends in the initial mass ratio and orbital separation \citep{2015ApJ...805...20S}. The maximum accretion rate onto a stellar accretor is, 
\begin{align}
    \dot{M}_a = \frac{C M_a}{\tau_{\rm KH, a}},
\end{align}
where the default factor $C=10$ is assumed to account for the expansion of the star due to mass loss \citep{1972AcA....22...73P, 1977PASJ...29..249N, 2002MNRAS.329..897H,2015ApJ...805...20S}. This is calculated for every mass transfer event and is non-constant. 
Accretion onto the compact objects, such as black holes and neutron stars, is assumed to be Eddington limited such that, 
\begin{align}
    \dot{M}_a = \dot{M}_{\rm edd} = 1.5 \times 10^{-8} \frac{R_*}{10 \rm km} \frac{M_\odot}{\rm yr}, 
\end{align}
where the black hole radius, $R_*$, is the Schwarzschild radius. The Eddington accretion limit occurs when the BH accretion disk reaches a maximum luminosity beyond which radiation pressure will overcome gravity, and the infall of material outside of the object will be halted. \citet{2020ApJ...897..100V} show that it is possible to break this limit via stable mass transfer or CE accretion, known as super-Eddington accretion. In the unstable mass transfer scenario, for simplicity, the CE phase is parameterised by \texttt{COMPAS} in the context of energy balance with the $\alpha_{\rm CE}$ formalism \citep{2010ApJ...716..114X} \citep[see also][and references therein]{2013A&ARv..21...59I}. The CE efficiency, $\alpha_{\rm CE}$, describes the efficiency of converting the orbital energy into kinetic energy to unbind the CE. This unbinding energy is defined as, 
\begin{align}
    E_{\rm bind} = \alpha_{\rm CE}\left(\frac{GM_d M_a}{2a_i} - \frac{GM_{\rm d, core}M_{\rm a, core}}{2a_f} \right) \label{eq:CE_alpha},
\end{align}
where G is the gravitational constant, $M_d$ is the mass of the pre-CE donor, $M_a$ is the mass of the accretor, $M_{\rm d, core}$ is the mass of the donor’s core, $a_i$ and $a_f$ are the pre- and post-CE orbital separations respectively. Increasing $\alpha_{\rm CE}$ increases the unbinding energy, so the CE can be ejected more efficiently at wider binary separations. Too small values of $\alpha_{\rm CE}$ would require the binary to shrink in order to obtain enough unbinding energy \citep{1976IAUS...73...35V,1979IAUS...83..401T, 1984ApJ...277..355W, 2013A&ARv..21...59I, 2021A&A...650A.107M,2022ApJ...937L..42H, 2022MNRAS.517.4034S}. \citet{1990ApJ...358..189D} proposed an alternative version to the $\alpha$-formalism for the binding energy, introducing a new parameter, $\lambda$, to characterise the central concentration of the donor's envelope,
\begin{align}
    E_{\rm bind} = -\frac{GM_dM_{\rm env}}{\lambda a_i r_L}, \label{eq:CE_lambda}
\end{align}
where $M_{\rm env} = M_\textrm{d} - M_{\rm d, core}$ is the mass of the envelope and $r_L = R_L/a_i$ is the ratio of the Roche lobe radius and pre-CE orbital separation. The post-CE orbital separation is determined by substituting Eq.~\ref{eq:CE_lambda} into Eq.~\ref{eq:CE_alpha}, 
\begin{align}
    \frac{a_f}{a_i} = \frac{M_{\rm d, core}M_a}{M_d} \frac{1}{M_a + 2M_{\rm env}/\alpha_{\rm CE} \lambda r_L}.
\end{align}
Due to lack of understanding and convenience in population synthesis calculations, $\alpha_{\rm CE}$ and $\lambda$ are set to constants, but \textit{should} be varied depending the stellar and binary parameters \citep{2010ApJ...716..114X}. For $0 \leq \alpha_{\rm CE} \leq 1$, the only source of energy to unbind the CE is the orbital energy. However, various hydrodynamical simulations over the years have shown that the orbital energy alone might not be sufficient for envelope removal \citep{2013A&ARv..21...59I,2023LRCA....9....2R}. Additional energy sources are required ($\alpha_{\rm CE} > 1$) such as recombination energy from evolved giant star envelopes \citep{2015MNRAS.447.2181I, 2022MNRAS.516.4669L} or the giant stars can launch jets as they accrete mass via an accretion disk \citep{2016ApJ...816L...9O,2017MNRAS.470.2929M}. 
In this paper, we explore $0.1 \leq \alpha_{\rm CE} \leq 100$. Although our upper limit seems extreme, it should be noted that \citet{2022ApJ...937L..42H} implements a two-step CE formalism which is equivalent to $\alpha_{\rm CE} > 10$ in most cases (see their Fig.~5). This emphasises the need for additional sources of energy, but it may also suggest that the current $\alpha_{\rm CE}$-formalism in \texttt{COMPAS} is wrong. As such, we choose to include $\alpha_{\rm CE} = 100$ as an exploratory parameter. See further discussion of this in Section~\ref{Summary}. 

Progenitors of GW sources provide an alternative path for comparison between models and observations. \citet{2014MNRAS.442.1980Z} uses the binary population synthesis code by \citet{2000MNRAS.315..543H} and \citet{2002MNRAS.329..897H} to study high-mass X-ray binaries (HMXB) and the correlation between their apparent luminosity and displacement from star clusters. A high value of  $\alpha_{\rm CE} = 0.8 - 1.0$ is preferred when comparing to observations by \citet{2004MNRAS.348L..28K}. However, the ambiguity in the definition of the donor core mass boundary introduces a large uncertainty (almost two orders of magnitude), which translates directly to the expected value of $\alpha_{\rm CE}$. \citet{2023MNRAS.524..245R} uses \texttt{COMPAS} to study black hole high-mass X-ray binaries (BH-HMXB). They find for the case where $f_{\rm MT} \leq 0.1$ (and $f_{\rm WR} = 0.1$, explained in Section \ref{fWR}), that $8\%$ of BH-HMXBs end up as BBHs or BHNSs. 
\citet{2007A&A...467.1181D} attempts to fit the mass transfer efficiency fraction for massive close binaries. They find a wide spread in $f_{\rm MT}$ and conclude that a single value is not valid. Therefore, more electromagnetic observations are required to constrain modelling degeneracies of massive binaries. 

There have been many studies on the mass transfer and CE phase. Many of these have explored how varying these parameters affect the BBH merger rate, the BH mass distribution range and which binary formation channels dominate these BBH mergers \citep{2020ApJ...897..100V,2022ApJ...931...17V, 2022ApJ...940..184V}. Many of these studies also find that additional energy sources are necessary to produce BBH mergers and other double compact objects (DCO) \citep{2021MNRAS.502.4877S, 2021ApJ...910..152Z,2021A&A...650A.107M,2023MNRAS.523..221G,2024ApJ...963...80D}. 
To infer the formation channels of BBH mergers, many studies further explore the connection between the models and data \citep{2023ApJ...955..127C}. \citet{2021MNRAS.505.3873B} find the preferred parameter range for the CE and mass transfer efficiency is $[4,7]$ and $[0.7,0.8]$ respectively. They also showed that increasing the CE efficiency increases the uncertainty on the local BBH merger rate density. \citet{2019ApJ...883L..45F} find that the final orbital separation for their BNS's to merger within a Hubble time translates to $\alpha_{\rm CE} \sim 5$. \citet{2021ApJ...920L..13B} and \citet{2021A&A...649A.114G} also find a preference for $\alpha_{\rm CE} > 1$ for producing merger rates compared to the observed rates. However studies such as \citet{2021A&A...647A.153B},\citet{2021A&A...645A..54K} and \citet{2023MNRAS.523.5565C} find there is a preference for lower $\alpha_{\rm CE}$ ($0.2 < \alpha_{\rm CE} < 1$, $\alpha_{\rm CE} \leq 0.7$ and $\alpha_{\rm CE} = 0.5,1$ respectively). \citet{2021A&A...645A..54K} further questions the validity of the CE channel for producing massive BHs. They find that successful ejection of the CE is possible if the donor is a massive convective-envelope giant. 

We briefly mention the various studies that explore the uncertainties in parameters. But another uncertainty to note is in how these parameters are implemented in population synthesis codes. \citet{2021ApJ...922..110G} highlights the uncertainty in stellar evolution tracks by comparing the mass transfer and CE treatments in the detailed stellar evolution code, \texttt{MESA} \citep{2011ApJS..192....3P,2013ApJS..208....4P,2015ApJS..220...15P,2018ApJS..234...34P,2019ApJS..243...10P}, and rapid binary population synthesis code, \texttt{COSMIC} \citep{2020ApJ...898...71B}. The \texttt{COSMIC} models primarily form BBH mergers through the CE phase, while the stable mass transfer formation channel is preferred with \texttt{MESA}. Increasing $\alpha_{\rm CE}$ also increases the variation in the BBH merger rates between the models, due to variations in merger times. \citet{2022MNRAS.512.5717A} and \citet{2023MNRAS.525..706R} have compared population synthesis codes to explore the uncertainty in radial expansion of massive stars and the variation in the remnant mass. 

\subsubsection{Wolf--Rayet Winds ($f_{\rm WR}$)} \label{fWR}
Stellar winds are outflows of gas from the atmosphere of a star. Wolf--Rayet stars are evolved stars with high mass loss rates, where the hydrogen envelope of the stars are removed. 
To account for uncertainties in the mass-loss rates of Wolf--Rayet stars, in \texttt{COMPAS}, the Wolf--Rayet multiplier, $f_{\rm WR}$, is used to parameterise the mass loss rate \citep{2010ApJ...714.1217B}, 
\begin{align}
    \frac{dM}{dt} = f_{\rm WR}\times 10^{-13}L^{1.5}\left(\frac{Z}{Z_\odot}\right)^m \rm M_\odot yr^{-1},
\end{align}
where $M$ is the mass of the helium star, $L$ is the luminosity, $Z$ is the metallicity, $Z_\odot = 0.014$ \citep{2009ARA&A..47..481A} and $m=0.86$ \citep{2005A&A...442..587V}. Wolf--Rayet stars have been studied to understand their explosion mechanisms and determine if they are progenitors of gamma ray bursts \citep{2008A&A...484..831D}. 
More recently, Wolf--Rayet stars have been used to study GW sources \citep{2017A&A...603A.120V, 2018NewA...58...33B}. 
Decreasing $f_{\rm WR}$ leads to an increase in the BH masses \citep{2022arXiv220708837D, 2023MNRAS.524..245R}. In \texttt{COMPAS}, we choose the Wolf--Rayet multiplier range provided in \citet{2022MNRAS.517.4034S}, which accounts for the range of uncertainties in Wolf--Rayet mass-loss rates.

HMXBs are usually ideal candidates to study and constrain the effects of stellar mass loss in interacting binaries, as radiation driven winds are expected for giant stars. \citet{2021arXiv210709675S} finds that the orbital evolution of HMXBs depends on the wind velocity and mass ratio. Slow winds and high mass objects results in the binary shrinking, due to drag forces. In the case where there is no drag, the binary separation widens. GWs can provide an indirect approach to study the separation of binaries, as their progenitors include HMXBs. However, \citet{Liotine:2023ApJ} find that no currently observed HMXB is predicted to form a merging BBH with high probability. They use \texttt{COSMIC} \citep[][]{2020ApJ...898...71B} to conclude that the BHs in the detectable HMXBs are not likely to have masses $>35M_\odot$, the detectable HMXBs will not merge as BBHs within the Hubble time and the distribution of redshifts, masses and metallicities of detectable HMXBs and BBHs are different. \citet{2022ApJ...929L..26F} argue that discrepancies in the BH masses in BBHs and HMXBs are due to GW observational selection effects.  

\citet{2021ApJ...908..118N} and \citet{2023MNRAS.524..245R} find a preference for low $f_{\rm WR}$ ($<0.2$) for HMXB systems such as Cygnus X-1 \citep{2021Sci...371.1046M}. \citet{2022MNRAS.517.4034S} explores the correlated impact of model parameters on the merger rate. They identified a correlation between the Wolf-Rayet factor and metallicity, as increasing both parameters increases mass loss via stellar winds. They find most models that match the current observations have $f_{\rm WR} > 1$. This contradicts the previously mentioned studies due to \texttt{COMPAS} over-predicting the rates and masses of BBHs formed through chemically homogeneous evolution (see Section \ref{CHE}).  

\citet{2021MNRAS.505..663R} studies the CHE binaries as a formation channel for various $f_{\rm WR}$ with \texttt{COMPAS}. They find a correlation between the delay time and $f_{\rm WR}$, as increasing mass loss leads to wider binaries. This results in the merger rate decreasing with increasing $f_{\rm WR}$. For $f_{\rm WR}=1$, the CHE channel contributes $20$ $\rm Gpc^{-3} yr^{-1}$ to the total rate of $50$ $\rm Gpc^{-3} yr^{-1}$, which is an overestimation compared to GWTC-3. Due to the short delay times, the BBH merger rate peaks at higher redshifts ($z\approx 3-4$) compared to the peak SFR.   

\subsubsection{Specific AM loss ($f_\gamma$)} 
Mass transfer from a massive donor to a lower-mass accretor adds specific AM to the transferred mass, so conservation of AM requires that binary’s orbit shrinks. If mass transfer is non-conservative, angular momentum, $J_{\rm orb}$, is carried away by mass lost from the system. The change in AM is parameterised by $\gamma$,
\begin{align}
    \frac{\dot{J}_{\rm orb}}{\dot{M}_d} = \gamma \frac{J_{\rm orb}}{M_d}.
\end{align}
The orbital evolution of the binary is given by, 
\begin{align*}
    \frac{\dot{a}}{a} &= -2\frac{\dot{M}_d}{M_d}\left[1-f_{\rm MT}\frac{M_d}{M_a}-(1-f_{\rm MT})\left(\gamma + \frac{1}{2}\right)\frac{M_d}{M_d + M_a}\right] \\ 
    &= -2\frac{\dot{M}_d}{M_d}\left[1-\frac{f_{\rm MT}}{q}-(1-f_{\rm MT})\left(\gamma + \frac{1}{2}\right)\frac{1}{1+q}\right], \numberthis
\end{align*}
where $q=M_a/M_d$ is the mass ratio. \citet{2023ApJ...958..138W} introduces a new parametrisation, $a_\gamma$,  
\begin{align}
    a_\gamma = a_{\rm acc} + f_\gamma(a_{\rm L2} - a_{\rm acc}),
\end{align}
which is the effective decoupling radius between the accretor, $a_{\rm acc}$, and the L2 point, $a_{\rm L2}$. 
\texttt{COMPAS} default model assumes isotropic re-emission of matter during non-conservative stable mass transfer ($f_\gamma = 0$). In this model, mass lost from the donor is transported near the accretor and ejected via fast isotropic winds. Hydrodynamical simulations indicate that this mass is lost between the accretor and L2 point \citep{2018ApJ...863....5M, 2018ApJ...868..136M}. $f_\gamma$ determines the point where this AM is lost. If the donor star is growing rapidly with respect to its Roche Lobe, the system will lose mass through the outer Lagrangian point (L2), which causes the orbit to shrink. This corresponds to $f_\gamma=1$. AM loss directly from the accretor corresponds to $f_\gamma = 0$ \citep{2023arXiv231115778C, 2023MNRAS.524..245R}. We consider $0 \leq f_\gamma \leq 1$. It is possible for cases outside of this range such as intense feedback radiation from the accretor for mass loss at or near the the inner Lagrangian point, L1, ($f_\gamma < 0$) or magnetic fields that keep material in co-rotation beyond the L2 point ($f_\gamma \geq 1$) \citep{2020ApJ...895...29M,2023ApJ...958..138W}.

\citet{2023MNRAS.519.1409L} and \citet{2021A&A...650A.107M} find that their models predicted high mass transfer rates, which lead to mass loss through the L2 point. \citet{2018ApJ...863....5M, 2020ApJ...895...29M} argue that the specific AM is ejected between the accretor and the L2 point. \citet{2023ApJ...958..138W} uses \texttt{COMPAS} to study the stable mass transfer channel. They find that increasing $f_\gamma$ leads to a reduction in the number of systems that undergo stable mass transfer. Increasing $f_{\rm MT}$ while decreasing $f_\gamma$ increases the likelihood of stable mass transfer. For $f_\gamma \gtrsim 0.7$, nearly all systems experience CE evolution. 

\subsection{Chemically homogeneous evolution} \label{CHE}
In the chemically homogeneous evolution (CHE) channel, the stars do not expand. They are metal poor and rapidly rotating in tight binaries. The rotation efficiently transports hydrogen to the core and helium to the envelope. This evolves and forms a Wolf--Rayet star, which contracts rather than expands. Wind-driven mass loss does not widen the binary significantly for low metallicity systems and there is no mass transfer, with the exception of massive over-contact binaries \citep{2016A&A...588A..50M}. Highly spinning DCOs would produce GWs in the age of the Universe \citep{CHE, 2022PhR...955....1M, 2023arXiv231115778C}. \citet{2009A&A...497..243D, 2010ASPC..435..179D} introduces the concept of CHE to justify short period massive binaries such as the HMXB M33 X-7 \citep{2007Natur.449..872O}.

\citet{2016MNRAS.458.2634M} find that the CHE channel significantly contributes to the BBH merger rate ($10.5 \pm 0.5$ $\rm Gpc^{-3} yr^{-1}$ at redshift 0). These systems typically have a total mass ranging between $50-110$ $M_\odot$. \citet{2016A&A...588A..50M} explores the parameter space for forming CHE binaries. They find that at low metallicity, BBHs will merge within the age of the Universe due to weaker winds ($t_{\rm delay, min} = 0.4$\,Gyr for $Z_\odot/50$). There is strong evidence for increasing masses and mergers with decreasing metallicity \citep{2010ApJ...714.1217B, 2010MmSAI..81..302B, 2015A&A...574A..58K, 2015MNRAS.451.4086S, 2019MNRAS.490.3740N}, and for masses in the pair instability SN gap \citep{2020ApJ...897..100V,2024MNRAS.529.2980W}.\footnote{Due to the production of electron positron pairs in the core of massive stars, the reduction of pressure followed by gravitational collapse leads to oxygen burning and results in a partial or complete explosion, which leaves behind no compact remnant \citep{1964ApJS....9..201F,PhysRevLett.18.379}.} If these mergers were detected, they could probe the evolution of massive stars in the early Universe. In addition to calculating the merger rate and mass distribution for the CHE channel, \citet{2020MNRAS.499.5941D} finds that the cosmological merger rate ($5.8$ $\rm Gpc^{-3} yr^{-1}$ at redshift 0) can vary up to 40\% for extreme deviations in the metallicity-specific SFR. 

In \texttt{COMPAS}, there are two CHE modes, `optimistic' and `pessimistic', which refer to how a CH star evolves on the MS \citep{2021MNRAS.505..663R}. 
In the `optimistic' mode, if the star begins the main sequence with fast enough rotation to evolve as a CHE star, it is assumed the star continues rotating fast enough to remain CH throughout the MS and the spin is not checked. For the CHE mode `pessimistic' the spin is checked at every timestep, and if it has slowed below the CH metallicity-dependent rotational frequency threshold, it is immediately converted to a regular MS star and evolve to the end of the MS \citep{2022ApJS..258...34R}. We choose to study both variations in \texttt{COMPAS}, as well as no CHE, to determine if the CHE can be ruled out as a formation channel for GW sources. 

\subsection{Remnant mass prescription}
GWs can be used to infer the underlying mass distribution of BBHs, which depends on the metallicity and the type of explosion mechanism. Their formation can be described in three stages; stellar collapse and bounce, convective engine, and post-explosion fallback. Massive stars undergo a series of successful burning stages, building an iron core until electron degeneracy pressure can no longer support this. Electron capture and dissociation of the core elements into alpha particles accelerates the core collapse. This collapse is halted when the core reaches nuclear densities and nuclear forces cause the core to bounce and emit a shock outwards. This results in a mixing of the hot and cold layers in the star, known as convection. The convective engine provides the means to convert potential energy released in the collapse into explosion energy in the form of a supernova \citep{fryer2012compact, 2021ARep...65..937F}. The commonly accepted supernova explosion mechanism is neutrino driven, where some of the neutrinos emitted during the core collapse are reabsorbed, powering an explosion \citep{2007PhR...442...38J, 2009ApJ...694..664M,2012ARNPS..62..407J,2017hsn..book.1095J}. We choose to explore all remnant mass prescriptions available in the \texttt{COMPAS} version used in this paper, and describe them below. 

\citet{fryer2012compact} provide analytic prescriptions for estimating the remnant mass in population synthesis codes, based on at the time recent understanding of supernova and gamma-ray burst explosions. In their model/fit, the remnant mass depends on the core, fallback mass and explosion mechanism. They explore two neutrino-driven explosion mechanisms-- the rapid convection, which leads to explosion in the 250 ms after bounce, and the delayed convection, which can occur over a longer timescale. \citet{2022ApJ...931...94F} altered the remnant mass prescription from \citet{fryer2012compact} to study the growth rate of convection. They introduce a new parameter to describe the mixing growth time for convective instabilities in their remnant mass models, denoted $f_{\rm mix}$. A fast growth time, $f_{\rm mix}=4.0$, and a slow growth time, $f_{\rm mix}=0.5$, correspond to the rapid and delayed models respectively from \citet{fryer2012compact}. The growth time determined the supernova explosive energy \citep[see Fig.~4 in][]{Fryer:2015fis}. Whether a star undergoes a successful or failed supernova provides clues on the final remnant object. However, there are uncertainties surrounding these mechanisms \citep{2021hgwa.bookE..16M}. GW detection from a core collapse SN can be used to distinguish between explosion mechanisms \citep{2016PhRvD..94l3012P, 2023arXiv231118221P}

\citet{mandel2020simple} develop probabilistic prescriptions for the remnant mass based on results from 3D supernovae simulations in \citet{2016MNRAS.460..742M}. Instead of a single value prescription, the BH mass branches over a range of core masses, depending on the fallback strength. If the BH is not formed by complete fallback, the mass follows as normal distribution. These models are more realistic as they capture the stochasticity of stellar evolution. 

\citet{schneider2021pre} also applies the  \citet{mandel2020simple} semi-analytic SN prescription, but the resulting remnant mass prescription differs. The model focuses more on the removal of the hydrogen-rich envelope of the progenitor star and the evolution steps before core collapse. They also explore the amount of fallback. Their models overestimate the BH masses by 10\% if the BHs are not formed by direct collapse. They do not account for any convective boundary mixing beyond the core helium burning. 

\citet{2023ApJ...950L...9S} uses the models in \citet{schneider2021pre} and the stellar evolution code \texttt{MESA} \citep[][]{2011ApJS..192....3P} to explore the black hole mass distribution and chirp mass distribution of BBH mergers. The ability of stars to explode in the delayed neutrino-driven SN mechanism is characterised by the compactness of the progenitor cores \citep{2011ApJ...730...70O, 2016ApJ...818..124E}. This depends on the mass, radius and a nuclear reaction network, and a high compactness indicates unsuccessful explosions and collapse into a BH. Thus, the chirp mass distribution is bimodal and the model predicts a lack of chirp masses in the range $9-13M_\odot$. Mass accretion can influence the final outcome of core collapse and final remnant mass. \citet{2024arXiv240303984S} studies models with \texttt{MESA} that allow for multiple mergers. They find BH masses in the pair instability SN gap, which can explain events such as GW190521. They note that for a more complete picture of stellar mergers, further detailed multi-dimensional, hydrodynamic simulations and a systematic exploration of the relevant parameter space is needed.

\subsection{Summary} \label{Summary}
We outline the parameters and model assumptions varied in \texttt{COMPAS} in Table \ref{tab:parameters}. We note that the initial conditions and distributions for the binary systems are set to default (except for the metallicity where we use \textsc{Shark}), as shown in Table~\ref{tab:fid_case}. Future studies may choose to vary these parameters. 
\begin{table*}
    \centering
    \begin{tabular}{c|c|c}
        \hline
        Parameters & Value & Default \\
        \hline
        Remnant mass prescription & [MULLERMANDEL, SCHNEIDER2020, FRYER2012, FRYER2022] & MULLERMANDEL \\
        Chemical homogeneous evolution & [NONE, PESSIMISTIC,OPTIMISTIC] & PESSIMISTIC \\
        Mass transfer fraction ($f_{\rm MT}$) &  [0.1,0.2,0.5,1] & 0.5 \\
        Common envelope efficiency ($\alpha_{\rm CE}$) & [0.1,1,10,100] & 1.0 \\
        L2 point AM loss linear fraction ($f_\gamma$) & [0,0.25,0.5,0.75,1] & 0.5 \\
        Wolf-Rayet winds multiplier ($f_{\rm WR}$) & [0.1,0.5,1,5,10] & 1.0 \\
        \hline 
    \end{tabular}
    \caption{\texttt{COMPAS} hyper-parameters explored in this paper. Note that some prescriptions are only used in conjunction with others. The mass transfer fraction is only applied if the mass transfer accretion efficiency prescription is set to FIXED\_FRACTION. The default is THERMAL. Similarly, the AM loss linear fraction is only applied when the mass transfer AM loss prescription is MACLEOD\_LINEAR. The default is ISOTROPIC ($f_\gamma = 0$). Refer to Tables \ref{tab:all_models},\ref{tab:all_models_2} and \ref{tab:all_models_3} in Appendix \ref{Appendix} for all relevant prescriptions used for each model.}
    \label{tab:parameters}
\end{table*}
As discussed, we are aware of the limitations and approximations made in \texttt{COMPAS} and other population synthesis codes, that may cause discrepancies in the outputs. We summarize these limitations and the key questions below:
\begin{enumerate}
    \item The binding energy formalism utilises a fixed $\alpha_{\rm CE}$, but is this a valid approach? The parametrisation of $\alpha_{\rm CE}$ is a point of contention, particularly when additional energy sources are available. It is also not clear how these would scale with the orbital separation and mass, so the $\alpha$-formalism is not an accurate representation of CE ejection. \citet{2022ApJ...937L..42H} proposes a two-step CE process, which depends on the evolutionary stage of the donor and the companion's mass. They find that the predicted post-CE separation is wider compared to the $\alpha_{\rm CE}$ formalism. This may pose a problem for BBHs merging within Hubble time. \citet{2024arXiv240213180P} has recently implemented fits for the convective envelope mass and associated values of $\lambda$ to be used in conjunction with the two-step CE formalism. They find the effective combined value of $\alpha_{\rm CE} \lambda$ can range from $\sim 10^{-3}$ to $10^3$ when converted to the $\alpha$-formalism.
    \item Are the masses in \texttt{COMPAS} overestimated and how does this impact the CHE binaries? CHE stars evolve from the hydrogen-ZAMS directly to the helium-ZAMS. When a CHE star is close to the end of the MS, it is already rich in helium, and may start to resemble a helium/ Wolf--Rayet star. However, \texttt{COMPAS} implements the same wind loss prescription throughout the MS, instead of a specific Wolf--Rayet wind prescription with higher mass loss rates. This change is applied to the \texttt{MESA} CHE modelling \citep{2016A&A...588A..50M,2020MNRAS.499.5941D,2023arXiv231114041M}.
    \item Mass loss via radiation-driven winds is still an uncertainty in massive star progenitor models and accurate predictions are necessary for predicting their evolutionary paths. \citet{2023A&A...676A.109B} derive radiation-driven mass-loss from hot, massive stars depending on their fundamental stellar parameters and implement this in \texttt{MESA}. They find that lower mass-loss rates potentially allow for the creation of high-mass black holes even at higher metallicities compared to previous models.
    \item At ZAMS masses $> 50M_\odot$, \citet{2000MNRAS.315..543H} smoothly extrapolate the \citet{1998MNRAS.298..525P} model grid up to $150M_\odot$, which is not well tested outside of the prescribed mass range. We find the contribution of $M_{\rm ZAMS} > 50M_\odot$ to BBH mergers in \texttt{COMPAS} is non-negligible. This mass range could be significant as LVK detects more GW events in the pair instability mass gap \citep{2024MNRAS.529.2980W}. Significant improvement on this extrapolation method is far from trivial, as the evolution of high mass binaries is still very uncertain. This is reflected by the discrepancies between population synthesis codes in their stellar tracks \citep{2020MNRAS.497.4549A, 2022MNRAS.512.5717A, 2022A&A...668A..90A}. 
    \item There are many uncertainties in the mass transfer efficiency and AM loss, which influence the Roche lobe response to mass loss. Uncertainties in how a donor star responds to mass loss, combined with the Roche lobe response uncertainties contributes to the lack of understanding around the mass transfer stability boundary as a function of donor mass and radius \citep{2010ApJ...717..724G, 2015ApJ...812...40G, 2020ApJ...899..132G,2023A&A...669A..45T}.
    \item Is the AM loss occurring closer to the accretor or the L2 point? And should the implementation of $f_\gamma$ vary for degenerate and non-degenerate accretors? When performing the analysis in this paper, split functionality for AM loss from degenerate and non-degenerate accretors was yet to be implemented in \texttt{COMPAS}. Physically, it is more accurate for $f_\gamma$ to be implemented differently for the primary and secondary star, particular when the accretor is a compact object. It is not clear how this would impact our current results, where we have only considered the AM loss near the non-degenerate object. \citet{2023arXiv230813146G} derive an analytic model for mass loss from the binary that is informed by observations of X-ray binary accretion disks and hydrodynamical simulations, which considers AM loss from a disk wind around a compact object accretor. They find their model predicts more AM loss compared to the standard case of isotropic re-emission and the orbital separation of a binary can be considerably affected by mass loss via a disk wind.
    \item The \texttt{COMPAS} version used in this paper does not consider tidal evolution, but it is believed to play a significant role in the evolution of massive stars \citep{2022ApJS..258...34R}. Tidal forces can bring the compact object into the envelope of the secondary which is now a giant star. This implies that \texttt{COMPAS} underestimates the number of BH-giant system mergers, while overestimating the number of BBH mergers \citep{2023MNRAS.523..221G}.  When performing the analysis in this paper, the newest version of \texttt{COMPAS} implemented tides (Kapil et al., in prep.). 
    \item \texttt{COMPAS} has two modes; single stellar evolution (SSE) and binary stellar evolution (BSE). The \citet{2000MNRAS.315..543H,2002MNRAS.329..897H} SSE algorithms are implemented in the BSE mode at all evolutionary phases and provides the stellar attributes for both components of the binary. However, the fitting formulae are not well behaved outside of the metallicity range $Z \in [10^{-4}, 0.03]$.
    \item The SSE tracks in \texttt{COMPAS} are based on non-rotating stars and do not account for effects on stellar evolution due to mild rotation. Thus, the modelling of rotational mixing, core overshooting (convection beyond the Schwarzschild criterion) and donor stripping is not accurate. Many studies had investigated the impact of these parameters on stellar structures using \texttt{MESA} and \texttt{PARSEC} \citep{2018ApJ...859..100C,2019MNRAS.485.4641C,2020A&A...637A...6L,2024A&A...682A.123T}. 
\end{enumerate}

\section{Predicting the GW merger rate for different COMPAS models} \label{Section 3}
In this section we demonstrate our method for rescaling the merger rate from a fiducial, carefully cross-matched, simulation of galaxy formation (via \textsc{Shark}) and binary population synthesis (from \texttt{COMPAS}). This allows us to produce predictions for the rate and properties of an ensemble of GW events for any model without the large computational cost of redoing the cross-matching from scratch.  From \citet{10.1093/mnras/stad1757}, the number of GW events per \textsc{Shark} galaxy that occur at time $t_{j}$ from stars born at time $t_{i}$ using \texttt{COMPAS} model $\alpha$ was written as
\begin{align}
    \frac{dN_{\rm GW}^\alpha}{dN_{\rm gal}}(Z,t_{i},t_{j}) &= \frac{\psi(Z,t_{i})}{M_{\rm tot}^\alpha} \Delta t_i \sum_n^{N_{\mathrm{binary}}} \vartheta(t_{i} + \tau_{n}^\alpha(Z), t_j, \Delta t_j), \label{f_eff}
\end{align}
where 
\begin{align}
\Delta t_j &= t_j - t_{j-1}, \\ 
\vartheta (t, t_j, \Delta t_j) &= \left\{ \begin{array}{cc} 
                 1 & \hspace{1.5mm} \text{if $t_j - \frac{\Delta t_j}{2} < t < t_j + \frac{\Delta t_j}{2}$,} 
\\
                0 & \hspace{1.5mm} \text{otherwise.} \\
\end{array} \right\} \label{eq:weight}
\end{align}
and $\psi(Z,t_{i})$ is the metallicity ($Z$) dependent SFR in the galaxy when the progenitor stars were born. 
$M_{\rm tot}^\alpha$ is a normalisation constant for a given model $\alpha$\footnote{$\alpha$ refers to some combination of hyper-parameters. In \citet{10.1093/mnras/stad1757}, it refers to changing the remnant mass prescription (see Table~\ref{tab:parameters}).} (see Eq 10 in \citealt{10.1093/mnras/stad1757}), which normalises the SFR based on the chosen stellar Initial Mass Function (IMF), in such a way that it cancels out the dependence of the number of GW events on the number of binaries that we simulate $N_{\mathrm{binary}}$, and any lower or upper mass cuts we apply when running \texttt{COMPAS}. $\vartheta (t, t_j, \Delta t_j)$ is a counter for the number of BBH mergers in \textsc{Shark} snapshot, $j$. $\tau_n^\alpha (Z)$ is the coalescence time of each binary outputted from the model $\alpha$. If the merger time, $t = t_i + \tau_n^\alpha (Z)$, falls in the $j^\mathrm{th}$ bin, a count is added to that bin, otherwise it does not contribute. Overall, this equation encapsulates that the rate consists of two contributions, one from the \textsc{Shark} host galaxy and the other from population synthesis (in our case from \texttt{COMPAS}).

\subsection{The effective factor for different models}
\begin{figure}
    \centering
    \includegraphics[width=\columnwidth]{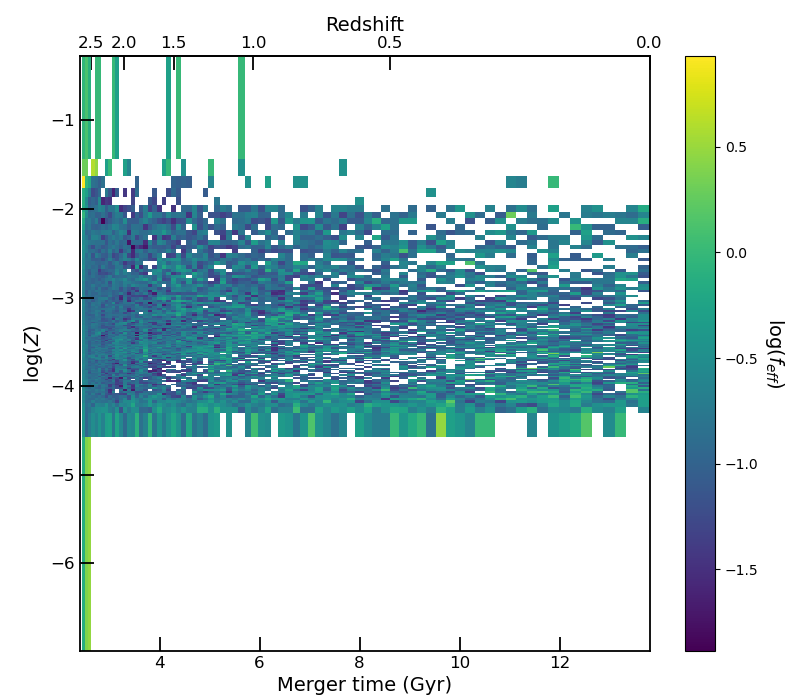}
    \caption{An example of the effective factor between two \texttt{COMPAS} models with different remnant mass prescriptions $\alpha=$ \protect\citet{fryer2012compact} and $\beta=$ \protect\citet{mandel2020simple} in bins of merger time and metallicity for binaries forming at redshift 2.5 (corresponding to $t_{i}\approx 2.62 \rm Gyr$). In this plot and the rest of our work, an un-equal metallicity binning is chosen such that the number of \textsc{Shark} galaxies in each metallicity bin in the same. Unshaded bins indicate regions of the parameter space where there are no gravitational waves forming in either one or both of the models.}
    \label{fig:feff_grid}
\end{figure}

Cross-matching the star formation history of every galaxy in a simulation with a COMPAS run is computationally expensive. To simplify this, we now demonstrate how we can rescale the rate.
If instead of considering GW events on a per galaxy basis we consider them in terms of number per metallicity and time $t_{i}$ and $t_{j}$ bins, we can write
\begin{align*}
    \frac{d^{3}N_{\rm GW}^\alpha}{dZ dt_i dt_j} &= \frac{d^{3}N_{\rm gal}}{dZ dt_i dt_j} \frac{dN_{\rm GW}^\alpha}{dN_{\rm gal}} \\
    &= \frac{d^{3}N_{\rm gal}}{dZ dt_i dt_j} \frac{\overline{\psi}(Z,t_i)}{M_{\rm tot}^\alpha} \Delta t_i \overline{ \sum_n \theta(t_i + \tau_{n}^\alpha(Z, t_i), t_j, \Delta t_j)}. \numberthis
\end{align*}
In the second line we have substituted in Eq.~\ref{f_eff} and use $\overline{\psi}$, $\overline{\theta}$ to acknowledge that this is the average of the star-formation history and Heaviside function over all galaxies that fall into the the bin $dZ dt_i dt_j$ (i.e., we assume that all galaxies in this bin produce the same number of gravitational waves, which is strictly true only for infinitesimally small bins). We now can introduce a new population synthesis model, $\beta$\footnote{$\beta$ now refers to the new \texttt{COMPAS} models in this paper and is some combination of the hyper-parameters in Table~\ref{tab:parameters}.}, where the merger rate is unknown. If we have the merger rate from model $\alpha$, we can take a ratio of the two models and rearrange,
\begin{align*}
    \frac{d^{3}N_{\rm GW}^\beta}{dZ dt_i dt_j} &= \frac{M_{\rm tot}^\alpha}{M_{\rm tot}^\beta} \frac{d^{3}N_{\rm GW}^\alpha}{dZ dt_i dt_j} \frac{\overline{ \sum_n \theta(t_i + \tau_{n}^\beta(Z, t_i), t_j, \Delta t_j)}}{\overline{ \sum_n \theta(t_i + \tau_{n}^\alpha(Z, t_i), t_j, \Delta t_j)}} \\
    &= \frac{M_{\rm tot}^\alpha}{M_{\rm tot}^\beta} \frac{d^{3}N_{\rm GW}^\alpha}{dZ dt_i dt_j} f^{\alpha \beta}_{\rm eff}(Z, t_i, t_j, \Delta t_j). \numberthis \label{eq:f_eff_2} 
\end{align*}
We hence find that retaining the information from \textsc{Shark} is not required as the star formation history, $\overline{\psi}(Z,t_i)$, and the galaxy distribution are independent of the population synthesis model. In practice we also expect that for most use cases $\frac{M_{\rm tot}^\alpha}{M_{\rm tot}^\beta} \approx 1$ since changes to the IMF between different literature values primarily affect stars that are too small to become binary black holes or neutron stars, although we retain it here for completeness.\footnote{Note this factor also needs to be included if one were to rescale between models where different lower or upper mass limits have been used to draw stars from the IMF, where a different mass ratio distribution for the initial binaries is assumed, or the overall binary fraction is changed} The only caveat then is on the free choice of binning used to evaluate what we call the `effective factor' $f^{\alpha\beta}_{\mathrm{eff}}$ between the two models --- although the dependence on galaxy simulation has been absorbed into our fiducial $\frac{d^{3}N_{\rm GW}^\alpha}{dZ dt_i dt_j}$, there is still the implicit assumption that the rate in each galaxy in the bin is the same. We find that this approximation remains valid over a range of metallicity binning choices, even for the quite wide bins we adopt as our default (100 un-evenly spaced bins for $\log Z \in [-7.0, -0.3]$). We show an example distribution of $f_{\rm eff}$ in these metallicity and merger time bins in Fig.~\ref{fig:feff_grid} for a single choice of binary formation redshift of 2.5 (corresponding to $t_{i}\approx 2.62 \rm Gyr$).

\subsection{Generating the merger rate}
Here we give a qualitative overview of our algorithm for predicting the GW merger rate for our different \texttt{COMPAS} models. 
\begin{enumerate}
    \item The starting point of our algorithm is a high-fidelity set of GW events simulated from a combination of \textsc{Shark} and \texttt{COMPAS} model $\alpha$. The detailed steps used for generating these are covered in \citet{10.1093/mnras/stad1757} and so will not be repeated here. Briefly, this is done by cross-matching the galaxy metallicities to the binary population metallicities outputted from \texttt{COMPAS}. We inject this population into the \textsc{Shark} galaxies and track their evolution until they merge using Eq.~\ref{eq:weight}. 
    \item From this, $\frac{d^{3}N_{\rm GW}^{\alpha}}{dZ dt_i dt_j}$ is calculated by taking the fiducial merger rate outputs and binning in terms of metallicity and birth and merger time. For the metallicity we use quantiles to determine the bin edges such that there is an even number of galaxies in each bin. For the birth and merger time, we simply use the 170 snapshots from \textsc{Shark} to define the 170 bins.
    \item Having established our metallicity, birth time and merger time bins, we take our full list of \texttt{COMPAS} binaries, read in the metallicity and evolution time, which is added to the birth time ($t_{\rm birth} + t_{\rm BBH} + t_{\rm coalescence}$), and add these to the corresponding bins. We apply this to both the fiducial and new model, take the ratio and multiply by the merger rate in (ii) to estimate the new merger rate, shown in Eq.~\ref{eq:f_eff_2}.
\end{enumerate}
\begin{figure}
    \centering\includegraphics[width=\columnwidth]{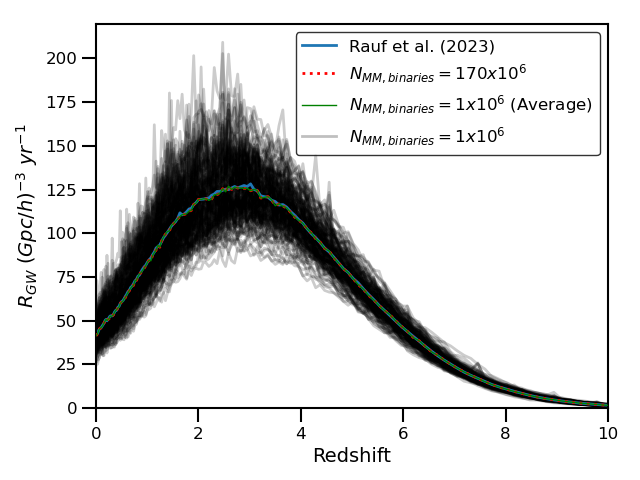}
    \caption{Volumetric BBH merger rate with remnant mass prescription from \citet{mandel2020simple} as a function of redshift. The blue line is the \citet{10.1093/mnras/stad1757} merger rate from running \texttt{COMPAS} and \textsc{Shark}. The red dashed line is the merger rate rescaled with the effective factor calculated using the full \texttt{COMPAS} run containing $170 \times 10^6$ binaries. The grey lines (whose overlapping nature visually appears as a shaded black region) are the rescaled merger rate where the effective factor has been calculated using 170 unique sub-samples of $10^6$ binaries from the full \texttt{COMPAS} run. The green line is the average over these sub-samples.}
    \label{fig:Merger_rate_err}
\end{figure}
Once we have our estimate for the number of gravitational waves in our new model $\beta$ we evaluate the volumetric merger rate at time $t_{j}$ by simply summing over the metallicity bins and birth time bins, 
\begin{align}
    R_{\rm GW}^\beta(t_{j}) = \frac{1}{V_{\mathrm{sim}}}\sum_{i = 0}^{170} \sum_{k = 0}^{100} \frac{d^{3}N_{\rm GW}^\beta}{dZ_{k} dt_i dt_j} \label{f_eff_total}, 
\end{align} 
where $V_{\mathrm{sim}}$ is the volume of our original \textsc{Shark} simulation.

For efficiency, we choose a sample $\tilde{N}_{\rm GW}^{\beta} = 10^6$ binaries from our $\beta$ \texttt{COMPAS} run to estimate the effective factor while using the full sample for our fiducial model, $\alpha$, where $N_{\rm GW}^{\alpha} =170\times10^6$.\footnote{Note that we draw only $\sim 10^6$ unique masses for this process from the IMF, but each one is randomly assigned a different metallicity from \textsc{Shark}, resulting in $\sim 170\times10^6$ unique \textit{combinations} of initial masses and metallicities, and so $\sim 170\times10^6$ unique binaries.} The merger rate is reduced by a factor of 170 but this is compensated for by the new normalisation in Eq.~\ref{eq:f_eff_2}, $\frac{M_{\rm tot}^{\alpha}}{M_{\rm tot}^{\beta}} \sim 170$. Thus, this formalism removes the dependence on the sample size. 
We test the validity of our rescaling procedure and using only $10^6$ binaries for model $\beta$ by splitting one of our fiducial \texttt{COMPAS} runs from \citet{10.1093/mnras/stad1757} (using the remnant mass model of \citealt{mandel2020simple}) into 170 by $10^6$ sub-samples. We then use these to estimate the rescaled merger rate for this model given $\alpha=\,$\citet{fryer2012compact}, and look at the spread in the merger rate. We show in Fig.~\ref{fig:Merger_rate_err} that overall the rescaling method using equal numbers of binaries for models $\alpha$ and $\beta$ recovers the \citet{10.1093/mnras/stad1757} merger rate from our high-fidelity cross-matching, but also that using only $10^6$ binaries for model $\beta$ returns, on average, an unbiased estimate of the merger rate. The use of a smaller number of binaries in the rescaling procedure does introduce additional Poisson scatter in the merger rate, which we propagate through into our model comparison/evaluation (see Section~\ref{Section 4}).

\begin{figure}
    \centering
    \includegraphics[width=\columnwidth]{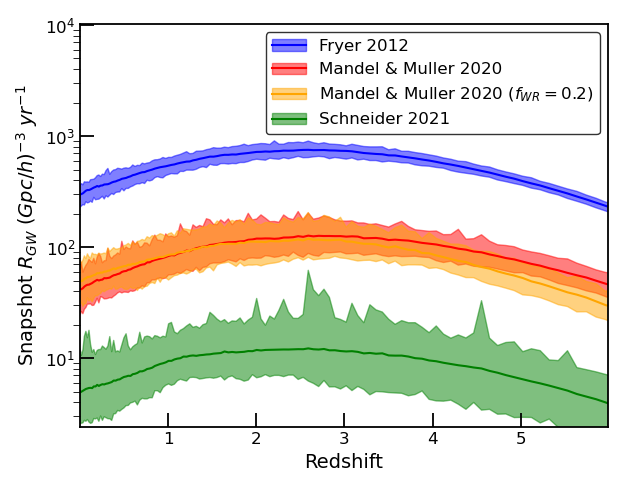}
    \caption{Volumetric rate plot as a function of redshift from \citet{10.1093/mnras/stad1757} recreated with the effective factor scaling method. The error on the plots are the range of merger rates from sampling from each \texttt{COMPAS} run 170 times. We use a logarithmic scale for the y-axis, as the merger rates for the models spans several orders of magnitude.}
    \label{fig:RGW_vol}
\end{figure}

In Fig.~\ref{fig:RGW_vol}, we show the volumetric BBH merger rates in the \textsc{Shark} simulations using the effective factor approximation for four different models. In all cases we find that we are able to recover rates from \citet{10.1093/mnras/stad1757} very well. We note that the plot is noisier than in \citet{10.1093/mnras/stad1757} due to the smaller sample of BBHs taken from each \texttt{COMPAS} run, but again this noise arises only from Poisson sampling error. 




\section{Bayesian inference and model comparison} \label{Section 4}
In this section we outline our procedure for comparing different \texttt{COMPAS} models to data from GWTC-1 \citep{2019PhRvX...9c1040A}, GWTC-2 \citep{2021PhRvX..11b1053A} and GWTC-3 \citep{2023PhRvX..13d1039A} including our rescaling methodology. We validate this method by comparing the likelihoods given the data for the four models from \citet{10.1093/mnras/stad1757} for which we have both `brute-force' measurements of the GW rate and our rescaled calculations.
\subsection{Selection criteria} \label{Selection_criteria}
The Bayesian analysis will depend strongly on $N_{\rm obs}$, which is the number of \textit{confident} GW detections. How confident we are in the GW signals and the validity of the sources is determined by the following ranking statistics: 
\begin{enumerate}
    \item Signal-to-noise ratio (SNR): This is proportional to the signal amplitude divided by the noise amplitude. The higher the SNR, the stronger the signal compared to the noise, and more likely that the signal will be detected. This is the main ranking statistic when the signal shape is known from theory, such as the signal for a BBH merger. We use the network SNR, which is the square root of the sum of the squares of SNRs from the triggers which form the GW event (only from the detectors that produce triggers contributing to the event; \citealt{2022CQGra..39u5012C}). The SNR, $\rho$, in a single detector is defined as,
    \begin{align}
       \rho^2 = \langle h|h \rangle = 4\textrm{Re}\left[\int_{f_{\rm min}}^{f_{\rm max}}\frac{h^*(f)h(f)}{S(f)} df\right],
    \end{align}
    where $h(f)$ is the waveform measured by the detector, $h^*(f)$ is its Hermitian conjugate, $S(f)$ is the power spectrum density, $\textrm{Re}[]$ is the real component of the inner product, and $f_{\rm min}$ and $f_{\rm max}$ are the minimum and maximum limits for the considered frequency range \citep{2015PhRvD..91b3005F,2018MNRAS.477.4685B}. 
    \item False alarm rate (FAR): Once the threshold for the SNR is set, the FAR can be determined, providing an alternative way to estimate the significance of the event \citep{zheng2021needle}. The FAR is the number of noise events with a signal equal to or greater than the GW event, divided by the total duration of the background data. The false alarm rate of single-detector candidates given $\rho$ is,
    \begin{align}
        \textrm{FAR}(\rho) = \int_\rho^{\rho_{\rm max}} \Lambda_np_n(\rho')d\rho', 
    \end{align}
    where $\Lambda_n$ is the mean Poisson rate of noise triggers and $p_n(\rho')$ is the probability densities describing the distribution of $\rho$ \citep{2017CQGra..34o5007C}. 
    \item $p_{\rm astro}$: The probability that a particular event is of astrophysical origin, as opposed to noise fluctuations or terrestrial, is defined by $p_{\rm astro}$. It is measured by comparing the GW event rate to the background event rate for a fixed SNR and FAR. The probability of a trigger, $x$, being astrophysical is,
    \begin{align}
        p_{\rm astro}(x) = \frac{p(x|\mathcal{S})\pi(\mathcal{S})}{p(x|\mathcal{S})\pi(\mathcal{S})+p(x|\mathcal{N})\pi(\mathcal{N})}, \label{eq:pastro}
    \end{align}
\end{enumerate}
where $\mathcal{S}$ and $\mathcal{N}$ describe the signal and noise model, as well as the distribution of signal and noise sources respectively \citep{2017CQGra..34o5007C, 2023arXiv230317628B}. 

For our most rigorous selection, we combine $\rm FAR \leq 0.25\,yr^{-1}$ and $p_{\rm astro} \geq 0.9$, as the FAR alone will not provide information on the sources being astrophysical, but rather the behaviour of the signal background. The $p_{\rm astro}$ reveals more events in the population rich areas of the parameter space, and is a more realistic metric for setting threshold for electromagnetic follow-up \citep{2018ApJ...861L..24L, 2022CQGra..39e5002A}. The caveat of $p_{\rm astro}$ is that that independent pipelines will obtain different $p_{\rm astro}$ due to a difference in data reduction techniques. \citet{2023PhRvD.108h3043B} proposes an unified $p_{\rm astro}$ by combining triggers from all pipelines, which incorporates extra information about a signal. \citet{2020PhRvD.102h3026G} also emphasises the need for a pipeline independent $p_{\rm astro}$, as they show that the posterior distribution of the population hyper-parameters can change with an astrophysically motivated prior. Given these caveats, we also investigate a simpler, broader selection using only an $SNR\geq8$ cut, which matches that used in evaluating our detection probability function $p_{\rm det}(\theta)$ for our models and provides us with a larger samples of events. However, the downside of this is that it potentially allows events into our sample that may not be astrophysical. In these cases, we allow our Bayesian framework to decide if these events should be ruled out. 

Our supersample of GW events for querying the probability distribution function (PDF) for a set of parameters in each \texttt{COMPAS} model (see Section~\ref{model_selection}) consists of confirmed BBHs and BH-mass-gap events such as GW190521 \citep{2020PhRvL.125j1102A} and GW190814 \citep{2020ApJ...896L..44A}. This gives a total of 93 catalog GW detections. Of these, the following GW events are \textit{not} included in our analysis for the following reasons:
\begin{itemize}
    \item GW170817: Confirmed BNS \citep{2017PhRvL.119p1101A}
    \item GW190425: Confirmed BNS \citep{2020ApJ...892L...3A}
    \item GW190426\_152155: Not astrophysical \citep{2021PhRvX..11b1053A}
    \item GW190909\_114149: Not astrophysical \citep{2021PhRvX..11b1053A}
    \item GW191219\_163120: Confirmed NSBH \citep{2023PhRvX..13d1039A}
    \item GW200115\_042309: Confirmed NSBH \citep{2021ApJ...915L...5A} 
\end{itemize}
We then apply our two sets of selection criteria. We have 79 GW events with $\rm SNR \geq 8$\footnote{We use the network SNR provided in \url{https://gwosc.org/eventapi/html/GWTC/?page=1&pagesize=100}.} and 60 GW events with $p_{\rm astro} \geq 0.9$ and $\rm FAR \leq 0.25 yr^{-1}$. Table~\ref{tab:GW_list} lists the events used with each selection criteria.  
Out of the 87 BBH GW events we could have included (93 less the 6 listed above) the following GW events do not fit either criteria: 
\begin{multicols}{2}
\begin{itemize}
    \item GW190403\_051519
    \item GW190514\_065416
    \item GW191113\_071753
    \item GW200208\_222617 
    \item GW200220\_061928
    \item GW200306\_093714 
    \item GW200308\_173609
    \item GW200322\_091133
\end{itemize}
\end{multicols}

\subsection{Fitting LVK posterior for various \texttt{COMPAS} models} \label{model_selection}
Our basic fitting methodology is based on the widely-used Bayesian framework \citep[e.g.,][]{2015ApJ...810...58S, 2019ApJ...886...25B, 2019MNRAS.486.1086M, 2022hgwa.bookE..45V,2023arXiv230101312P}. Given the BBH GW data from some observing runs, $\mathcal{H}$, the posterior for a model specified via hyper-parameters $\lambda$ (for instance a certain mass transfer efficiency $f_{\rm MT}$ or specific AM loss $f_{\gamma}$), takes the form of an inhomogeneous Poisson distribution,
\begin{equation}
\mathcal{L}(\lambda, N_\lambda|\mathcal{H}) = \pi(\lambda) e^{-\mu_\lambda}N_\lambda^{N_{\rm obs}}\prod_k^{N_{\rm obs}} \mathcal{I}^k. \label{eq:P_unmarginalised}
\end{equation}
$N_{\lambda}$ is the number of GW events predicted by the model in a given time-frame, while $\mu_{\lambda}$ is the number the model suggests we would actually detect. $N_{\rm obs}$ is the actual number of detected events in our dataset.
We assume that \textit{a priori} all hyper-parameter combinations are equally likely, so $\pi(\lambda)=1$. Finally, $\mathcal{I}^k$ denotes the likelihood of observing GW event $h^{k}$ given the hyper-parameters $\lambda$, written as a marginalisation over the binary properties predicted by that model, 
\begin{align}
    \mathcal{I}^k = \int \mathcal{L}(h^k|\theta)p_{\rm sim}(\theta|\lambda)d\theta = \frac{1}{N_s^k} \sum_{i=1}^{N_s^k} \frac{p_{\rm sim}(\theta_i^k|\lambda)}{\pi^k(\theta_i^k)},
\end{align}
where $\theta=\{\mathcal{M}_c, q, z\}$ are the GW event parameters chirp mass, mass ratio and redshift respectively and $p_{\rm sim}(\theta|\lambda)$ is the probability distribution function for the event-level parameters generated by \texttt{COMPAS} using the \textsc{Shark} metallicity as input. The latter sum arises from the use of the $N_s^k$ publicly available posterior samples to represent the likelihood of event $k$, $\mathcal{L}(h^{k}|\theta)$. Formally $p_{\rm sim}(\theta|\lambda)$ is obtained by marginalising over all the metallicities and birth times used that can give rise to an event with parameters $\theta$:
\begin{equation}
p_{\rm sim}(\theta|\lambda) = \int_{0}^{\infty} \int_{0}^{t_{j}}p_{\rm sim}(\mathcal{M}_c, q, z(t_{j}) | Z, t_{i},\lambda)\,dt_{i} dZ.
\end{equation}

$N_\lambda$ is the number of GW events in the Universe within an observing time that is expected from each \texttt{COMPAS} model. We can evaluate this given our rescaled merger rate density (expressed in individual metallicity and birth time bins, and in units of $\rm Mpc^{-3} yr^{-1}$) and $p_{\rm sim}$,
\begin{align*}
    N_\lambda &= 4\pi T_{\rm obs} \int_{\theta \in V_{\theta}} \int_{0}^{\infty} \int_{0}^{t_{j}(z)} p_{\rm sim}(\theta | Z, t_{i},\lambda)   \\
        & \qquad \qquad R_{\rm GW}^{\lambda}(Z,t_{i},t_{j}(z))\,\chi^2(z)\frac{c}{H(z)} \frac{1}{1+z} dt_{i} dZ d\theta, \numberthis \label{N}
\end{align*}
where $\chi(z)$ is the comoving distance to the redshift $z$, $c$ is the speed of light, $H(z)$ is the Hubble parameter, and $T_{\rm obs}$ is the total LVK observing time (duty cycle). This is 118.0, 138.8 and 142.0 days for O1, O2 and O3 respectively, which we converted to years. 

We can extend this to the predicted \textit{detected} number of GW events using a detection probability for the event-level parameters $p_{\rm det}(\mathcal{M}_c, q, z)$ given a particular detector configuration and signal-to-noise threshold.
\begin{align*}
    \mu_\lambda &= 4\pi T_{\rm obs} \int_{\theta \in V_{\theta}} \int_{0}^{\infty} \int_{0}^{t_{j}(z)} p_{\rm sim}(\theta | Z, t_{i},\lambda) p_{\rm det}(\theta)  \\
        & \qquad \qquad R_{\rm GW}^{\lambda}(Z,t_{i},t_{j}(z))\,\chi^2(z)\frac{c}{H(z)} \frac{1}{1+z} dt_{i} dZ d\theta. \numberthis \label{eq:mGW} 
\end{align*}

For Eq.~\ref{eq:P_unmarginalised}, we are assuming that $N_{\lambda}$ events are all formed via the isolated formation channel. There is some evidence for some events in GWTC-3 \citep{2023PhRvX..13d1039A} forming via the dynamical channel \citep{2021ApJ...908L..38A,2022ApJ...940..171R}. To avoid any assumptions on the formation channel and discount information on the predicted vs. observed GW rate, we can marginalise over $N_{\lambda}$ to remove its dependence by multiplying with a prior $\pi(N_{\lambda})=\frac{1}{N_{\lambda}}$ and integrating \citep{2018ApJ...863L..41F}.
\begin{align}
    \mathcal{L}(\lambda|\mathcal{H}) &= \int \frac{1}{N_{\lambda}}e^{-N_{\lambda}\beta(\lambda)}N_{\lambda}^{N_{\rm obs}}\prod_k^{N_{\rm obs}} \mathcal{I}^k dN_{\lambda} \notag \\ 
    &= (N_{\rm obs} - 1)!\prod_k^{N_{\rm obs}} \frac{\mathcal{I}^k}{\beta(\lambda)}, 
    \label{eq:P_marginalised} \numberthis
\end{align}
For convenience we can use $N_\lambda$ and $\mu_\lambda$ to define some detection efficiency $\beta(\lambda) = \frac{\mu_\lambda}{N_\lambda}$. We make model inferences using both of these posteriors in our work and compare how our conclusions change when $N_{\lambda}$ is included as information.


In order to evaluate our posteriors, we first calculate the model volumetric rate using a variant of Eq.~\ref{f_eff_total} but kept in separate metallicity and birth time bins. Given $T_{\rm obs}$, the merger rate and the appropriate volumetric factors, we then find the number of GW events that would fall in each metallicity and time bin, $N_{\lambda, i, j}$, and draw that many events (characterised by a chirp mass and mass ratio) from our full list of $10^6$ \texttt{COMPAS} binaries.\footnote{We find $N_{\rm COMPAS, i, j}$ in each metallicity and snapshot bins from the \texttt{COMPAS} run. We calculate a weight defined as $N_{\lambda, i, j}/N_{\rm COMPAS, i, j}$ to avoid repeatedly drawing from the \texttt{COMPAS} run. We apply this weight to the PDF. When calculating the detected PDF, we multiply this weight by $p_{\rm det}$.} We store the chirp mass, mass ratio and merger redshift of the events across all metallicity and birth time bins before then building the PDF $p_{\rm \textsc{sim}}(\theta|\lambda)$ using Kernel Density Estimation (KDE), in particular using the fast \texttt{KDEpy.FFTKDE} package. The bandwidth for our KDE is calculated using Silverman's rule of thumb \citep{1986desd.book.....S} for each hyper-parameter distribution and averaged to obtain the single band-width for the full 3D distribution. As a by-product of method we can also estimate $N_{\lambda}$ and $\mu_{\lambda}$ at the same time by simply keeping track of how many COMPAS binaries we are drawing.

For $p(\theta_i^k|\lambda)$, we query the KDE-built PDF using the chirp mass, mass ratio and redshift from the public LVK posterior samples. We construct $\pi^k(\theta_i^k)$ by building another KDE with the public LVK prior samples and querying this again with the posterior samples at each point $\theta_i^k$. 
In the case where $p_{\rm sim}(\theta|\lambda) = 0$ for all posterior samples, this implies that the \texttt{COMPAS} model cannot reproduce the BBHs in the posterior samples, as they fall outside of the KDE. We flag these events and ultimately remove them from the analysis for that specific model comparison.
For each of the $N_{\lambda, i, j}$, we calculate the detection probability, $p_{\rm det}$, based on the masses, redshift and luminosity distance. The detection probability comes from a selection effect function in \texttt{COMPAS}, which takes the BH masses, redshift and luminosity distance as inputs \citep{2018MNRAS.477.4685B, 2022ApJS..258...34R}. The SNR is calculated by computing the source waveform using the waveform model from the LAL software suite \citep{lalsuite}.

\begin{figure*}
    \centering
    \includegraphics[width=\textwidth]{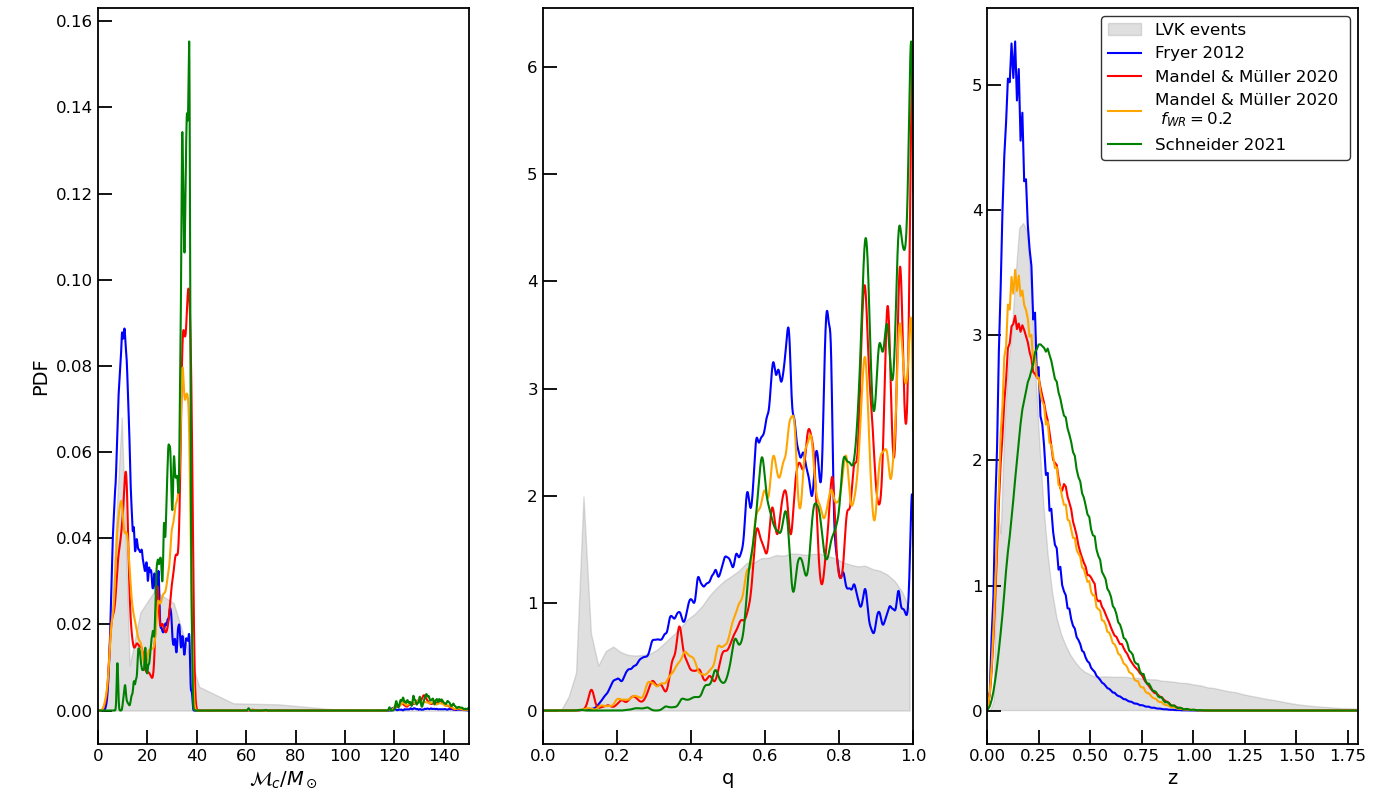}
    \caption{Distribution of chirp mass, mass ratio and redshift for different models. Different colours refer to the different remnant mass prescriptions investigated in \citet{10.1093/mnras/stad1757}. The grey shaded regions are the inferred distribution from the LVK collaborations, calculated using the posterior samples for the $N_{\rm obs} = 79$ GW events based on the SNR $\geq$ 8 cut.}
    \label{fig:PDF_dist}
\end{figure*}

We apply the same procedure also to estimate marginalised, observed 1D distributions $p_{\rm \textsc{sim\_det}}(\mathcal{M}_c|\lambda)$, $p_{\rm \textsc{sim\_det}}(q|\lambda)$ and $p_{\rm \textsc{sim\_det}}(z|\lambda)$.\footnote{$p_{\rm \textsc{sim\_det}}$ refers to the detected PDF, which is marginalised over the other two fitting parameters \textit{after} the detection probability is applied.} We show these 1D distributions for our four fiducial models in Fig.~\ref{fig:PDF_dist} alongside the GWTC data, with our main observations being:

\begin{itemize}
    \item With the exception of \citet{schneider2021pre}, the models replicate the first peak of the observed chirp mass distribution well. No models fit the second peak around $30M_\odot$. For all models, there is zero probability of producing chirp masses $> 40M_\odot$ until 100--130\,M$_\odot$. This implies these models predict high mass BHs that either may have not been detected due to the rarity of finding these types of events with current detectors, as they may be associated with noise in the low frequency band, or as a result of inaccuracies in the models resulting in an incorrect representation of the observed population.
    \item The mass ratio distribution is quite noisy but overall relatively flat, similar to the observed distribution. However for \citet{mandel2020simple} and \citet{schneider2021pre} there is a peak in the distribution after $q \approx 0.8$ indicating a preference for equal mass binaries. \citet{mandel2020simple} with $f_{\rm WR}=0.2$ has a preference for $q \sim 0.6$ and \citet{fryer2012compact} for $q \sim 0.7-0.8$. 
    \item The shape of the redshift distributions for the models fit the observed distribution quite well. Overall, the \citet{fryer2012compact} model is the best fit, while the other models have a wider distribution. This indicates BBHs from these models are more likely to be detected at higher redshifts (closer to $z \sim 1$) compared to \citet{fryer2012compact}. 
\end{itemize}

\subsection{Quantifying the uncertainty in $N_\lambda$, $p(\lambda, N_\lambda|\mathcal{H})$ and $p(\lambda|\mathcal{H})$} 
\begin{figure}
    \centering
    \includegraphics[width=\columnwidth]{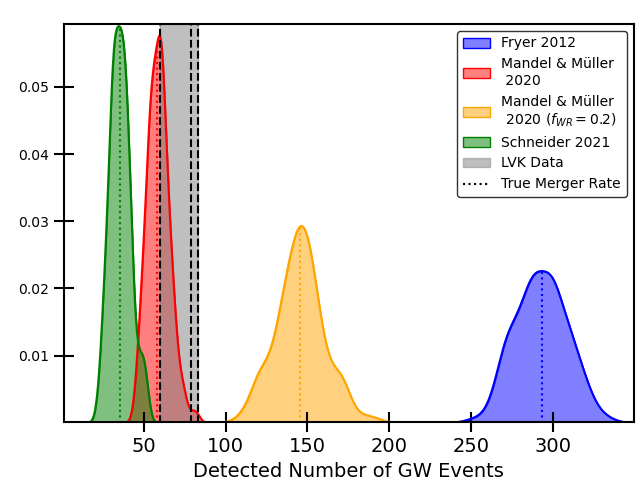}
    \caption{Normalised histograms of the detected number of BBH merger events calculated using Eq.~\ref{eq:mGW} over the LVK observing time for the last three observing runs. The coloured distributions are obtained from the predictions from the model for 170 sub-samples. The grey band represents the uncertainty in the observed number of BBH mergers depending on the selection criteria adopted (see Section \ref{Selection_criteria}). The dotted lines are the number of events predicted using the \citet{10.1093/mnras/stad1757} merger rate from \citet{10.1093/mnras/stad1757}.}
    \label{fig:NGW}
\end{figure}

\begin{figure}
    \centering
    \includegraphics[width=\columnwidth]{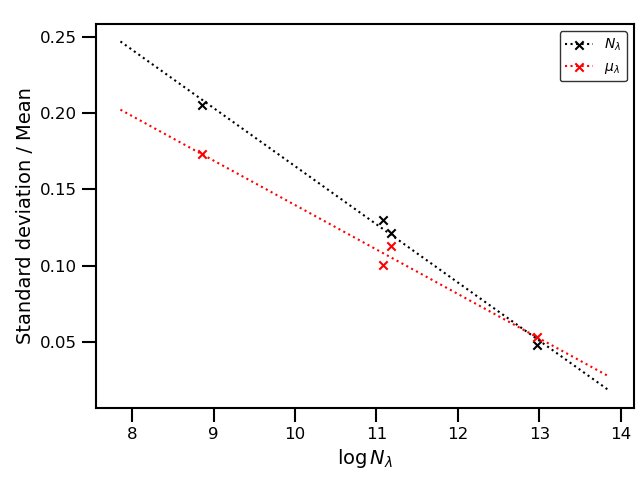}
    \caption{Relative error on the numbers of generated and detected events versus the average $\log N_\lambda$ for the four \citet{10.1093/mnras/stad1757} models from 170 sub-samples. The crosses correspond to the relative error from the models and the dashed lines are linear fits to the points. The black points and fit are for $N_\lambda$ and the red points and fits are for $\mu_\lambda$.}
    \label{fig:NGW_stdev}
\end{figure}

For our four models in \citet{10.1093/mnras/stad1757} we were able to split \texttt{COMPAS} outputs into 170 sub-samples to compute 170 effective factors and produce 170 theoretical, $N_\lambda$, and detected, $\mu_\lambda$, numbers of GW events. We show the distribution of the detected number of events in Fig.~\ref{fig:NGW}. For the error on the total number of events, we can use the standard deviation across our 170 subsamples. 
However, if we have only one sample of $N_{\rm sample}$ binaries for a \texttt{COMPAS} output, we require an alternative approach to assign an error for these models. Given the number of models we test in this work, the alternatives of simply using more samples or bootstrapping using the $N_{\rm sample}=10^6$ samples would be computationally challenging. Furthermore, subsampling or bootstrapping to generate an error would fail to capture the impact of progenitors of rare events that are not represented in the $N_{\rm sample}=10^6$, but which are present in the distributions obtained from our four models with $N_{\rm sample}=170\times 10^6$, and from which our errors are extrapolated. We find that the ratio of the standard deviation to the mean number of events as a function of $\log N_\lambda$ for the four models has a linear trend such that the relative error increases as the total number of GW events generated by the model decreases, in a roughly Poissonion fashion. In Fig.~\ref{fig:NGW_stdev} we show the linear fits for the standard deviation in $N_\lambda$ and $\mu_\lambda$, which we show in the following equations: 
\begin{align}
    \frac{\Delta N_\lambda}{N_\lambda} &\approx -0.04 \log N_\lambda + 0.55 \label{eq:NGW_stdev} \\ 
    \frac{\Delta \mu_\lambda}{\mu_\lambda} &\approx -0.03 \log N_\lambda + 0.43 \label{eq:muGW_stdev}
\end{align}

We can now apply this to single samples, assuming similar Gaussian distributions for other \texttt{COMPAS} models. To justify this, we find that the mean $\mu_\lambda$ over the 170 sub-samples in Fig.~\ref{fig:NGW} is equivalent to the $\mu_\lambda$ we obtain using the \citet{10.1093/mnras/stad1757} merger rate from the full \texttt{COMPAS} output (one sample) for all four models that we test.

\begin{figure}
    \centering
    \includegraphics[width=\columnwidth]{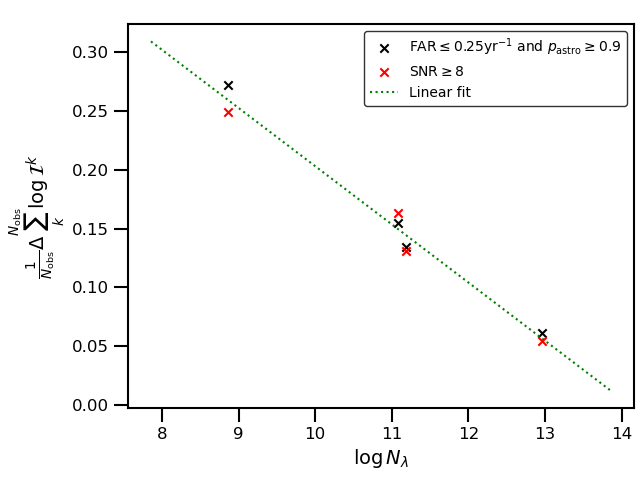}
    \caption{The standard deviation in the sum of the log-likelihood per event, $\log \mathcal{I}/N_{\rm obs}$, versus the mean $\log N_\lambda$ over the 170 sub-samples for the four \citet{10.1093/mnras/stad1757} models. We show the linear fit (green dashed line) that applies to both sets of selection criteria; $ \rm FAR \leq 0.25 yr^{-1}$ and $p_{\rm astro} \geq 0.9$ (black) and $\rm SNR \geq 8$ (red).}
    \label{fig:I_k_err}
\end{figure}

Now to address the uncertainty in $\mathcal{I}^k$; We find another linear trend for the standard deviation in $\sum_{k}^{N_{\rm obs}} \log \mathcal{I}^k$ over the 170 sub-samples as a function of the mean $\log N_\lambda$ for both our selection criteria, as shown in Fig.~\ref{fig:I_k_err}. 

The linear fits for the standard deviation in the likelihood sums are given by:
\begin{align}
    \frac{1}{N_{\rm obs}}\Delta \sum_k^{N_{\rm obs}} \log \mathcal{I}^k \approx -0.05\log N_\lambda + 0.70 \label{eq:I_k_stdev_2}
\end{align}
For new \texttt{COMPAS} models shown in Section \ref{Section 5}, we are able to use Eq.~\ref{eq:I_k_stdev_2} and to calculate the error on $\sum_{k}^{N_{\rm obs}} \log \mathcal{I}^k$ given $N_\lambda$ without needing to resample. We can now propagate the error in $N_\lambda$, $\mu_\lambda$ and $\sum_{k}^{N_{\rm obs}} \log \mathcal{I}^k$ into the error on the log-likelihood. 
\begin{align}
    \Delta \log \mathcal{L}(\lambda|\mathcal{H}) &= \sqrt{\left(N_{\rm obs} \frac{\Delta \mu_\lambda}{\mu_\lambda}\right)^2 + \left(N_{\rm obs}\frac{\Delta N_\lambda}{N_\lambda}\right)^2 + \left(\Delta \sum_{k}^{N_{\rm obs}} \log \mathcal{I}^k\right)^2} \nonumber \\
    \Delta \log \mathcal{L}(\lambda, N_\lambda|\mathcal{H}) &= \sqrt{(\Delta \mu_\lambda)^2  + \left(N_{\rm obs} \frac{\Delta N_\lambda}{N_\lambda}\right)^2 + \left(\Delta \sum_{k}^{N_{\rm obs}} \log \mathcal{I}^k\right)^2}
\end{align} 
Although only rudimentary given the small number of models available to us to generate fitting formulae, even adopting this simple uncertainty on the likelihood allows us to consider the statistical variations in model evaluation using population synthesis. As such it enables us to draw more confident conclusions on which hyper-parameters cause major changes in the likelihood, and whether these changes are physically significant. 

\subsection{Comparing posteriors with and without rescaling merger rate} 

\begin{figure*}
    \centering
    \includegraphics[width=\textwidth]{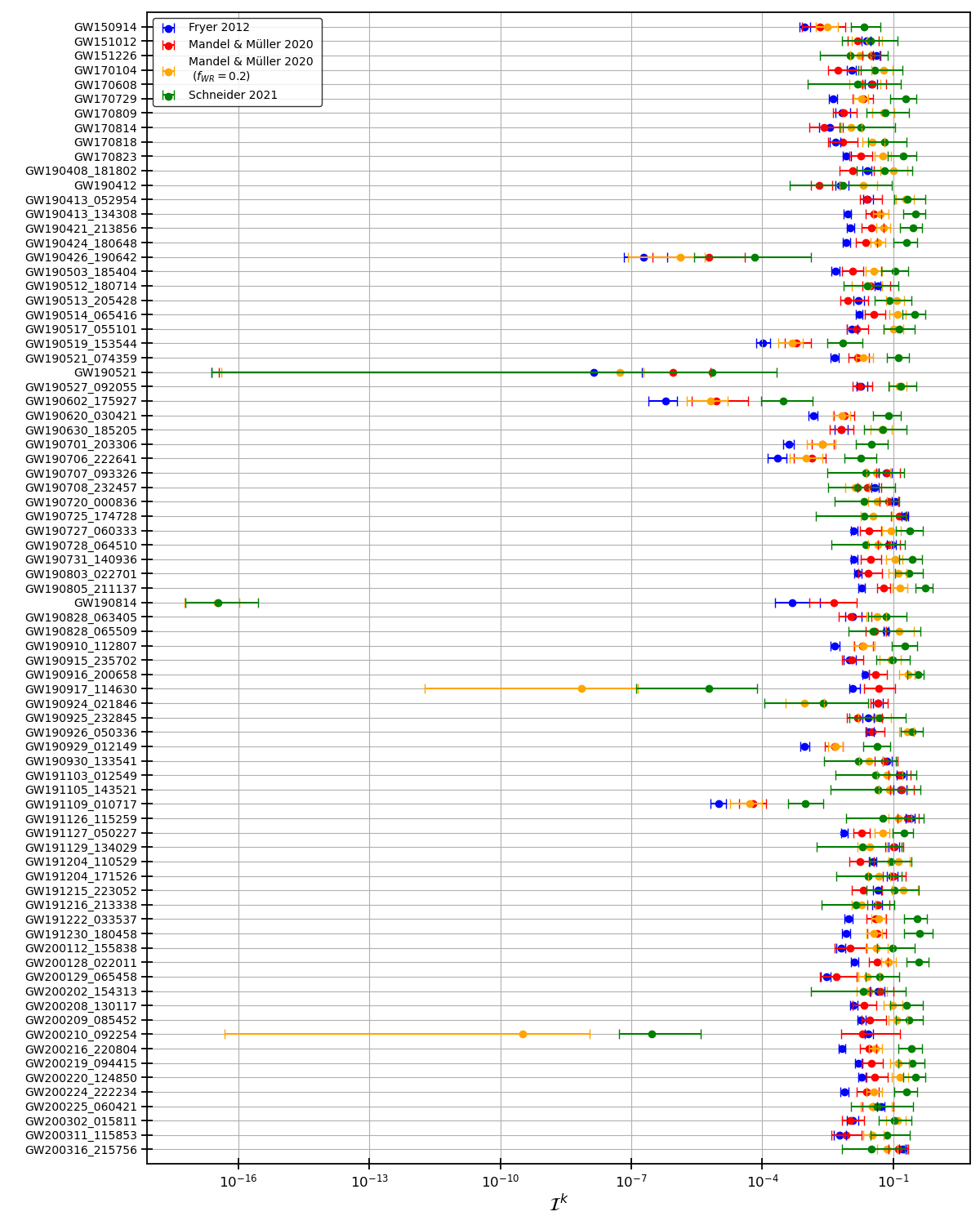}
    \caption{The likelihood, $\mathcal{I}^k$, for each GW event with $\rm SNR \geq 8$. Each colour represents a different remnant mass prescription from \texttt{COMPAS} and the error bar represents the sampling error over 170 samples when building the simulated PDF of $\mathcal{M}_c$, $q$ and $z$ given the model.}
    \label{fig:I_k_old_models}
\end{figure*}
Fig.~\ref{fig:I_k_old_models} shows the $\mathcal{I}^k$ term for individual GW events. This is a practical approach to determine the probability of individual GW events occurring in the populations generated by the \texttt{COMPAS} models. The \citet{schneider2021pre} model generally returns high probabilities when querying the PDF, while the \citet{fryer2012compact} model is disfavoured. We find some events are outliers and are not well reproduced by some or all of the models relative to the other events. As an example, \citet{mandel2020simple} with $f_{\rm WR}=0.2$ and \citet{schneider2021pre} are less likely to output the progenitors of GW190814, GW190917\_114630 and GW200210\_092254 compared to \citet{fryer2012compact} and \citet{mandel2020simple}. For GW190426\_190642, GW190602\_175927 and GW191109\_010717 \citet{schneider2021pre} is distinctly preferred over the other models, although in these cases, and especially that of GW190521, the models are generally less able to produce the required event-level properties. This is expected for GW190521 as both masses, $\sim 85M_\odot$ and $\sim 66M_\odot$, most likely lie in the pair-instability supernovae mass gap, making it atypical of binaries formed via the isolated formation channel \citep{2020PhRvL.125j1102A}.
\begin{figure}
    \centering
    \includegraphics[width=\columnwidth]{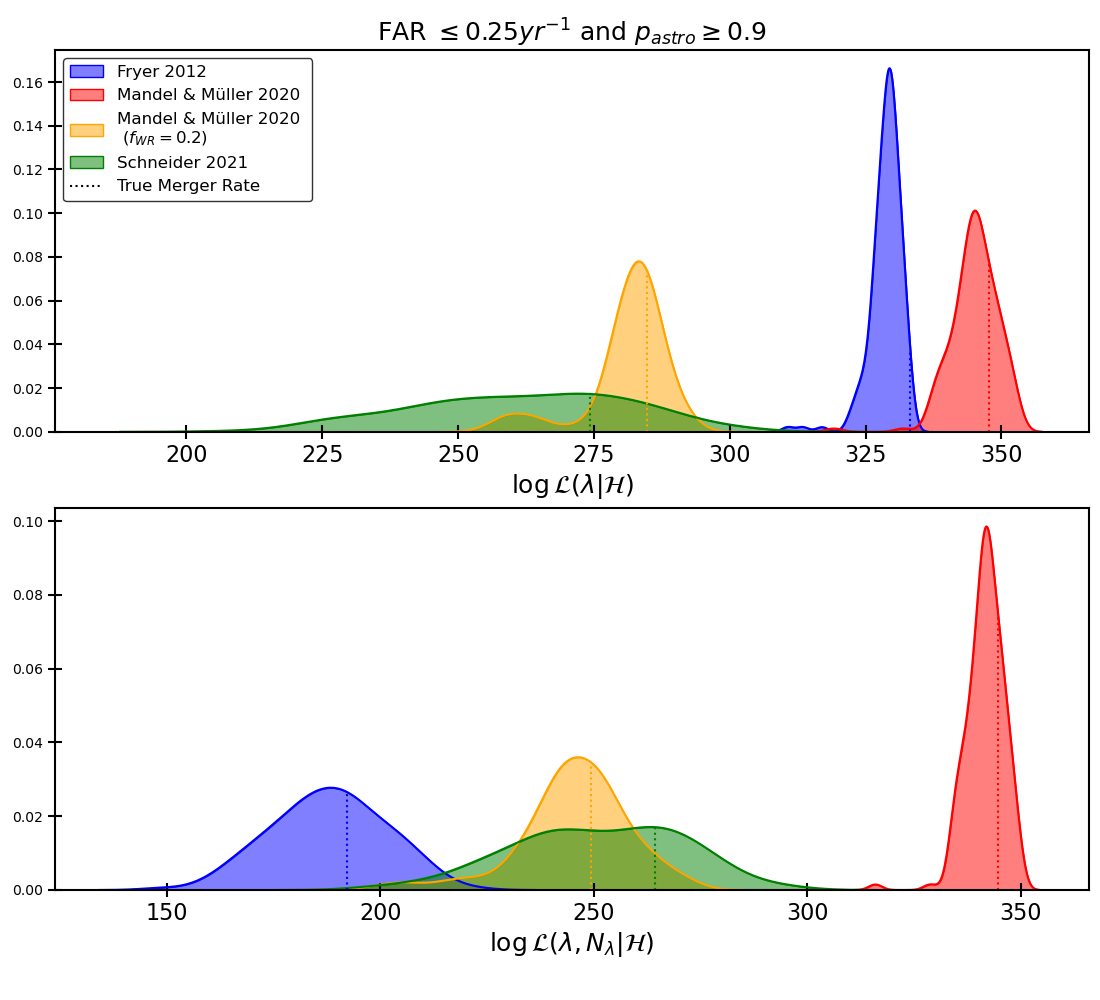}
    \includegraphics[width=\columnwidth]{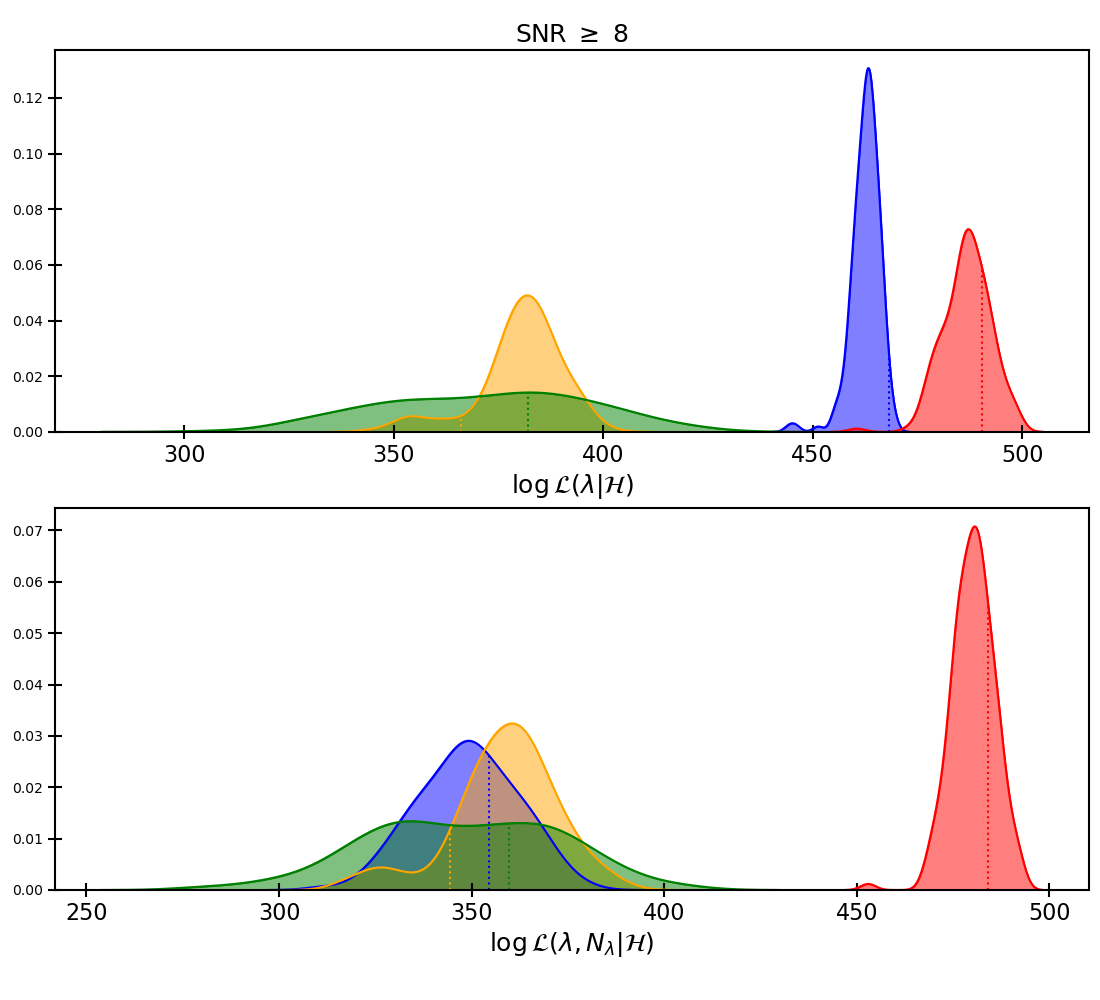}
    \caption{Normalised histograms of the log likelihoods obtained using 170 sub-samples for the models explored in \citet{10.1093/mnras/stad1757} with and without the merger rate dependence. The top two panels are the likelihoods comparing the models to the GW events that fit the $ \rm FAR \leq 0.25 yr^{-1}$ and $p_{\rm astro} \geq 0.9$ criteria, while the bottom two panels compares to the $\rm SNR \geq 8$ criteria. The dotted lines are the likelihoods using the \citet{10.1093/mnras/stad1757} merger rate. In all cases we see that the true likelihood is well represented within our distribution of 170 subsamples, indicating our use of a relatively smaller number of $10^6$ binaries is not expected to bias our results.}
    \label{fig:Likelihood_old_models}
\end{figure}

For a more comprehensive assessment of the models, we can analyse the likelihoods. In Fig.~\ref{fig:Likelihood_old_models} we now incorporate the detection probability and show the likelihoods with and without $N_\lambda$ dependence. The \citet{mandel2020simple} appears to the best fit to the observed population. This can be justified with Fig.~\ref{fig:PDF_dist}. The \citet{mandel2020simple} model has two peaks in the chirp mass distribution and a relatively flat mass ratio distribution, similar to the LVK population distribution. If we assume that the models output accurate populations, then incorporating the number of GW events makes the likelihood worse – \citet{fryer2012compact} greatly overestimates the number of detected events, while \citet{schneider2021pre} underestimates it. In general the latter is preferred as it leaves `room' for some of the LVK observed events to arise from the dynamical formation channel. The exception is the \citet{mandel2020simple} model, which performs well since the model predicts the same number of GW events within the observed range. The true number of events predicted is $\sim 60$, comparable to $N_{\rm obs}$ for the $p_{\rm astro} \geq 0.9$ and $ \rm FAR \leq 0.25 yr^{-1}$ selection criteria. For $\rm SNR \geq 8$, we see an increase in the likelihood for all models, but the distributions remain the same. The biggest difference in the unmarginalised likelihood for our two selections occurs for the \citet{fryer2012compact} model, which produces the largest number of detected events out of the four models. This is due to $N_{\rm obs}$ increasing to 79, such that the difference between the predicted $\mu_\lambda$ and the number of observed LVK events is slightly reduced.

\section{Results} \label{Section 5}
Having validated our methodology using the four models analysed in \citet{10.1093/mnras/stad1757}, this section discusses the outputs from our new set of \texttt{COMPAS} models where we vary the range of hyper-parameters described previously in Section~\ref{Section 2} (see also Table~\ref{tab:parameters}). We start with describing the trends in the number of detected events when varying the \texttt{COMPAS} hyper-parameters. We calculate the Bayes factor to identify groups of models that are more preferred. We then show our likelihood analysis, where we discuss the best and worst models when compared to the observed population parameters. We follow up this analysis and look at specific events and their $\mathcal{I}^k$ value to determine which models are more and less likely to generate the progenitors of these events. See Tables \ref{tab:all_models}, \ref{tab:all_models_2} and \ref{tab:all_models_3} in Appendix \ref{Appendix} for a detailed description of the hyper-parameters for all 304 \texttt{COMPAS} models in this analysis. 

\subsection{$\mu_{\rm GW}$ versus \texttt{COMPAS} parameters}
\begin{figure*}
    \centering
    \begin{multicols}{2}
        \includegraphics[width=\columnwidth]{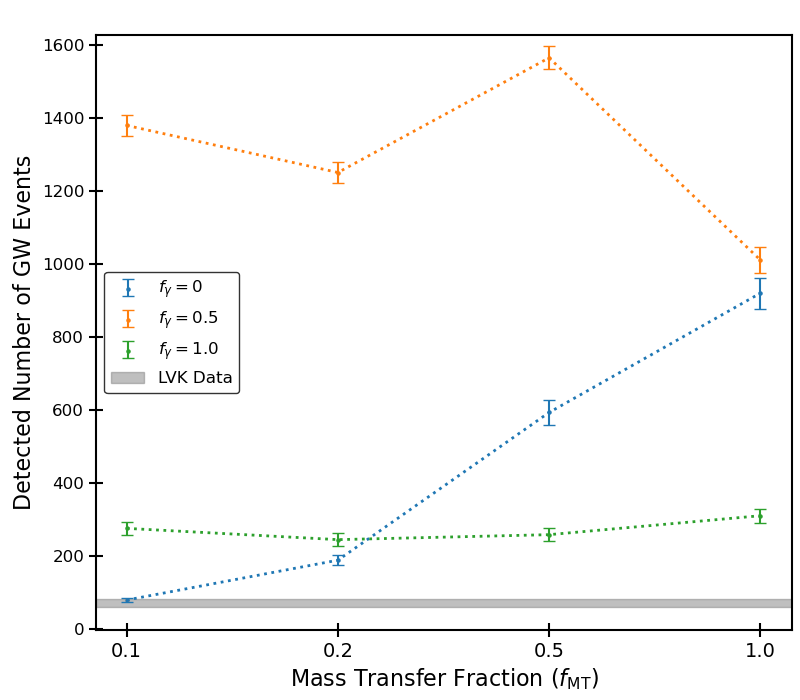} \par
        \includegraphics[width=\columnwidth]{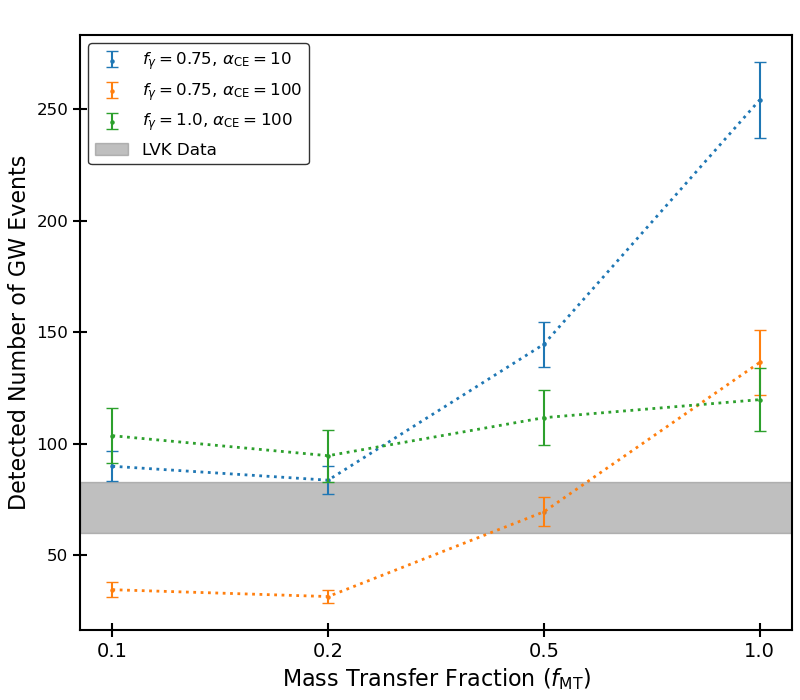}
    \end{multicols}
    \caption{Detected number of BBH merger events versus the mass transfer fraction for no CHE at fixed $f_\gamma$. \textit{Left}: We fix the CE evolution parameter $\alpha_{\rm CE}=0.1$ and show results for different specific AM loss parameters. \textit{Right}: We show the models for any $f_\gamma$ and $\alpha_{\rm CE}$ which output $\mu_\lambda$ close to the number of observed LVK events given by the grey band. 
    }
    \label{fig:NGW_vs_fMT}
\end{figure*}

Since we explore the number of GW events as a parameter in our posterior analysis, in this section we comment briefly on the trends between the detected number of GW events, $\mu_\lambda$, and the \texttt{COMPAS} hyper-parameters. 


In the left plot of Fig.~\ref{fig:NGW_vs_fMT}, we see $\mu_{\rm GW}$ increase with mass transfer fraction, $f_{\rm MT}$, when the corresponding AM is lost directly from the accretor ($f_\gamma = 0$). Increasing the mass transfer can lead to more equal mass objects, which we find most of our \texttt{COMPAS} models producing. This trend disappears for higher $f_\gamma$. We credit this to non-trivial evolutionary effects such as the stability prescriptions, that determine any future interactions within the binary. This preference for intermediate values of $f_\gamma$ agrees with the findings by \citet{2021MNRAS.502.4877S}. Irrespective of $f_{\rm MT}$, $\mu_{\rm GW}$ peaks for moderate values of $f_\gamma=0.5$ but drops again for high $f_\gamma$ when AM loss occurs closer to the L2 point. At this point, mass is lost at higher velocities so more AM is lost from the system, which in turn can induce CE evolution and merger during the mass transfer stage, stopping the formation of a BBH. These results are in agreement with \citet{2023MNRAS.519.1409L,2022ApJ...940..184V} that the number of stable mass channel objects increases for decreasing $f_\gamma$, as opposed to the objects formed by CE evolution. \citet{2021A&A...650A.107M} also find that the stable channel dominates the BBH merger rate. We note that when $f_{\rm MT} = 1$, $f_\gamma$ should have no impact. Any variations in the number of events at $f_{\rm MT} = 1$ is hence due to Eddington-limited accretion occurring after the first mass-transfer event(s), once the accretor has become a BH. \footnote{As a result of this finding, dual functionality in the AM loss was later implemented in \texttt{COMPAS}, as discussed in Section \ref{Summary}. If specific AM loss is set at the compact object as well as prior to the formation of the first compact object we expect the number of events will no longer vary for $f_{\rm MT} =1$, although at time of writing this functionality is yet to be comprehensively tested.}
   

We show the models that produce the closest $\mu_\lambda$ to the observed LVK events in the right plot of Fig.~\ref{fig:NGW_vs_fMT}. We find that for $\mu_{\rm GW}$ to be below 100 events (closer to the number of detected events from all observing runs), a CE efficiency $\alpha_{\rm CE}=10,100$ and $f_\gamma = 0.75,1.0$ with no CHE is required. This implies that the CE is ejected earlier in the evolution and the orbit tightens at a wider separation, which means longer merger times and less events in the same observed time. This result validates the studies by \citet{2021MNRAS.502.4877S}, \citet{2021MNRAS.505.3873B} and \citet{2023MNRAS.523..221G}. As discussed, $\alpha_{\rm CE} = 100$ is extreme, which suggests an alternative approach for the CE is required for forming populations with a comparable number of GW events to LVK. With the exception of these models, the CE efficiency has no significant impact on the populations, as the range for the number of events does not vary significantly with $\alpha_{\rm CE}$.

\begin{figure}
    \centering
    \includegraphics[width=\columnwidth]{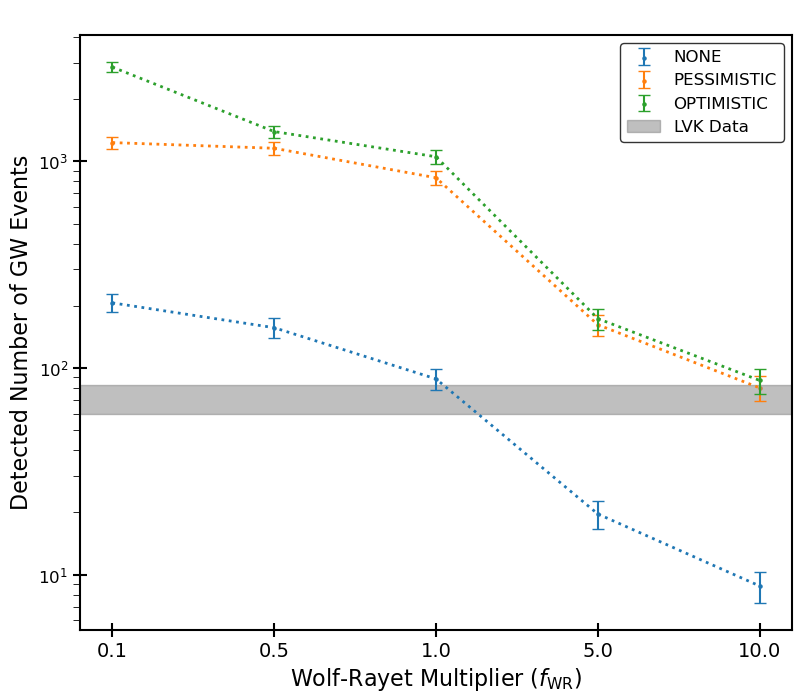}
    \caption{Detected number of BBH merger events versus the Wolf--Rayet multiplier for fixed CHE modes and the \citet{schneider2021pre} remnant mass prescription. The grey band is the observed number of LVK events.}
    \label{fig:muGW_vs_fWR}
\end{figure}

In Figure \ref{fig:muGW_vs_fWR}, $\mu_{\rm GW}$ decreases with increasing $f_{\rm WR}$. We find this for any CHE mode and remnant mass prescription, but we show this for \citet{schneider2021pre}. This is due to increasing the mass loss rates leading to smaller BHs and wider separations, so less mergers occur in the same observing time. These predictions agree with the work by  \citet{2021MNRAS.505..663R} and \citet{2022MNRAS.517.4034S} using \texttt{COMPAS}. We find that for no CHE, the rate of change in $\mu_\lambda$ across $f_{\rm WR}$ is smaller compared to the pessimistic and optimistic cases. 

The pessimistic and optimistic CHE modes produced $> 1000$ events due to their tight orbits. This suggests the CHE channel is not ideal for generating the current LVK detected population. We verify this with the likelihoods in Section \ref{likelihoods}.

\subsection{Bayes Factor and Likelihood Analysis} \label{likelihoods}
\begin{figure*}
    \centering
    \includegraphics[width=\textwidth]{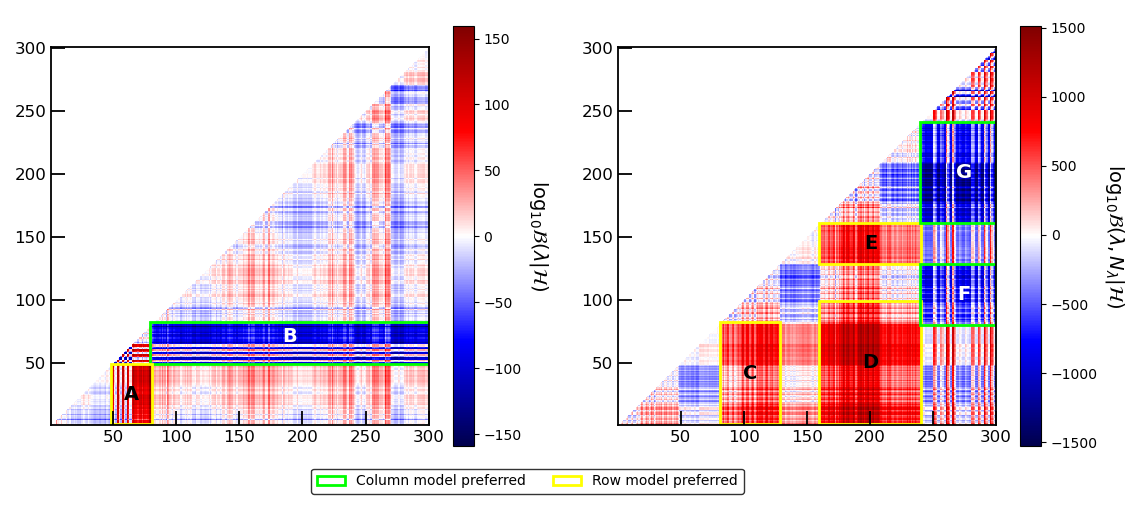}
    \caption{Comparison of models with $\rm SNR \geq 8$ using the Bayes factor. Each grid is coloured by $\log_{10}(\mathcal{B})$. For $\log_{10}(\mathcal{B}) > 0$, the model in the row (y-axis) is favoured over than the model in the column (x-axis), while for $\log_{10}(\mathcal{B}) < 0$ the model in the row is disfavoured over the model in the column. Left plot is the Bayes factor for the marginalised likelihood and right plot is the Bayes factor for the unmarginalised likelihood.}
    \label{fig:Bayes_factor}
\end{figure*}

To identify groups of models that are more preferred than others, we use the Bayes factor, $\mathcal{B}$. Since the priors for all models are the same, the Bayes factor is the ratio of the likelihoods between two models, here denoted $\alpha$ and $\kappa$:
\begin{equation} \label{eq:B_eq}
    \mathcal{B(\lambda|\mathcal{H})} = \frac{\mathcal{L}(\lambda_\alpha| \mathcal{H})}{\mathcal{L}(\lambda_\kappa| \mathcal{H})}; \qquad \mathcal{B(\lambda,N_{\lambda}|\mathcal{H})} = \frac{\mathcal{L}(\lambda_\alpha,N_{\lambda_{\alpha}}| \mathcal{H})}{\mathcal{L}(\lambda_\kappa,N_{\lambda_{\kappa}}| \mathcal{H})}. 
\end{equation}
In Fig.~\ref{fig:Bayes_factor}, we show the Bayes factor with the likelihoods using the GW events with the $\rm SNR \geq 8$ cut. 

We note that the Bayes factor indicates there is strong evidence that some groups of models are preferred more than others, which we will further discuss in the likelihood analysis. We emphasise these with highlighted boxes in Fig.~\ref{fig:Bayes_factor} labeled A-G.
We can understand the range of Bayes factor by considering the different components that enter Eq.~\ref{eq:B_eq}. The difference in $\log \beta(\lambda)$ between different models ranges from $\sim 0.05-3.44$. However, we find that the $\mathcal{I}^k$ terms are distributed over 50 orders of magnitude (see bottom plot in Fig.~\ref{fig:I_k_new_models}) and therefore dominates the Bayes factor. This confirms that some of the models are distinguishable and strongly preferred (more likely to produce progenitors of particular GW events) over others purely based on the masses and redshift distributions they predict. When including $N_\lambda$ as a parameter in the likelihood, the $-\mu_\lambda + N_{\rm obs}\log(\mu_\lambda)$ term this introduces into the log-likelihood has significant weight, and comes to dominate the Bayes factor, meaning that a comparison of two models using this likelihood strongly favours those that produce a number of events close to the observed number of events in our data sample, even if the event-level distributions are less well matched. Hence, we carefully consider the results using both of our likelihoods.

\begin{figure}
    \centering
    \includegraphics[width=\columnwidth]{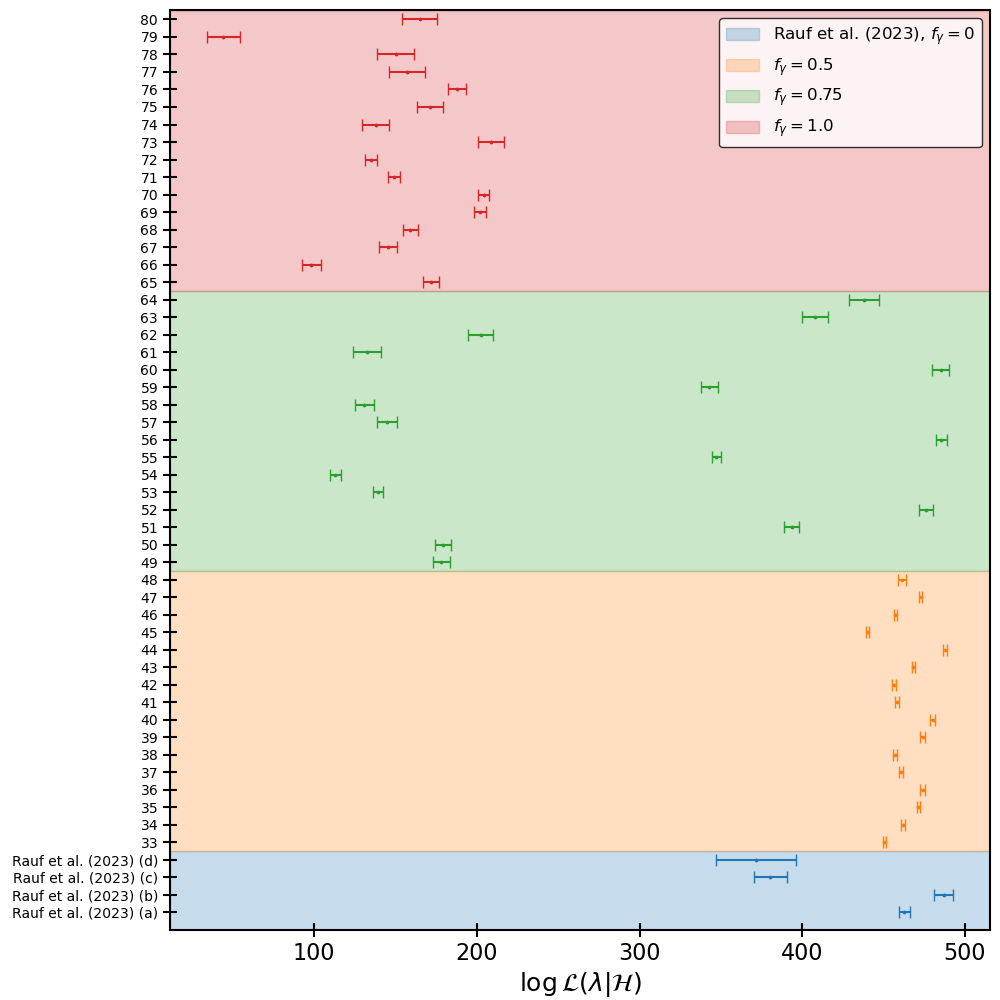}
    \caption{Likelihoods for models with no CHE. Each band corresponds to a fixed $f_\gamma$. Within each band, excluding the \citet{10.1093/mnras/stad1757} models in blue, the mass transfer fraction and CE efficiency are increasing with increasing model number, within the ranges in Table~\ref{tab:parameters}, with $f_{\rm MT}$ varying more rapidly.}
    \label{fig:L_f_gamma}
\end{figure}

\begin{figure}
    \centering
    \includegraphics[width=\columnwidth]{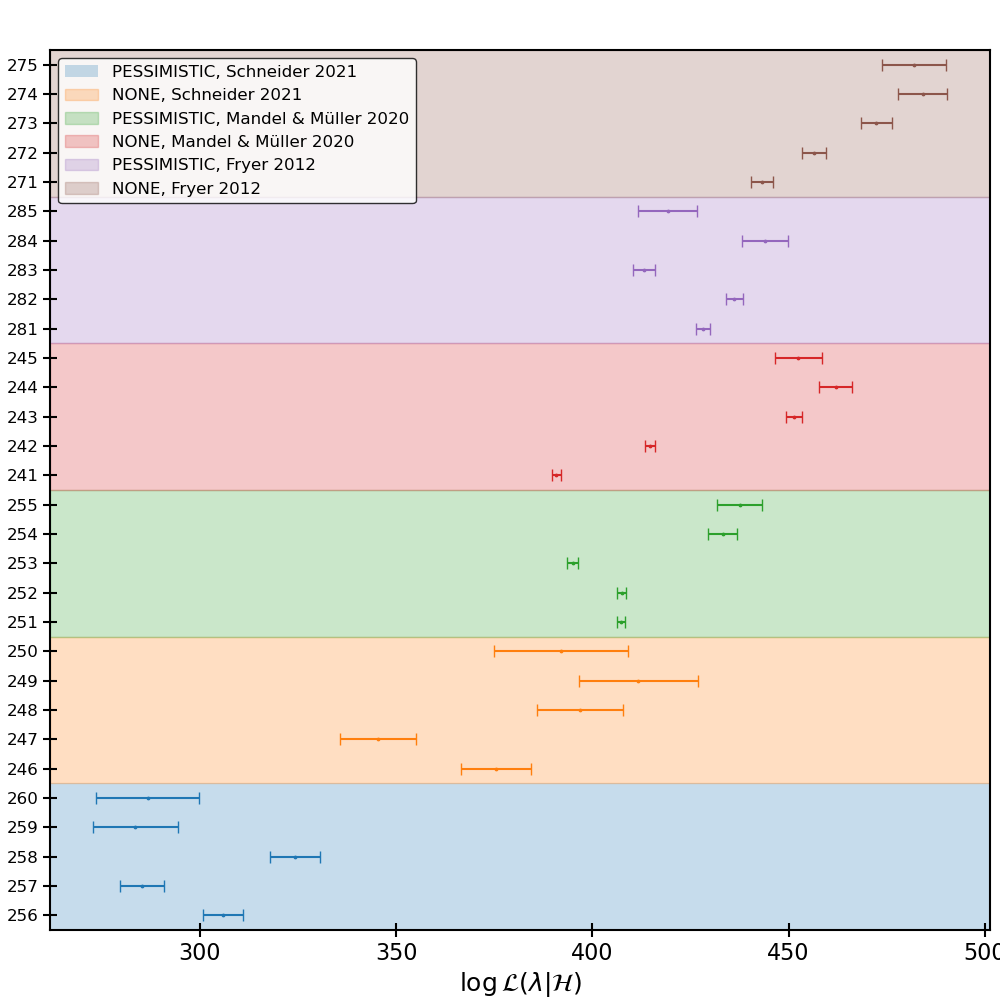}
    \caption{Marginalised likelihood for varying CHE and remnant mass prescription. In each band, the Wolf--Rayet multiplier is increased with increasing model number, across the values given in Table~\ref{tab:parameters}.}
    \label{fig:L_CHE_rem}
\end{figure}

Having established groups of models that are strongly preferred over other models, we now discuss the likelihoods for individual cases and the trends within these groups. The following results (Fig.~\ref{fig:L_f_gamma}, \ref{fig:L_CHE_rem}, \ref{fig:LGW_CHE}, \ref{fig:LGW_vs_FWR}, \ref{fig:Contour}, \ref{fig:I_k_new_models}) discussed use the $\rm SNR \geq 8$ selection cut. We see an increase in the likelihood, due to an increase in $N_{\rm obs}$, across all models compared to using $p_{\rm astro} \geq 0.9$ and $ \rm FAR \leq 0.25 yr^{-1}$, but the trends don't change substantially with the selection criteria and we find the differences are made more distinct when using the $\rm SNR \geq 8$ sample. We show this in Tables \ref{tab:all_models}, \ref{tab:all_models_2} and \ref{tab:all_models_3} with Bayes factors between our various models.

We see an increase in the likelihood from $f_\gamma = 0$ to $f_\gamma = 0.5$ and we find that the likelihoods are similar with no CHE and pessimistic and optimistic CHE. In Fig.~\ref{fig:L_f_gamma}, we show the likelihoods for fixed $f_\gamma$. It is clear that $f_\gamma = 0.5$ is preferred over $f_\gamma = 1.0$. Less BBH mergers are produced with $f_\gamma = 1.0$. Models with $f_\gamma = 0.5$ produce more equal mass binaries, which is more consistent with the LVK population \citep{2019ApJ...882L..24A}. However, $f_\gamma = 0.75$ is unique in that for low mass transfer $f_{\rm MT} < 0.5$ the population is disfavoured similar to $f_\gamma = 1.0$, but for high mass transfer $f_{\rm MT} > 0.5$ the model is more favourable. In all cases, we see little impact from the choice of CE efficiency. This highlights the degeneracy between stellar modelling assumptions and populations. Ultimately, we find that in order to reproduce the observed events from LVK, we need either momentum loss to occur close to the accretor, or, if occurring further away, for the mass transfer fraction to be high. These trends are also highlighted by boxes A and B in Fig.~\ref{fig:Bayes_factor}, where moderate values of $f_\gamma$ are strongly favoured when comparing population parameters. 

Fig.~\ref{fig:L_CHE_rem} then explores different combinations of the remnant mass prescription, CHE mode and Wolf--Rayet multiplier. Overall, we see weaker trends here than with the mass transfer fraction. However, the remnant mass prescription can be seen to impact the likelihood, with the \citet{fryer2012compact} or \citet{mandel2020simple} prescriptions preferred. It is worth reiterating that, of these two, the \citet{fryer2012compact} was seen to greatly over-predict the number of observed events to date. We find a weak trend of increasing likelihood in the absence of CHE, most evident for the \citet{schneider2021pre} model, but overall less significant an impact than was seen for changes in the mass transfer and AM loss parameters. We also see a weak preference for larger $f_{\rm WR}$ depending on the remnant mass model. 

\subsubsection{Likelihood including $N_{\lambda}$}

\begin{figure}
    \centering
    \includegraphics[width=\columnwidth]{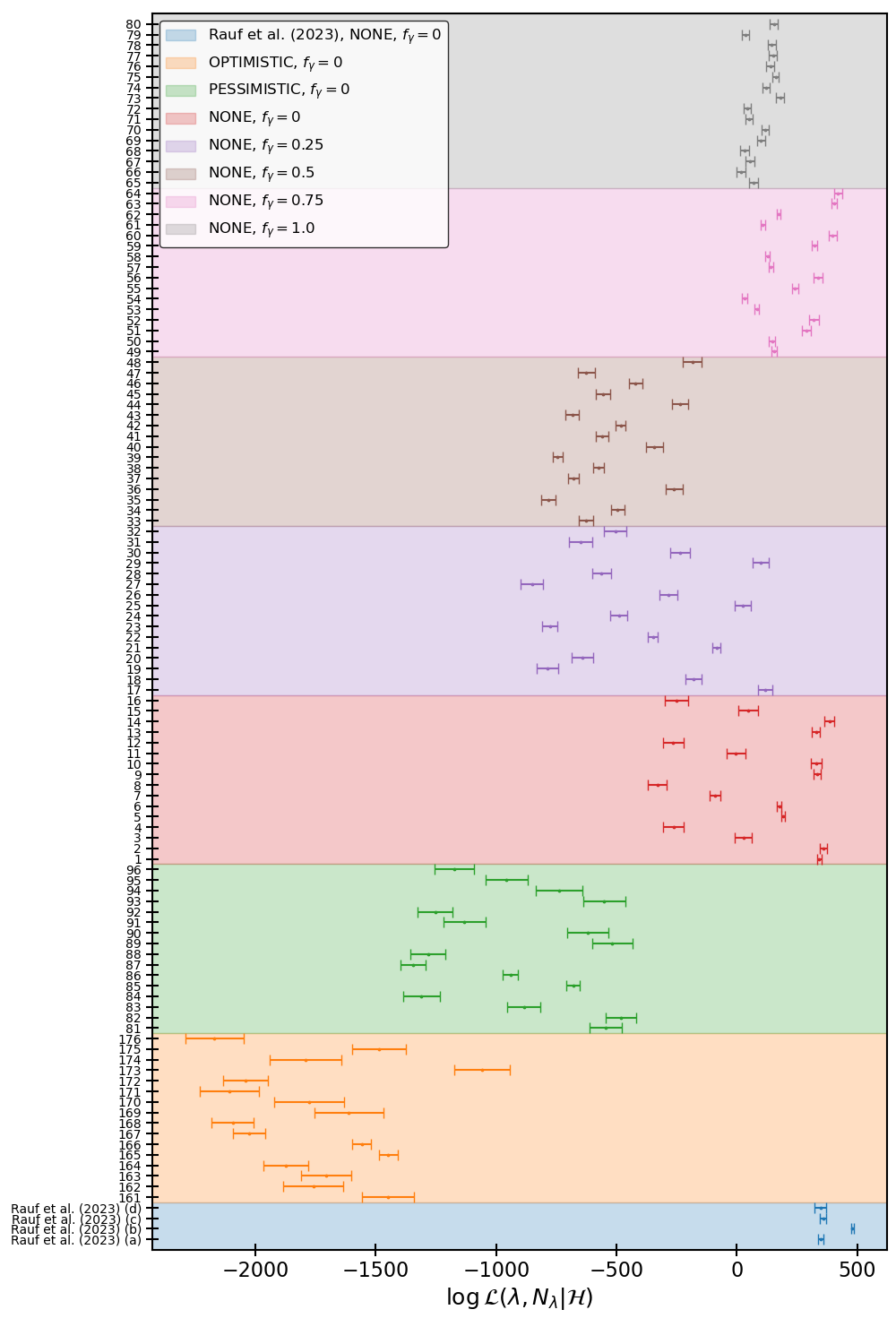}
    \caption{Model versus unmarginalised likelihood, where each coloured band corresponds to the a fixed CHE mode and $f_\gamma$. Within each band, excluding the \citet{10.1093/mnras/stad1757} models, the mass transfer fraction and CE efficiency are increasing with increasing model number, within the ranges in Table~\ref{tab:parameters}, with $f_{\rm MT}$ varying more rapidly.}
    \label{fig:LGW_CHE}
\end{figure}

When including the dependence on the number of GW events, we are able to identify more distinct groups of trends, particularly with the CHE modes as shown in Fig.~\ref{fig:LGW_CHE}. This is in contrast to the likelihoods without the number of events, indicating that the CHE mode more strongly affects just the number of BBH mergers, rather than the underlying distributions of mass or redshift. The likelihood generally reduces when switching from no CHE to Pessimistic or Optimistic, which is not as notable for the likelihood without $N_\lambda$ dependence. We also see this impact in Fig.~\ref{fig:Bayes_factor}, where box C and D correspond to no CHE being strongly preferred over Pessimistic and Optimistic CHE and box E implies Pessimistic is strongly preferred over Optimistic CHE. Allowing for CHE would mean tighter orbits and more mergers within Hubble time, which would over-predict the number of observed events when compared to $N_{\rm obs}$. In addition, the likelihood now decreases from $f_\gamma=0.25$ to $f_\gamma = 0.5$. This is also due to the increased number of GW events, shifting them further from the observed number of LVK events. It then increases again with further increasing $f_\gamma = 0.75, 1.0$ due to the reduced numbers of events. Within each choice of CHE, we see a preference for low mass transfer efficiency for $f_\gamma =0$. This is required to keep the predicted number of events low or comparable to the actual number of LVK events, as these produce unequal mass wide binaries with long merger times. 

Although no CHE is preferred, this can depend on $f_\gamma$. We show a subset of no CHE with varying $f_\gamma$ in Fig.~\ref{fig:LGW_CHE}. Again we note the degeneracy in the hyper-parameters and populations. Low values of $f_{\rm MT}$ and $f_\gamma=0.25$ produce similar populations to $f_\gamma = 0$, while models with higher values of $f_{\rm MT}$ at $f_\gamma = 0.25$ appear similar to Pessimistic CHE with $f_\gamma = 0$ populations. Conversely, further increases to $f_\gamma = 0.5$ and \textit{high} $f_{\rm MT}$ produce high-likelihood populations similar to $f_\gamma = 0$, with now a decreasing likelihood as $f_{\rm MT}$ decreases. The likelihood further increases with $f_\gamma = 0.75$ particularly again for high $f_{\rm MT}$. There is hence a flip in the behaviour of increasing mass transfer fraction depending on whether the momentum loss is occurring closer to the accretor or L2 point. These changes in likelihood closely reflect the trends seen in the number of predicted events seen in Fig.~\ref{fig:NGW_vs_fMT}, but further understanding of this complex behaviour warrants further study.

\begin{figure*}
    \begin{multicols}{2}
        \centering
        \includegraphics[width=\columnwidth]{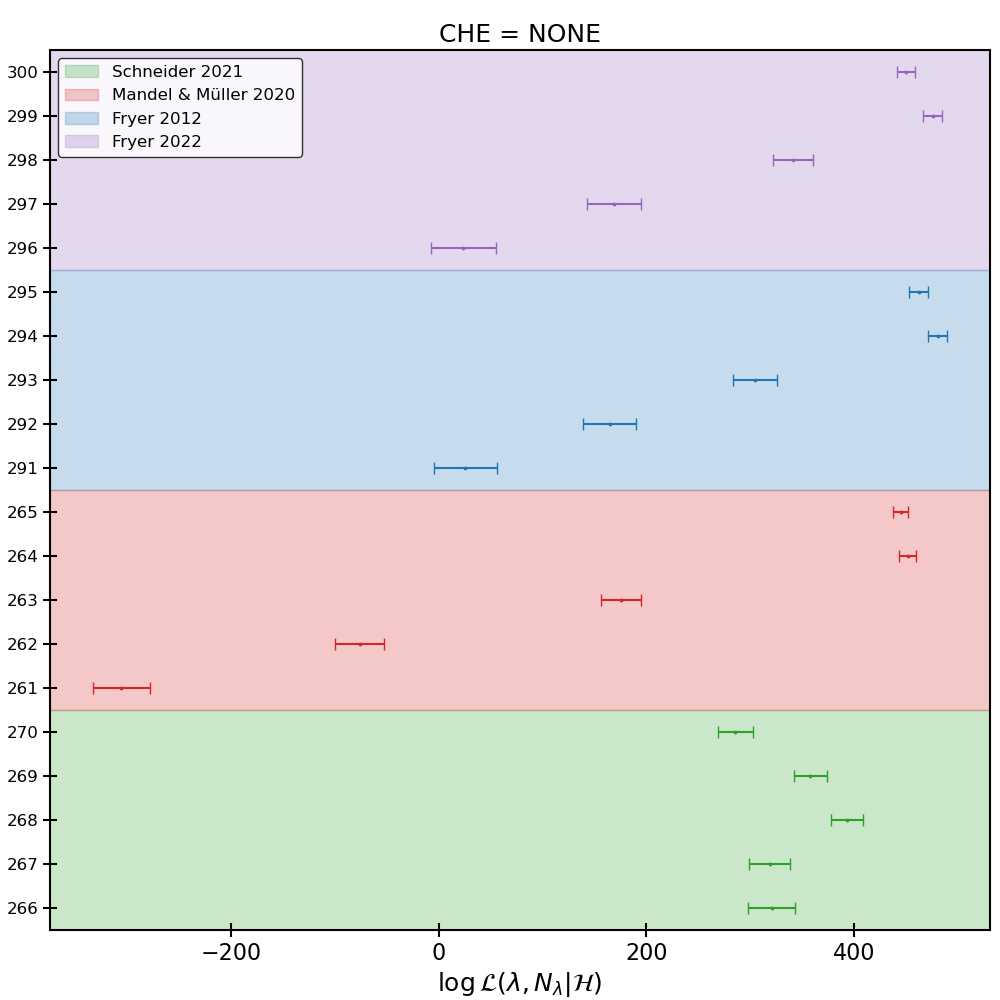} \par
        \includegraphics[width=\columnwidth]{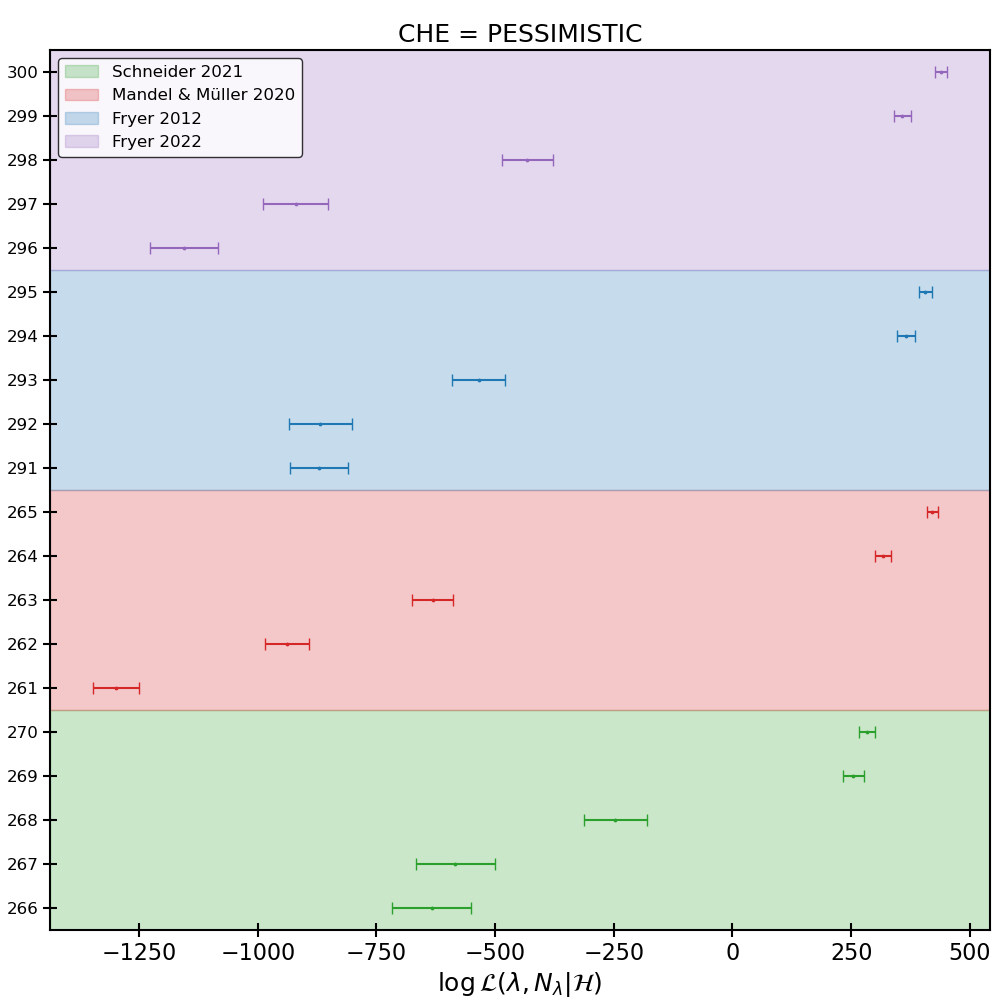} \par
    \end{multicols}
    \caption{Unmarginalised likelihood for No CHE (\textit{Left}) and pessimistic CHE (\textit{Right}) for fixed remnant mass prescriptions. Within each band we vary the Wolf--Rayet multiplier with values given in Table~\ref{tab:parameters}, increasing the value with increasing model number.}
    \label{fig:LGW_vs_FWR}
\end{figure*}

Finally, in Fig.~\ref{fig:LGW_vs_FWR}, we find a strong case for likelihood increasing with $f_{\rm WR}$, with the exception of \citet{schneider2021pre} with no CHE. This is also shown in Fig.~\ref{fig:Bayes_factor} by boxes F and G, where $f_\gamma = 0$ and thermal timescale mass transfer rates are strongly preferred over moderate values of $f_\gamma$ and fixed $f_{\rm MT}$. For fixed CHE and remnant mass prescription, there is a preference for models with $f_{\rm WR}=10$. This is because mass loss via strong winds can widen binaries. This would increase the merger time and decrease the number of GW events closer to the LVK detected number of events. Overall, again no CHE is preferred over pessimistic and optimistic CHE --- in the presence of CHE, one needs a rather finely-tuned set of: high Wolf--Rayet factor, high $f_{\gamma}$ and low mass transfer efficiency to recover a reasonable number of mergers.

Given the main assumption in this paper being all GW events originate from the isolated binary formation channel, the observed number of events in the LVK data places an upper bound on $\mu_\lambda$. Any models that output $\lesssim N_{\rm obs}$ cannot be ruled out. However, even in the case where the number of detected events is close to $N_{\rm obs}$, this is not a good indicator of if the model population will match the observed population. We propose checking the marginalised and unmarginalised likelihood simultaneously to find if there is overlap in models. This will determine the models with comparable populations \textit{and} detected number of events to the LVK data. We check the 30 largest likelihoods out of the 300 new models for both these cases separately, and identify the common models and their hyper-parameters. We again find a clear preference for no CHE for $f_{\rm WR} \geq 1$ and $f_\gamma = 0$,  in that these models often return representative numbers of detected events, \textit{and} populations of masses and redshifts, but the choice of remnant mass prescription varies between \citet{mandel2020simple}, \citet{fryer2012compact} and \citet{2022ApJ...931...94F}. 

\subsubsection{Comparisons of the best individual models to data}

\begin{figure*}
    \begin{multicols}{2}
    \centering
    \includegraphics[width=\columnwidth]{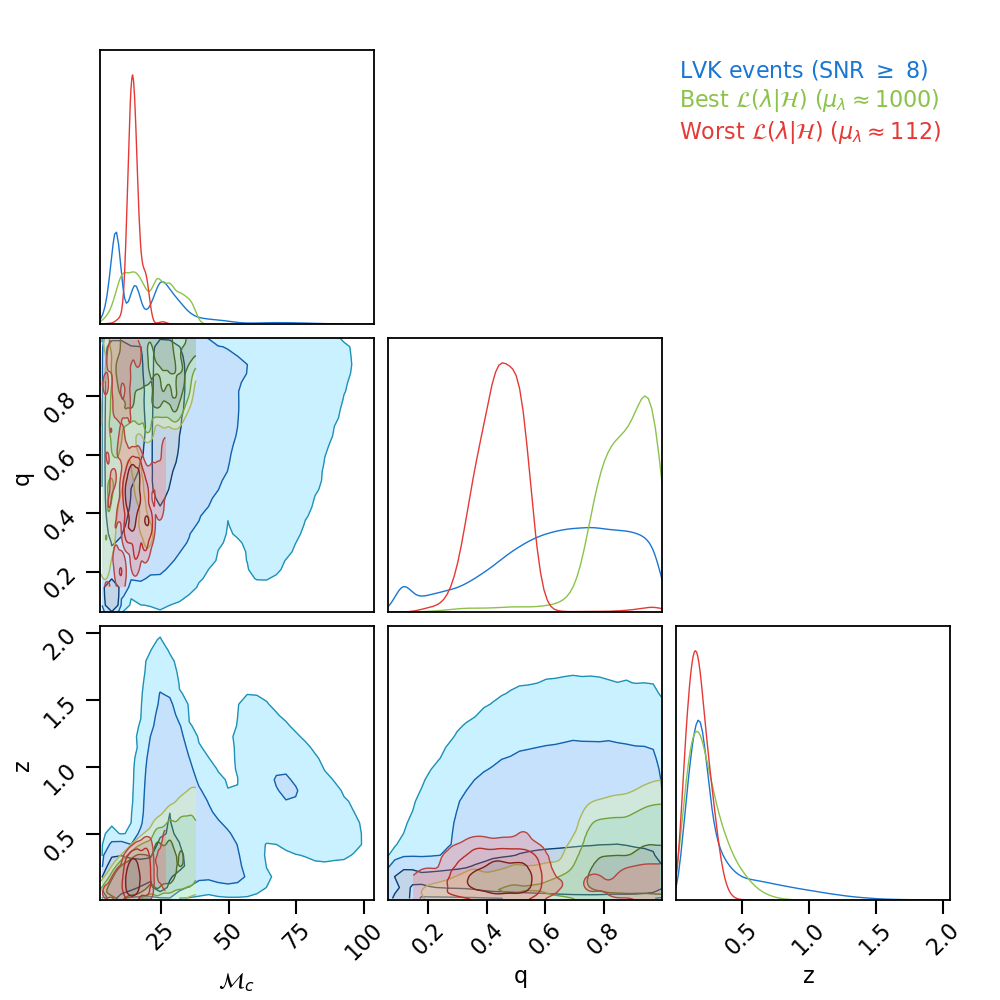}
    \par
    \includegraphics[width=\columnwidth]{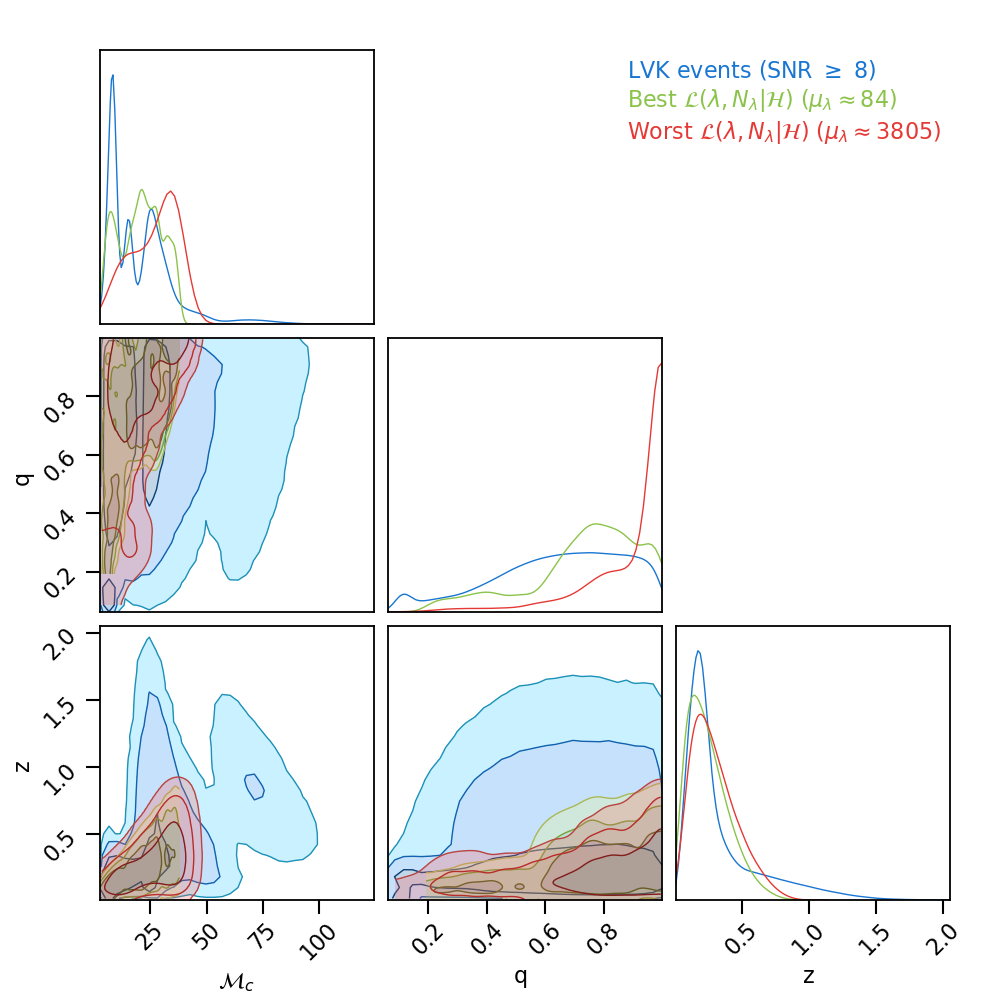}
    \end{multicols}
    \caption{$\mathcal{M}_c$, $q$ and $z$ $3\sigma$ distributions for the models with the best (green contour) and worst (red contour) likelihood. On the left we show the likelihood without the $N_\lambda$ dependence and the right plot shows the likelihood with the $N_\lambda$ dependence. The blue contour combines the posteriors for the $N_{\rm obs} = 79$ GW events based on the $\rm SNR \geq 8$ cut.}
    \label{fig:Contour}
\end{figure*}

We now compare the visual population parameter distributions to the GW events with $\rm SNR \geq 8$ posteriors for the 300 new models. 
Model  44 has the best likelihood ($\mathcal{L}(\lambda|\mathcal{H})$) which implies that the model population best matches the observed LVK population. As shown in Fig.~\ref{fig:Contour}, we find that the chirp mass and mass ratio distribution falls within the 2$\sigma$ contour of the GW events. The LVK chirp mass distribution has three distinct peaks. The best model does not replicate the first peak but it does for the second and third and generally matches the overall chirp mass distribution. That said, the mass ratio distribution is uniform for the LVK data but the model does prefer high mass ratios ($q > 0.6$). Model 79 outputs the worst likelihood. The chirp mass and mass ratio distributions have very narrow peaks corresponding to the second peak in the LVK mass distribution around $\mathcal{M}_{c}=20 M_{\odot}$ and at $q=0.5$. The redshift distribution range is also smaller compared to Model 44 and the data. However, the number of detected events, $\mu_\lambda$ is closer to $N_{\rm obs}=79$. This demonstrates the intrinsic difficulty in population synthesis of returning the correct event-level properties and overall GW rate.


The best likelihood with a dependence on the number of events ($\mathcal{L}(\lambda, N_\lambda|\mathcal{H})$) is Model 274. This model has a peak in the chirp mass distribution corresponding to the first peak in the LVK mass distribution and recovers the higher mass peaks reasonably well too, albeit with absence in the very high mass tail. The mass ratio distribution is not uniform, there is a preference for mass ratios $ > 0.5$, but $ < 0.5$ are also possible such that the model and data distributions agree reasonably well. This model outputs $\mu_\lambda \approx 83$ events. 
The worst likelihood is model 199. This model has has a rather high peak chirp mass around $40M_\odot$ and a strong preference for high mass ratios ($q > 0.8$). It is likely that as a result of producing higher mass BHs with smaller separations, this models outputs the largest detected number of events, $\sim 3805$, which negatively impacts the likelihood. 

In all cases, our models seem to underpredict the high chirp mass and high mass ratio events with $z \geq 0.5$ when compared to the LVK distribution. It is possible that these events are not produced frequently by \texttt{COMPAS}. This may be a limitation in the SSE implementation, as discussed in Section \ref{Summary}. 
It is likely that these systems are not produced efficiently through the isolated channel, which could indicate that some of our LVK events progenitors form dynamically. And finally, these events may just be rare or `exceptional' in any formation channel/ environment. In this case, approaches more like \citet{2024arXiv240509739P} --- which uses the ``normalised evidence'' within a statistical framework to determine if hierarchical mergers in active galactic nuclei or globular clusters can justify exceptionally large events such as GW190521 --- may be more appropriate.  

\subsection{Which data are the most informative?} \label{model_distinguishing}

\begin{figure}
    \centering
    \includegraphics[width=\columnwidth]{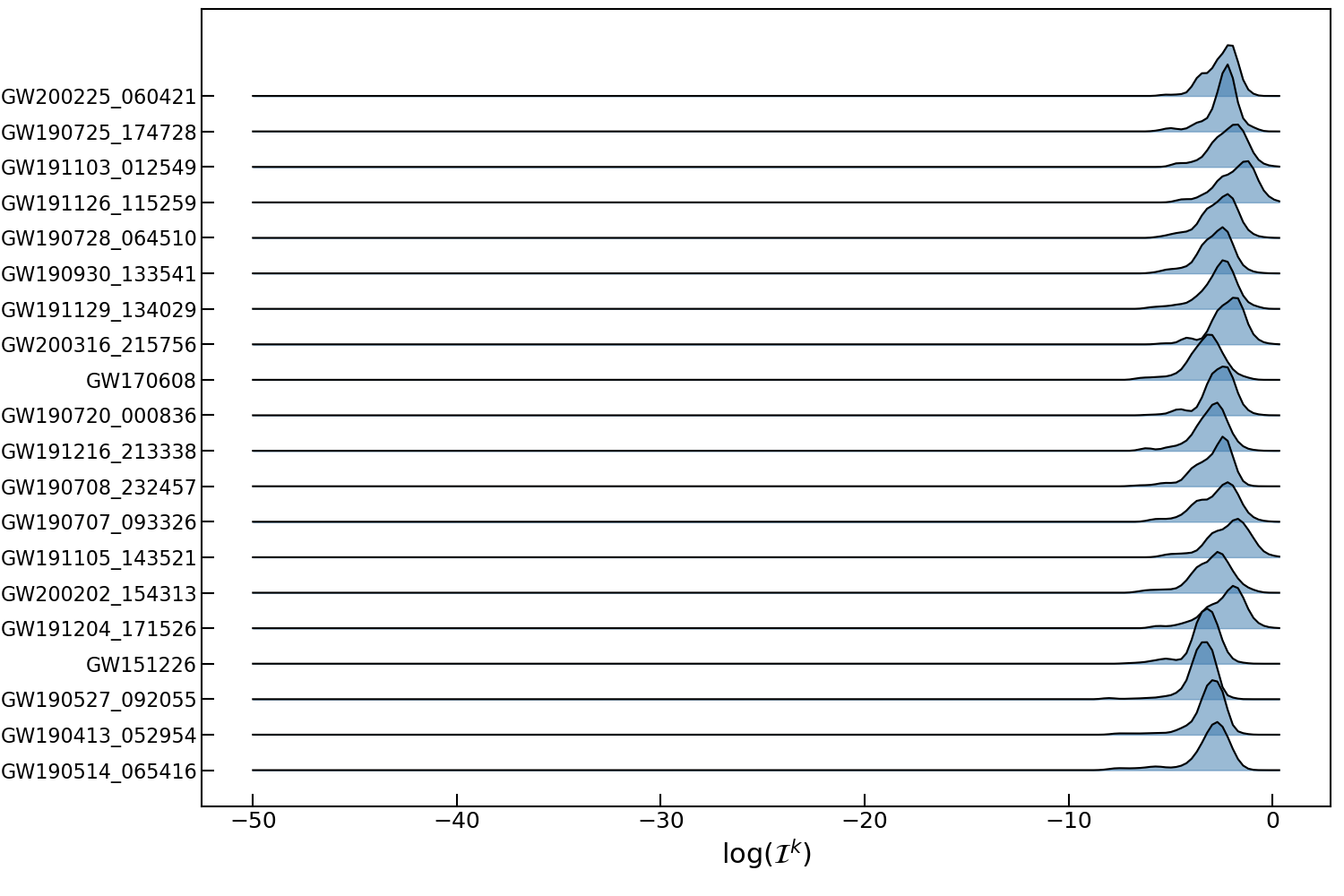}
    \includegraphics[width=\columnwidth]{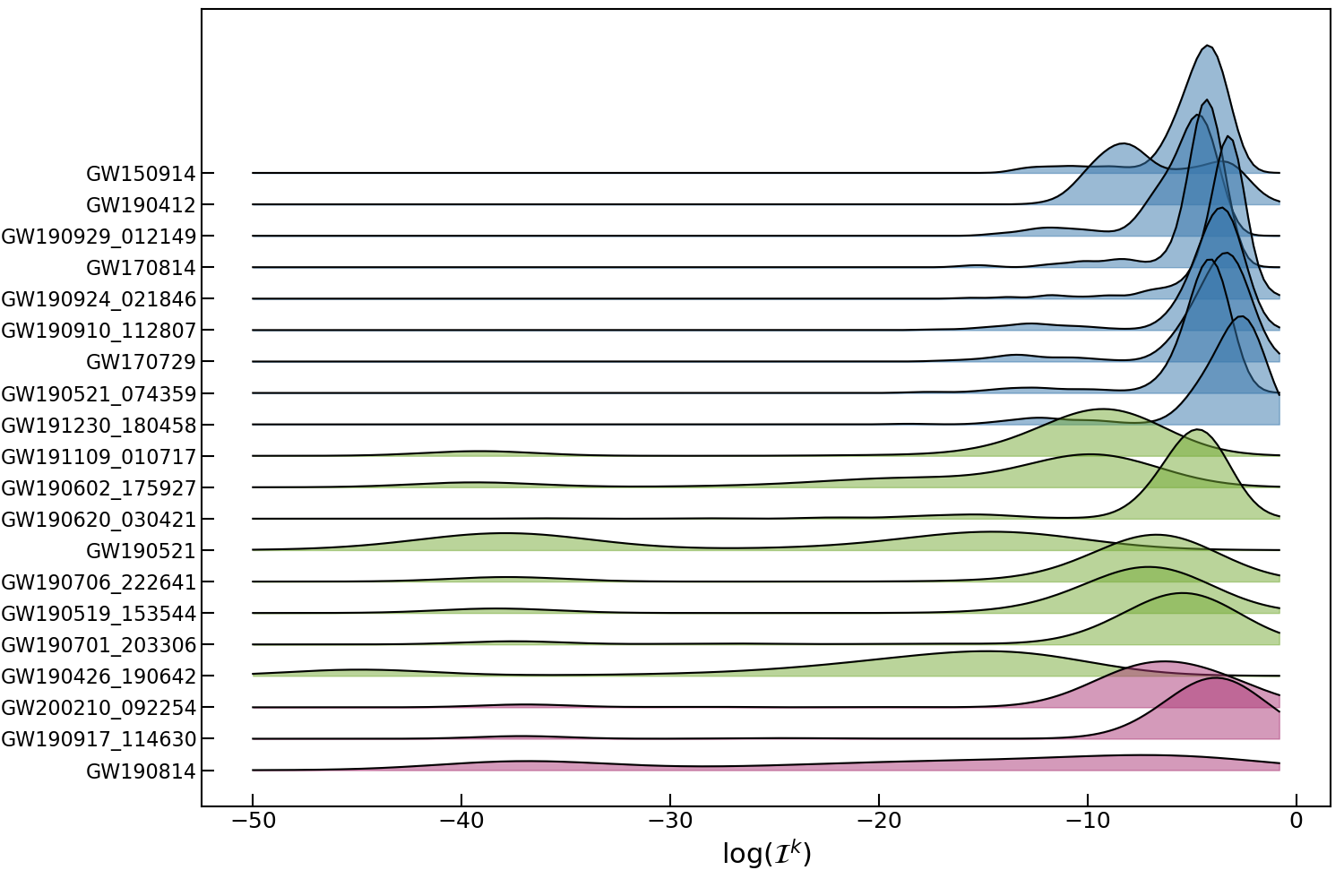}
    \caption{Distribution of $\log(\mathcal{I}^k)$ terms for the 300 new \texttt{COMPAS} models. The top plot shows the GW events where the distribution are constrained to high $\log \mathcal{I}^k$ values and the bottom plot shows GW events with broad distributions. Out of the distinguishable events, the pink distributions correspond to events with one component in the lower mass gap and the green distributions correspond to events with at least one component in the upper mass gap.}
    \label{fig:I_k_new_models}
\end{figure}

Given our range of models and the form with which the likelihood is evaluated by summing over each GW event, we can also ask the question of which events allow for the greatest distinction between our various \texttt{COMPAS} models. For each GW event, we evaluate the distribution of $\log(\mathcal{I}^k)$ terms for all 300 models. We use the range and mean of the distributions to determine if the progenitors of the GW events are distinguishable between models. We also look for skewed distributions, which indicate that some models are more or less likely to produce progenitors of particular GW events. We show an example of 20 of these --- the top with the least variance between models, and the ten with the most --- in Fig.~\ref{fig:I_k_new_models}.

The top plot of Fig.~\ref{fig:I_k_new_models} shows the GW events where their $\mathcal{I}^k$ terms are tightly constrained to high probabilities. This implies that all the models are very likely to produce the progenitors of these events. The distributions appear Gaussian. However, upon focusing on the $\log(\mathcal{I}^k) < -5$ region they appear to have a right skewed distribution, implying there a few models in this tail that are less preferred to form the progenitors of these events. 

\begin{figure}
    
    \centering
    \includegraphics[width=\columnwidth]{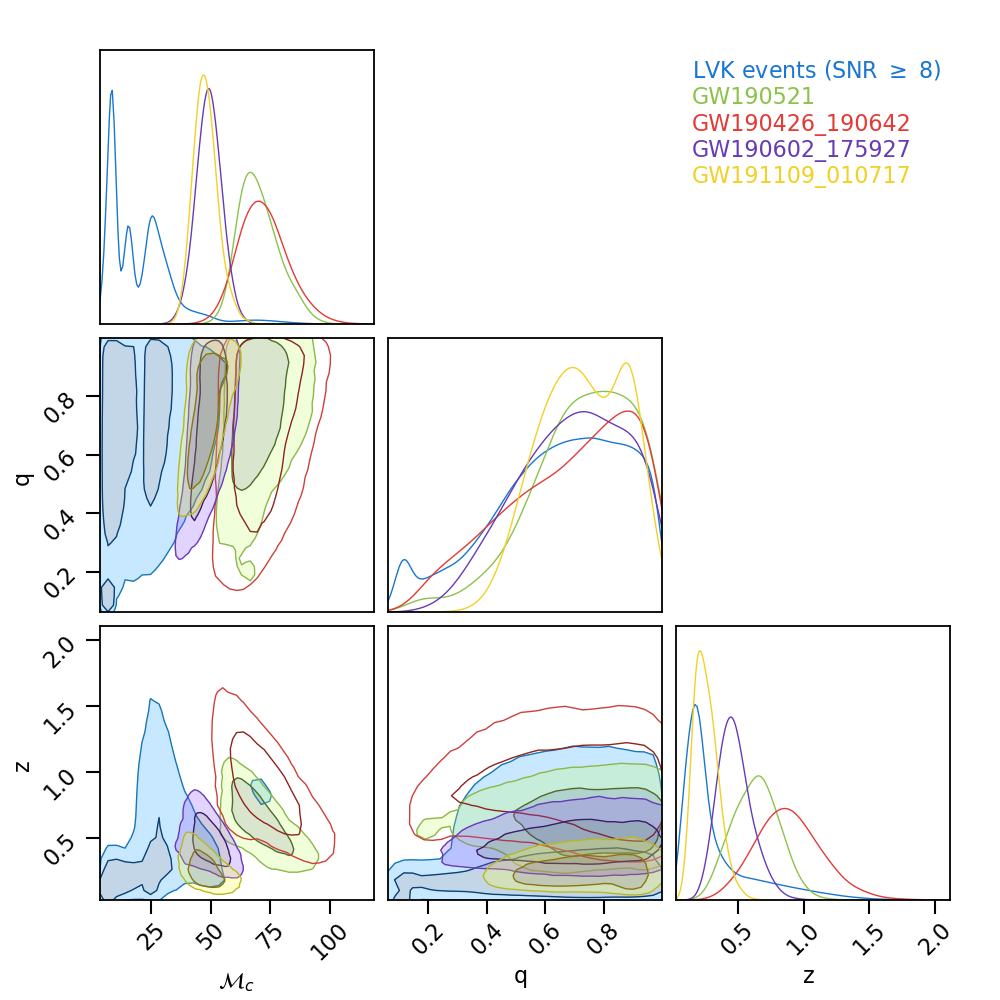}
    \par

    \caption{Posterior distributions for `outlier' events, (identified by the green distributions in Fig.~\ref{fig:I_k_new_models}) that are not well reproduced by the majority of our \texttt{COMPAS} models. The blue contour is the full posterior distribution for LVK data with SNR $\geq$ 8, which clearly indicates the majority of the data are at both lower mass and lower redshift. We show GW190426\_190642 as an unfilled contour, as it does not fit the $p_{\rm astro} \geq 0.9$ and $ \rm FAR \leq 0.25 yr^{-1}$ criteria.}
    \label{fig:Outlier}
\end{figure}

The bottom plot shows the GW events where the progenitors populations are more clearly distinguishable between models. In particular, GW190814 and GW190521 have wide bimodal distributions, apparently separating our \texttt{COMPAS} models into those that can, and those that cannot, reproduce the characteristics of these events. In Figure \ref{fig:Outlier} we show these events with low probability of their progenitors forming in some models. They appear to be close to, or in, the pair-instability mass gap $\sim 50 -130 M_\odot$ \citep{2021ApJ...912L..31W}. GW190426\_190642 has a total mass that supersedes the total mass of GW190521, and it could be argued that they are from the same formation channel. However, GW190426\_190642 is less certain to be a true event, having $p_{\rm astro} = 0.75$ and $\rm FAR = 4.1 yr^{-1}$ \citep{2022A&A...665A..20B}. GW190602\_175927 and GW191109\_010717 have at least one component in the mass gap. The spin-orbit misalignment of GW191109\_010717 provides further evidence it has a dynamical origin \citep{2023ApJ...954...23Z}. However, GW191109\_010717 has been flagged as problematic due to data quality issues, which may lead to unreliable inference of the source parameters such as the precession \citep{2022PhRvD.106j3019T}. GW190917\_114630 and GW200210\_092254 have a secondary component in the lower mass gap between the most massive neutron star and smallest BH, like GW190814. The low probability of generating these events by some models indicate that the isolated formation channel is not ideal for forming unequal mass binaries. However, the ability of some models to form the progenitors of these events suggests that this mass gap might be narrower or non-existent. More observations of similar events is paramount in confirming this \citep{2020ApJ...899L...1Z}. 

Overall, looking at the models that perform well for these events, they tend to produce a dearth of higher mass BHs, with chirp mass distributions that peak around $40M_\odot$ and mass ratio distributions that peak around $q \approx 0.9$. However, this often comes at the expense of producing enough lower mass or more extreme mass ratio binaries, to explain the mass and redshift distribution of the other GW events.

This is particularly emphasised for GW190521 where we find that the posterior samples often did not fall at all within the chirp mass or mass ratio range output from \texttt{COMPAS}. In these cases we assign a probability of zero to these events such that they make no difference to the sum in Eq.~\ref{eq:P_unmarginalised} and \ref{eq:P_marginalised}, but are also hence not represented in Fig.~\ref{fig:I_k_new_models}. Model choices that fail to produce any progenitor close to GW190521 include:
\begin{itemize}
    \item No CHE (i.e., models 1-80 and 241-250), implying that formation channels with mass transfer and CE evolution produce smaller BH masses. 
    \item The remnant mass prescriptions \citet{mandel2020simple} and \citet{schneider2021pre}, which tend to produce smaller BH masses. The exception being model 271-280, which use \citet{fryer2012compact} and \citet{2021ARep...65..937F}.
    \item Models 221 and 270, where CHE is set to Optimistic, but $f_\gamma = 0.75$ (Model 221) and/or $f_{\rm WR} =10$ (Model 270), which may be contributing to extreme mass loss from the progenitors again leading to a population of particularly low-mass BHs. 
\end{itemize}

Overall, the difficulty in simultaneously producing both these larger events and the bulk of LVK data even across the large range of models we explore in this work lends credence to the idea of multiple formation pathways for these events. 


\section{Conclusion} \label{Section 6}
In this work we present a thorough analysis of the ability of over 300 binary population synthesis models to reproduce the GW events observed so far within Gravitational Wave Transients Catalogues 1-3. To enable this, we developed an efficient method for predicting the BBH merger rate in the isolated channel for various \texttt{COMPAS} models (Section~\ref{Section 3}), validated based on previous work from \citet{10.1093/mnras/stad1757}. We then used this within a Bayesian inference method to calculate the likelihood of these models with and without the dependence on the number of GW events predicted by each model, $N_\lambda$ (Section~\ref{Section 4}). 

The \texttt{COMPAS} population synthesis code contains a number of free parameters --- we introduce some of the most salient of these in Section~\ref{Section 2}. These are the mass transfer fraction, CE efficiency, specific AM loss fraction, CHE, remnant mass prescription and Wolf--Rayet winds . We find that various combinations of these hyper-parameters impact the chirp mass, mass ratio and redshift distribution, as well as changing the orbital separations of binaries, which affects the number of BBH mergers within an observed time. Our results fitting these models to data in Section~\ref{Section 5} demonstrate there is a preference for subsets of models dictated by the type of isolated formation channel and specific AM loss. However, we frequently find opposing trends in the likelihoods with and without $N_\lambda$ dependence --- models that produce realistic event-level masses and redshifts frequently over-predict the number of events we should have observed to date. Looking at models which perform well with both likelihoods, we find that models with no CHE, low to moderate values of $f_\gamma$ (such that AM loss occurs closer to the donor) and higher rates of mass-loss from Wolf--Rayet winds ($f_{\rm WR} > 1$) are generally preferred by current data. We provide a quantitative summary of the hyper-parameters for the models that best reproduce the GW population excluding GW190521 at the end of Section~\ref{model_distinguishing}.

We find that the selection effects impact our likelihoods. Although the selection allows for potentially non-astrophysical sources in the sample, the $\rm SNR \geq 8$ cut on the GW events increases the likelihood for \textit{all} models compared to $ \rm FAR \leq 0.25 yr^{-1}$ and $p_{\rm astro} \geq 0.9$. Preference for models do not change with either selection criteria. However, these conclusions may vary with future GW detections, particularly if the BH masses are in the mass gaps. We also use the individual event likelihoods $\mathcal{I}^k$ across all our models term to distinguish individual events that are easy or more difficult to reproduce by our models. We identify the usual suspects GW190521 and GW190814 as typically having low probability of existing in our new models, as well as some outlier events in the the lower and upper mass gap. This motivates future studies to consider joint population inference and model selection studies across multiple formation channels \citep{2022MNRAS.511.1454G,2023arXiv230709097G}. 

Our method for efficiently calculating the merger rate can also be applied to other population synthesis codes. However, we note that it is only applicable in the case where the metallicity dependent SFR is the same for the fiducial and new model. We argue that using a different SFR may not impact the PDF significantly. \citet{2023ApJ...948..105V} finds that variation in the SFR do not impact the features in the primary BH mass distribution and have minimal impact on the low mass end of the distribution. For the redshift range studied in this paper, \textsc{Shark} predicts a metallicity dependent SFR density compared to the fiducial model explored by \citet{2019MNRAS.490.3740N}, which produces a BBH merger rate comparable to GWTC-1. This may be contributing to our models overestimating the number of events. Variations in stellar evolution in population synthesis \citep{2022MNRAS.516.5737B,2024AnP...53600170C} and modelling of galaxy distributions in galaxy simulations\footnote{We refer to Fig.~2 and 3 of \citet{10.1093/mnras/stad1757} for comparisons of \textsc{Shark} to observations.} \citep{2015MNRAS.450.4486F,2024MNRAS.531.3551L} can translate to large uncertainties in the SFR, that will impact the BBH merger rate. These uncertainties are non-negligible and should be accounted for in future studies of BBH model selection. 

We acknowledge this work relies on various assumptions and simplifications. For future work, we consider including the effective spin as a population parameter in the analysis. The covariance between the mass ratio and effective spin can arise due to CE physics, mass transfer efficiency and Eddington limited accretion \citep{2020ApJ...895L..28M, 2021A&A...647A.153B, 2022ApJ...933...86Z, 2022ApJ...938...45B, 2023ApJ...958...13A,2024arXiv240412426O}. This could be used to further distinguish between \texttt{COMPAS} models. Alternatively, we can build the PDF only with $\mathcal{M}_c$, mass ratio or redshift to test the validity of the model outputs. We could also consider including natal kick velocity as one of the hyper-parameters to vary in \texttt{COMPAS} in the future, as this is not well constrained for BHs \citep{2021ApJ...920..157C,2023MNRAS.523.5565C,2024PhRvL.132s1403V}, or more fundamental parameters such as the binary fraction. Varying the selection function cuts applied to current data to restrict the analysis to only the strongest detections may further inform us on trends but given we find consistent results with our two different selection functions this may not necessarily provide improved constraints on the hyper-parameters. 

Future GW detectors will observe GW events at a wider range of frequencies and redshifts, and will provide data with increasing precision \citep{2023arXiv230317628B}. This will answer a variety of astrophysical questions and provide a clearer picture of the BBH population distribution. In the absence of electromagnetic counterparts, BBH merger populations with the \textit{correct} modelling are crucial for an unbiased inference of cosmological parameters \citep{2021PhRvD.104f2009M, 2023JCAP...12..023G}. With deeper catalogues and higher SNR detections, our selection criteria will be more refined, which will hopefully allow us to further constrain the uncertainties in massive binary evolution models. However, as always, with more GW events comes greater change of observing rare exceptional events to challenge these models. Until then, in the case of limited number of GW events we must rely on observations of massive binary stars to constrain the different assumptions for massive binary evolution. Observations of X-ray binaries, red supergiants, luminous blue variables and Wolf--Rayet stars can help constrain binaries at different evolutionary stages \citep{2022ARA&A..60..455E, 2023arXiv231101865M,2024arXiv240511571L}. 


\section*{Acknowledgements}
We thank Carole Perigois and Michella Mapelli for useful discussion on their paper that motivated this work. We thank Ilya Mandel, Eric Thrane and Paul Lasky for useful discussions on the methodology and results. We also thank Floor Broekgaarden for useful discussions on the \texttt{COMPAS} hyper-parameters and relevant literature. We also thank the anonymous reviewer for their helpful comments, that improved the quality of the manuscript. 

The authors acknowledge support from the Australian Research Council (ARC) Centre of Excellence for Gravitational Wave Discovery (OzGrav), through project numbers CE170100004 and CE230100016. LR and CH acknowledge support through ARC Discovery Project DP20220101395. SS is a recipient of an ARC Discovery Early Career Research Award (DE220100241).
\section*{Data Availability}
Prior and posterior samples of GW events were provided by Carole Perigois. Simulations in this paper made use of the COMPAS rapid binary population synthesis code (version v02.38.08), which is freely available at  \href{http://github.com/TeamCOMPAS/COMPAS}{this repository}.



\bibliographystyle{mnras}
\bibliography{example} 




\appendix
\section{Model initial conditions and fixed parameters} \label{Appendix_initial}
In Table~\ref{tab:fid_case}, we list the distributions and ranges for the input parameters used to define the initial conditions of our \texttt{COMPAS} binaries. These are fixed for all models explored in this paper and \citet{10.1093/mnras/stad1757}. 
\begin{table*}
    \centering
    \begin{tabular}{c|c|c}
       \hline
       \hline
       Distribution & Model  & Range \\
       \hline
         \multirow{2}{*}{IMF} & \citet{kroupa2002initial} &  $[5,150]M_\odot$\\
        & & $[5,200]M_\odot$ \citep{10.1093/mnras/stad1757} \\
         \hline
         Semi-major axis & Flat-in-log & $[0.01,1000]$ AU\\
         \hline
         Orbital period & Flat-in-log & $[0.1,1000]$ days\\
         \hline
         Mass ratio & \citet{sana2012binary} & $[0.01,1]$ \\
         \hline
         \multirow{2}{*}{Eccentricity} & ZERO (circular) & 0  \\
         & \citet{sana2012binary} \citep{10.1093/mnras/stad1757} & $[0,1]$\\
         \hline
         Log Metallicity & \textsc{Shark} \citep{2018MNRAS.481.3573L} & $[-7.0,-0.3]$ \\
       \hline
    \end{tabular}
    \caption{Summary of COMPAS default initial conditions and range.}
    \label{tab:fid_case}
\end{table*}

\section{GW Data} \label{Appendix_GW}
In Table~\ref{tab:GW_list}, we list the GW events used as data for each selection criteria, as described in Section~\ref{Selection_criteria}.
\begin{table*}
    \small
    \centering
    \resizebox{0.80\columnwidth}{!}{\begin{tabular}{|c|c|c|}
    \hline
        GW events & $ \rm FAR \leq 0.25 yr^{-1}$ and $p_{\rm astro} \geq 0.9$ & $\rm SNR \geq 8$ \\ \hline
        GW150914 & \checkmark & \checkmark \\ \hline
        GW151012 & \checkmark & \checkmark \\ \hline
        GW151226 & \checkmark & \checkmark \\ \hline
        GW170104 & \checkmark & \checkmark \\ \hline
        GW170608 & \checkmark & \checkmark \\ \hline
        GW170729 & \checkmark & \checkmark \\ \hline
        GW170809 & \checkmark & \checkmark \\ \hline
        GW170814 & \checkmark & \checkmark \\ \hline
        GW170818 & \checkmark & \checkmark \\ \hline
        GW170823 & \checkmark & \checkmark \\ \hline
        GW190408\_181802 & \checkmark & \checkmark \\ \hline
        GW190412 & \checkmark & \checkmark \\ \hline
        GW190413\_052954 &  & \checkmark \\ \hline
        GW190413\_134308 & \checkmark & \checkmark \\ \hline
        GW190421\_213856 & \checkmark & \checkmark \\ \hline
        GW190424\_180648 &  & \checkmark \\ \hline
        GW190426\_190642 &  & \checkmark \\ \hline
        GW190503\_185404 & \checkmark & \checkmark \\ \hline
        GW190512\_180714 & \checkmark & \checkmark \\ \hline
        GW190513\_205428 & \checkmark & \checkmark \\ \hline
        GW190517\_055101 & \checkmark & \checkmark \\ \hline
        GW190519\_153544 & \checkmark & \checkmark \\ \hline
        GW190521\_074359 & \checkmark & \checkmark \\ \hline
        GW190521 & \checkmark & \checkmark \\ \hline
        GW190527\_092055 &  & \checkmark \\ \hline
        GW190602\_175927 & \checkmark & \checkmark \\ \hline
        GW190620\_030421 & \checkmark & \checkmark \\ \hline
        GW190630\_185205 & \checkmark & \checkmark \\ \hline
        GW190701\_203306 & \checkmark & \checkmark \\ \hline
        GW190706\_222641 & \checkmark & \checkmark \\ \hline
        GW190707\_093326 & \checkmark & \checkmark \\ \hline
        GW190708\_232457 & \checkmark & \checkmark \\ \hline
        GW190719\_215514 &  & \checkmark \\ \hline
        GW190720\_000836 & \checkmark & \checkmark \\ \hline
        GW190725\_174728 &  & \checkmark \\ \hline
        GW190727\_060333 & \checkmark & \checkmark \\ \hline
        GW190728\_064510 & \checkmark & \checkmark \\ \hline
        GW190731\_140936 &  & \checkmark \\ \hline
        GW190803\_022701 & \checkmark & \checkmark \\ \hline
        GW190805\_211137 &  & \checkmark \\ \hline
        GW190814 & \checkmark & \checkmark \\ \hline
        GW190828\_063405 & \checkmark & \checkmark \\ \hline
        GW190828\_065509 & \checkmark & \checkmark \\ \hline
        GW190910\_112807 & \checkmark & \checkmark \\ \hline
        GW190915\_235702 & \checkmark & \checkmark \\ \hline
        GW190916\_200658 & & \checkmark \\ \hline
        GW190917\_114630 & & \checkmark \\ \hline
        GW190924\_021846 & \checkmark & \checkmark \\ \hline
        GW190925\_232845 & \checkmark & \checkmark \\ \hline
        GW190926\_050336 &  & \checkmark \\ \hline
        GW190929\_012149 &  & \checkmark \\ \hline
        GW190930\_133541 & \checkmark & \checkmark \\ \hline
        GW191103\_012549 &  & \checkmark \\ \hline
        GW191105\_143521 & \checkmark & \checkmark \\ \hline
        GW191109\_010717 & \checkmark & \checkmark \\ \hline
        GW191126\_115259 &  & \checkmark \\ \hline
        GW191127\_050227 &  & \checkmark \\ \hline
        GW191129\_134029 & \checkmark & \checkmark \\ \hline
        GW191204\_110529 &  & \checkmark \\ \hline
        GW191204\_171526 & \checkmark & \checkmark \\ \hline
        GW191215\_223052 & \checkmark & \checkmark \\ \hline
        GW191216\_213338 & \checkmark & \checkmark \\ \hline
        GW191222\_033537 & \checkmark & \checkmark \\ \hline
        GW191230\_180458 & \checkmark & \checkmark \\ \hline
        GW200112\_155838 & \checkmark & \checkmark \\ \hline
        GW200128\_022011 & \checkmark & \checkmark \\ \hline
        GW200129\_065458 & \checkmark & \checkmark \\ \hline
        GW200202\_154313 & \checkmark & \checkmark \\ \hline
        GW200208\_130117 & \checkmark & \checkmark \\ \hline
        GW200209\_085452 & \checkmark & \checkmark \\ \hline
        GW200210\_092254 &  & \checkmark \\ \hline
        GW200216\_220804 &  & \checkmark \\ \hline
        GW200219\_094415 & \checkmark & \checkmark \\ \hline
        GW200220\_124850 &  & \checkmark \\ \hline
        GW200224\_222234 & \checkmark & \checkmark \\ \hline
        GW200225\_060421 & \checkmark & \checkmark \\ \hline
        GW200302\_015811 & \checkmark & \checkmark \\ \hline
        GW200311\_115853 & \checkmark & \checkmark \\ \hline
        GW200316\_215756 & \checkmark & \checkmark \\ \hline
    \end{tabular}}
    \caption{List of GW events included in the analysis for each selection criteria.}
    \label{tab:GW_list}
\end{table*}

\section{Model Hyper-Parameters and Bayes Factor Data} \label{Appendix}
Extending on Table~\ref{tab:parameters}, for each model in Tables~\ref{tab:all_models}, \ref{tab:all_models_2} and \ref{tab:all_models_3} we detail the hyper-parameters and Bayes factor for each selection criteria. \citet{10.1093/mnras/stad1757} (a)-(d) refer to the models explored in \citet{10.1093/mnras/stad1757}, where we vary the remnant mass prescription (with the exception of \citet{10.1093/mnras/stad1757} (c) where we also vary $f_{\rm WR}$). 
\begin{table*}
    \tiny
    \centering
     \resizebox{1.05\textwidth}{!}{\begin{tabular}{|c|c|c|c|c|c|c|c|c|c|c|c|c|}
    \hline
        \multirow{2}{*}{Model} & \multirow{2}{*}{CHE} & Remnant mass & \multirow{2}{*}{$f_{\rm WR}$} & \multirow{2}{*}{$\gamma$ prescription} & \multirow{2}{*}{$f_\gamma$} & \multirow{2}{*}{$\alpha_{\rm CE}$} & $\dot{M}_a$ & \multirow{2}{*}{$f_{\rm MT}$} & $\log_{10} \mathcal{B}(\lambda|\mathcal{H})$ & $ \log_{10} \mathcal{B}(\lambda|\mathcal{H})$ & $\log_{10} \mathcal{B}(\lambda, N_\lambda|\mathcal{H})$ & $ \log_{10} \mathcal{B}(\lambda, N_\lambda|\mathcal{H})$ \\ 
        & & prescription & & & & & prescription & & & ($\rm SNR \geq 8$) & & ($\rm SNR \geq 8$) \\\hline
  Rauf et al (2023) (a) & NONE & FRYER2012 & 1 & ISOTROPIC & 0 & 1 & THERMAL & 1 & -10.1 $\pm$ 1.7 & -10.9 $\pm$ 1.6 & -66.8 $\pm$ 6.1 & -57.3 $\pm$ 6.8 \\ \hline
\rowcolor{Green!30}
        \textbf{Rauf et al (2023) (b)} & \textbf{NONE} & \textbf{MULLERMANDEL} & \textbf{1} & \textbf{ISOTROPIC} & \textbf{0} & \textbf{1} & \textbf{THERMAL} & \textbf{1} & \textbf{-3 $\pm$ 2.2} & \textbf{-0.4 $\pm$ 2.5} & \textbf{0 $\pm$ 2.7} & \textbf{-0.3 $\pm$ 4.6} \\ \hline
        Rauf et al (2023) (c) & NONE & MULLERMANDEL & 0.2 & ISOTROPIC & 0 & 1 & THERMAL & 1 & -30.8 $\pm$ 3.4 & -46.7 $\pm$ 4.4 & -41.8 $\pm$ 5.9 & -53.2 $\pm$ 7.1 \\ \hline
        Rauf et al (2023) (d) & NONE & SCHNEIDER2020 & 1 & ISOTROPIC & 0 & 1 & THERMAL & 1 & -39.3 $\pm$ 8.6 & -50.5 $\pm$ 10.6 & -39.5 $\pm$ 8.8 & -57.8 $\pm$ 11.4 \\ \hline
        1 & NONE & MULLERMANDEL & 0.2 & MACLEOD\_LINEAR & 0 & 0.1 & FIXED & 0.1 & -37.7 $\pm$ 3.8 & -61.5 $\pm$ 2.8 & -35.6 $\pm$ 4.9 & -59.9 $\pm$ 5.5 \\ \hline
        2 & NONE & MULLERMANDEL & 0.2 & MACLEOD\_LINEAR & 0 & 0.1 & FIXED & 0.2 & -22.7 $\pm$ 3.6 & -37 $\pm$ 2.6 & -45.5 $\pm$ 7.2 & -53.1 $\pm$ 7.6 \\ \hline
        3 & NONE & MULLERMANDEL & 0.2 & MACLEOD\_LINEAR & 0 & 0.1 & FIXED & 0.5 & -26.8 $\pm$ 2.7 & -44.1 $\pm$ 1.9 & -195.6 $\pm$ 15.2 & -196.9 $\pm$ 15.4 \\ \hline
        4 & NONE & MULLERMANDEL & 0.2 & MACLEOD\_LINEAR & 0 & 0.1 & FIXED & 1 & -25.7 $\pm$ 2.2 & -43.3 $\pm$ 1.4 & -325.1 $\pm$ 18.8 & -322.9 $\pm$ 19.1 \\ \hline
        5 & NONE & MULLERMANDEL & 0.2 & MACLEOD\_LINEAR & 0 & 1 & FIXED & 0.1 & -62.3 $\pm$ 1.3 & -83.3 $\pm$ 0.7 & -114.4 $\pm$ 3.7 & -125.5 $\pm$ 5 \\ \hline
        6 & NONE & MULLERMANDEL & 0.2 & MACLEOD\_LINEAR & 0 & 1 & FIXED & 0.2 & -34 $\pm$ 1.3 & -52.7 $\pm$ 0.7 & -126.2 $\pm$ 4.7 & -132.1 $\pm$ 5.8 \\ \hline
        7 & NONE & MULLERMANDEL & 0.2 & MACLEOD\_LINEAR & 0 & 1 & FIXED & 0.5 & -27.7 $\pm$ 1.5 & -37 $\pm$ 0.8 & -257.3 $\pm$ 10 & -248.6 $\pm$ 10.6 \\ \hline
        8 & NONE & MULLERMANDEL & 0.2 & MACLEOD\_LINEAR & 0 & 1 & FIXED & 1 & -24.3 $\pm$ 1.8 & -34.3 $\pm$ 1.1 & -363 $\pm$ 17.3 & -352.3 $\pm$ 17.6 \\ \hline
        9 & NONE & MULLERMANDEL & 0.2 & MACLEOD\_LINEAR & 0 & 10 & FIXED & 0.1 & -43.9 $\pm$ 6.1 & -66.1 $\pm$ 4.8 & -42.3 $\pm$ 7.6 & -64.5 $\pm$ 7.5 \\ \hline
        10 & NONE & MULLERMANDEL & 0.2 & MACLEOD\_LINEAR & 0 & 10 & FIXED & 0.2 & -24.5 $\pm$ 4.9 & -41.8 $\pm$ 3.7 & -55.2 $\pm$ 10.4 & -64.8 $\pm$ 10.6 \\ \hline
        11 & NONE & MULLERMANDEL & 0.2 & MACLEOD\_LINEAR & 0 & 10 & FIXED & 0.5 & -23.9 $\pm$ 2.9 & -41 $\pm$ 2 & -210.2 $\pm$ 17.3 & -210.5 $\pm$ 17.5 \\ \hline
        12 & NONE & MULLERMANDEL & 0.2 & MACLEOD\_LINEAR & 0 & 10 & FIXED & 1 & -24.9 $\pm$ 2.1 & -33 $\pm$ 1.3 & -335.5 $\pm$ 18.4 & -323.6 $\pm$ 18.7 \\ \hline
        13 & NONE & MULLERMANDEL & 0.2 & MACLEOD\_LINEAR & 0 & 100 & FIXED & 0.1 & -34.9 $\pm$ 8.4 & -53.1 $\pm$ 6.8 & -38.2 $\pm$ 8.7 & -65.8 $\pm$ 8.1 \\ \hline
        14 & NONE & MULLERMANDEL & 0.2 & MACLEOD\_LINEAR & 0 & 100 & FIXED & 0.2 & -16.6 $\pm$ 6.1 & -34.4 $\pm$ 4.9 & -29 $\pm$ 10.1 & -41.9 $\pm$ 10.1 \\ \hline
        15 & NONE & MULLERMANDEL & 0.2 & MACLEOD\_LINEAR & 0 & 100 & FIXED & 0.5 & -24.7 $\pm$ 3.2 & -35.2 $\pm$ 2.3 & -194.1 $\pm$ 17.5 & -188.3 $\pm$ 17.7 \\ \hline
        16 & NONE & MULLERMANDEL & 0.2 & MACLEOD\_LINEAR & 0 & 100 & FIXED & 1 & -25.8 $\pm$ 2.5 & -38 $\pm$ 1.7 & -325.4 $\pm$ 21.7 & -317.7 $\pm$ 21.9 \\ \hline
\rowcolor{Green!30}
        17 & NONE & MULLERMANDEL & 0.2 & MACLEOD\_LINEAR & 0.25 & 0.1 & FIXED & 0.1 & -8.8 $\pm$ 2.5 & -15 $\pm$ 1.6 & -166.8 $\pm$ 13.3 & -157.2 $\pm$ 13.6 \\ \hline
        18 & NONE & MULLERMANDEL & 0.2 & MACLEOD\_LINEAR & 0.25 & 0.1 & FIXED & 0.2 & -12.2 $\pm$ 1.8 & -19.7 $\pm$ 1.1 & -299.1 $\pm$ 15.1 & -287.1 $\pm$ 15.5 \\ \hline
        19 & NONE & MULLERMANDEL & 0.2 & MACLEOD\_LINEAR & 0.25 & 0.1 & FIXED & 0.5 & -17.4 $\pm$ 1.5 & -24.9 $\pm$ 0.8 & -566.4 $\pm$ 19.6 & -550 $\pm$ 19.9 \\ \hline
        20 & NONE & MULLERMANDEL & 0.2 & MACLEOD\_LINEAR & 0.25 & 0.1 & FIXED & 1 & -11.6 $\pm$ 1.5 & -16.5 $\pm$ 0.8 & -505.8 $\pm$ 19 & -487.5 $\pm$ 19.3 \\ \hline
        21 & NONE & MULLERMANDEL & 0.2 & MACLEOD\_LINEAR & 0.25 & 1 & FIXED & 0.1 & -15.7 $\pm$ 1.3 & -23.3 $\pm$ 0.7 & -256.2 $\pm$ 7.8 & -245.5 $\pm$ 8.5 \\ \hline
\rowcolor{Green!30}
        22 & NONE & MULLERMANDEL & 0.2 & MACLEOD\_LINEAR & 0.25 & 1 & FIXED & 0.2 & -8.6 $\pm$ 1.2 & -15.7 $\pm$ 0.8 & -374 $\pm$ 9.1 & -359.9 $\pm$ 9.8 \\ \hline
        23 & NONE & MULLERMANDEL & 0.2 & MACLEOD\_LINEAR & 0.25 & 1 & FIXED & 0.5 & -10.1 $\pm$ 1.2 & -16.1 $\pm$ 0.8 & -564.2 $\pm$ 13.4 & -546.2 $\pm$ 13.9 \\ \hline
\rowcolor{Green!30}
        24 & NONE & MULLERMANDEL & 0.2 & MACLEOD\_LINEAR & 0.25 & 1 & FIXED & 1 & -7.4 $\pm$ 1.4 & -11.2 $\pm$ 0.8 & -440.3 $\pm$ 15.6 & -421.9 $\pm$ 16 \\ \hline
        25 & NONE & MULLERMANDEL & 0.2 & MACLEOD\_LINEAR & 0.25 & 10 & FIXED & 0.1 & -17.2 $\pm$ 2.4 & -26.4 $\pm$ 1.6 & -205.3 $\pm$ 14.5 & -197.6 $\pm$ 14.9 \\ \hline
        26 & NONE & MULLERMANDEL & 0.2 & MACLEOD\_LINEAR & 0.25 & 10 & FIXED & 0.2 & -26.4 $\pm$ 1.8 & -40.5 $\pm$ 1.1 & -338.9 $\pm$ 16.4 & -332.9 $\pm$ 16.7 \\ \hline
        27 & NONE & MULLERMANDEL & 0.2 & MACLEOD\_LINEAR & 0.25 & 10 & FIXED & 0.5 & -22.5 $\pm$ 1.4 & -30.9 $\pm$ 0.8 & -594.5 $\pm$ 20 & -578.6 $\pm$ 20.3 \\ \hline
        28 & NONE & MULLERMANDEL & 0.2 & MACLEOD\_LINEAR & 0.25 & 10 & FIXED & 1 & -14.9 $\pm$ 1.5 & -19.6 $\pm$ 0.8 & -470.7 $\pm$ 17.7 & -452.7 $\pm$ 18.1 \\ \hline
        29 & NONE & MULLERMANDEL & 0.2 & MACLEOD\_LINEAR & 0.25 & 100 & FIXED & 0.1 & -13.2 $\pm$ 2.9 & -31.7 $\pm$ 2 & -162.7 $\pm$ 14.8 & -165.7 $\pm$ 15.1 \\ \hline
        30 & NONE & MULLERMANDEL & 0.2 & MACLEOD\_LINEAR & 0.25 & 100 & FIXED & 0.2 & -24.4 $\pm$ 2 & -35.8 $\pm$ 1.2 & -319.4 $\pm$ 17.4 & -311.1 $\pm$ 17.7 \\ \hline
        31 & NONE & MULLERMANDEL & 0.2 & MACLEOD\_LINEAR & 0.25 & 100 & FIXED & 0.5 & -16 $\pm$ 1.6 & -23.6 $\pm$ 0.9 & -506.4 $\pm$ 20.6 & -490.8 $\pm$ 20.9 \\ \hline
        32 & NONE & MULLERMANDEL & 0.2 & MACLEOD\_LINEAR & 0.25 & 100 & FIXED & 1 & -14.4 $\pm$ 1.8 & -21.3 $\pm$ 1.1 & -443.5 $\pm$ 20.8 & -428.2 $\pm$ 21 \\ \hline
        33 & NONE & MULLERMANDEL & 0.2 & MACLEOD\_LINEAR & 0.5 & 0.1 & FIXED & 0.1 & -10.9 $\pm$ 1.2 & -16.2 $\pm$ 0.7 & -499.5 $\pm$ 13 & -481.7 $\pm$ 13.4 \\ \hline
\rowcolor{Green!30}
        34 & NONE & MULLERMANDEL & 0.2 & MACLEOD\_LINEAR & 0.5 & 0.1 & FIXED & 0.2 & -6.9 $\pm$ 1.2 & -11.2 $\pm$ 0.7 & -442 $\pm$ 12.1 & -424 $\pm$ 12.6 \\ \hline
\rowcolor{Green!30}
        35 & NONE & MULLERMANDEL & 0.2 & MACLEOD\_LINEAR & 0.5 & 0.1 & FIXED & 0.5 & -3.5 $\pm$ 1.2 & -7.1 $\pm$ 0.8 & -569.3 $\pm$ 13.7 & -548.8 $\pm$ 14.1 \\ \hline
\rowcolor{Green!30}
        36 & NONE & MULLERMANDEL & 0.2 & MACLEOD\_LINEAR & 0.5 & 0.1 & FIXED & 1 & -3.2 $\pm$ 1.7 & -6 $\pm$ 0.9 & -339.7 $\pm$ 15.6 & -321.9 $\pm$ 15.9 \\ \hline
\rowcolor{Green!30}
        37 & NONE & MULLERMANDEL & 0.2 & MACLEOD\_LINEAR & 0.5 & 1 & FIXED & 0.1 & -7.3 $\pm$ 1.2 & -11.7 $\pm$ 0.9 & -522.6 $\pm$ 9.4 & -503.5 $\pm$ 10 \\ \hline
\rowcolor{Green!30}
        38 & NONE & MULLERMANDEL & 0.2 & MACLEOD\_LINEAR & 0.5 & 1 & FIXED & 0.2 & -8.9 $\pm$ 1.2 & -13.4 $\pm$ 0.8 & -476.7 $\pm$ 9.2 & -458.3 $\pm$ 9.9 \\ \hline
\rowcolor{Green!30}
        39 & NONE & MULLERMANDEL & 0.2 & MACLEOD\_LINEAR & 0.5 & 1 & FIXED & 0.5 & -3.1 $\pm$ 1.2 & -6 $\pm$ 0.9 & -553 $\pm$ 9.9 & -532 $\pm$ 10.5 \\ \hline
\rowcolor{Green!30}
        40 & NONE & MULLERMANDEL & 0.2 & MACLEOD\_LINEAR & 0.5 & 1 & FIXED & 1 & -2.4 $\pm$ 1.5 & -3.3 $\pm$ 0.9 & -378.3 $\pm$ 15.4 & -358 $\pm$ 15.8 \\ \hline
\rowcolor{Green!30}
        41 & NONE & MULLERMANDEL & 0.2 & MACLEOD\_LINEAR & 0.5 & 10 & FIXED & 0.1 & -8.6 $\pm$ 1.2 & -12.9 $\pm$ 0.8 & -470.3 $\pm$ 10.9 & -451.8 $\pm$ 11.4 \\ \hline
\rowcolor{Green!30}
        42 & NONE & MULLERMANDEL & 0.2 & MACLEOD\_LINEAR & 0.5 & 10 & FIXED & 0.2 & -8.8 $\pm$ 1.2 & -13.7 $\pm$ 0.8 & -435.8 $\pm$ 9.4 & -418.4 $\pm$ 10 \\ \hline
\rowcolor{Green!30}
        43 & NONE & MULLERMANDEL & 0.2 & MACLEOD\_LINEAR & 0.5 & 10 & FIXED & 0.5 & -5.6 $\pm$ 1.2 & -8.5 $\pm$ 0.8 & -526.7 $\pm$ 12.2 & -506 $\pm$ 12.7 \\ \hline
\rowcolor{Green!30}
        \textbf{44} & \textbf{NONE} & \textbf{MULLERMANDEL} & \textbf{0.2} & \textbf{MACLEOD\_LINEAR} & \textbf{0.5} & \textbf{10} & \textbf{FIXED} & \textbf{1} & \textbf{0 $\pm$ 1.6} & \textbf{0 $\pm$ 0.9} & \textbf{-331.9 $\pm$ 14.2} & \textbf{-311.5 $\pm$ 14.6} \\ \hline
        45 & NONE & MULLERMANDEL & 0.2 & MACLEOD\_LINEAR & 0.5 & 100 & FIXED & 0.1 & -13.6 $\pm$ 1.2 & -20.6 $\pm$ 0.7 & -466.2 $\pm$ 12.7 & -450.6 $\pm$ 13.2 \\ \hline
\rowcolor{Green!30}
        46 & NONE & MULLERMANDEL & 0.2 & MACLEOD\_LINEAR & 0.5 & 100 & FIXED & 0.2 & -8.9 $\pm$ 1.3 & -13.2 $\pm$ 0.7 & -408.9 $\pm$ 11.9 & -391.5 $\pm$ 12.4 \\ \hline
\rowcolor{Green!30}
        47 & NONE & MULLERMANDEL & 0.2 & MACLEOD\_LINEAR & 0.5 & 100 & FIXED & 0.5 & -3.7 $\pm$ 1.3 & -6.4 $\pm$ 0.7 & -500.9 $\pm$ 15.1 & -480.4 $\pm$ 15.5 \\ \hline
\rowcolor{Green!30}
        48 & NONE & MULLERMANDEL & 0.2 & MACLEOD\_LINEAR & 0.5 & 100 & FIXED & 1 & -7.8 $\pm$ 2 & -11.6 $\pm$ 1.2 & -305.2 $\pm$ 17 & -289.2 $\pm$ 17.3 \\ \hline
        49 & NONE & MULLERMANDEL & 0.2 & MACLEOD\_LINEAR & 0.75 & 0.1 & FIXED & 0.1 & -110.2 $\pm$ 3.2 & -134.4 $\pm$ 2.3 & -122 $\pm$ 5.7 & -141.4 $\pm$ 6.3 \\ \hline
        50 & NONE & MULLERMANDEL & 0.2 & MACLEOD\_LINEAR & 0.75 & 0.1 & FIXED & 0.2 & -109.7 $\pm$ 3.1 & -134.1 $\pm$ 2.2 & -126.6 $\pm$ 5.9 & -145.2 $\pm$ 6.5 \\ \hline
        51 & NONE & MULLERMANDEL & 0.2 & MACLEOD\_LINEAR & 0.75 & 0.1 & FIXED & 0.5 & -23.2 $\pm$ 3 & -41 $\pm$ 2.1 & -75.9 $\pm$ 8.4 & -83.7 $\pm$ 8.9 \\ \hline
\rowcolor{Green!30}
        52 & NONE & MULLERMANDEL & 0.2 & MACLEOD\_LINEAR & 0.75 & 0.1 & FIXED & 1 & -2 $\pm$ 2.8 & -5.1 $\pm$ 2 & -78.1 $\pm$ 9.6 & -69.5 $\pm$ 10 \\ \hline
        53 & NONE & MULLERMANDEL & 0.2 & MACLEOD\_LINEAR & 0.75 & 1 & FIXED & 0.1 & -123.3 $\pm$ 2.3 & -151.5 $\pm$ 1.5 & -152.2 $\pm$ 5.2 & -172.9 $\pm$ 6 \\ \hline
        54 & NONE & MULLERMANDEL & 0.2 & MACLEOD\_LINEAR & 0.75 & 1 & FIXED & 0.2 & -133.8 $\pm$ 2.3 & -162.7 $\pm$ 1.5 & -174.6 $\pm$ 5.9 & -194.7 $\pm$ 6.7 \\ \hline
        55 & NONE & MULLERMANDEL & 0.2 & MACLEOD\_LINEAR & 0.75 & 1 & FIXED & 0.5 & -40.8 $\pm$ 2.3 & -61 $\pm$ 1.4 & -93.2 $\pm$ 6.5 & -103.4 $\pm$ 7.2 \\ \hline
        56 & NONE & MULLERMANDEL & 0.2 & MACLEOD\_LINEAR & 0.75 & 1 & FIXED & 1 & -1.3 $\pm$ 2.4 & -1 $\pm$ 1.6 & -73.9 $\pm$ 8.1 & -62.1 $\pm$ 8.7 \\ \hline
        57 & NONE & MULLERMANDEL & 0.2 & MACLEOD\_LINEAR & 0.75 & 10 & FIXED & 0.1 & -119.8 $\pm$ 3.7 & -148.9 $\pm$ 2.7 & -119 $\pm$ 5 & -147.6 $\pm$ 5.6 \\ \hline
        58 & NONE & MULLERMANDEL & 0.2 & MACLEOD\_LINEAR & 0.75 & 10 & FIXED & 0.2 & -126.1 $\pm$ 3.6 & -155.1 $\pm$ 2.6 & -124.6 $\pm$ 4.7 & -153.5 $\pm$ 5.4 \\ \hline
        59 & NONE & MULLERMANDEL & 0.2 & MACLEOD\_LINEAR & 0.75 & 10 & FIXED & 0.5 & -42.5 $\pm$ 3.4 & -62.9 $\pm$ 2.4 & -53.1 $\pm$ 5.7 & -69 $\pm$ 6.3 \\ \hline
\rowcolor{Green!30}
        60 & NONE & MULLERMANDEL & 0.2 & MACLEOD\_LINEAR & 0.75 & 10 & FIXED & 1 & -2.4 $\pm$ 3.2 & -1.2 $\pm$ 2.3 & -45.9 $\pm$ 8.2 & -35.5 $\pm$ 8.7 \\ \hline
        61 & NONE & MULLERMANDEL & 0.2 & MACLEOD\_LINEAR & 0.75 & 100 & FIXED & 0.1 & -124.4 $\pm$ 5 & -154.3 $\pm$ 3.8 & -124.5 $\pm$ 5.4 & -161.8 $\pm$ 5.7 \\ \hline
        62 & NONE & MULLERMANDEL & 0.2 & MACLEOD\_LINEAR & 0.75 & 100 & FIXED & 0.2 & -99.8 $\pm$ 4.5 & -123.9 $\pm$ 3.4 & -101 $\pm$ 4.9 & -133.2 $\pm$ 5.3 \\ \hline
        63 & NONE & MULLERMANDEL & 0.2 & MACLEOD\_LINEAR & 0.75 & 100 & FIXED & 0.5 & -19 $\pm$ 4.7 & -34.6 $\pm$ 3.6 & -16.1 $\pm$ 5.7 & -33.3 $\pm$ 6 \\ \hline
        64 & NONE & MULLERMANDEL & 0.2 & MACLEOD\_LINEAR & 0.75 & 100 & FIXED & 1 & -16.4 $\pm$ 5.3 & -21.6 $\pm$ 4.2 & -25 $\pm$ 8.4 & -26.2 $\pm$ 8.5 \\ \hline
        65 & NONE & MULLERMANDEL & 0.2 & MACLEOD\_LINEAR & 1 & 0.1 & FIXED & 0.1 & -112.6 $\pm$ 3.2 & -137.3 $\pm$ 2.2 & -163.4 $\pm$ 8.7 & -178.3 $\pm$ 9.1 \\ \hline
        66 & NONE & MULLERMANDEL & 0.2 & MACLEOD\_LINEAR & 1 & 0.1 & FIXED & 0.2 & -141.1 $\pm$ 3.6 & -169.2 $\pm$ 2.6 & -181.7 $\pm$ 8.8 & -200.8 $\pm$ 9.2 \\ \hline
        67 & NONE & MULLERMANDEL & 0.2 & MACLEOD\_LINEAR & 1 & 0.1 & FIXED & 0.5 & -121.5 $\pm$ 3.3 & -148.7 $\pm$ 2.4 & -166.5 $\pm$ 8.6 & -184.4 $\pm$ 9 \\ \hline
        68 & NONE & MULLERMANDEL & 0.2 & MACLEOD\_LINEAR & 1 & 0.1 & FIXED & 1 & -115.7 $\pm$ 3 & -142.7 $\pm$ 2.1 & -178.6 $\pm$ 9 & -194.7 $\pm$ 9.4 \\ \hline
        69 & NONE & MULLERMANDEL & 0.2 & MACLEOD\_LINEAR & 1 & 1 & FIXED & 0.1 & -99.5 $\pm$ 2.6 & -124.2 $\pm$ 1.7 & -150.3 $\pm$ 7.2 & -165.1 $\pm$ 7.8 \\ \hline
        70 & NONE & MULLERMANDEL & 0.2 & MACLEOD\_LINEAR & 1 & 1 & FIXED & 0.2 & -99.3 $\pm$ 2.5 & -123.2 $\pm$ 1.7 & -142.9 $\pm$ 6.6 & -157.5 $\pm$ 7.3 \\ \hline
        71 & NONE & MULLERMANDEL & 0.2 & MACLEOD\_LINEAR & 1 & 1 & FIXED & 0.5 & -121.1 $\pm$ 2.6 & -147.1 $\pm$ 1.7 & -170.1 $\pm$ 7.1 & -186.5 $\pm$ 7.7 \\ \hline
        72 & NONE & MULLERMANDEL & 0.2 & MACLEOD\_LINEAR & 1 & 1 & FIXED & 1 & -126.2 $\pm$ 2.5 & -153.3 $\pm$ 1.7 & -172 $\pm$ 6.8 & -189.7 $\pm$ 7.5 \\ \hline
        73 & NONE & MULLERMANDEL & 0.2 & MACLEOD\_LINEAR & 1 & 10 & FIXED & 0.1 & -96.1 $\pm$ 4.6 & -121.1 $\pm$ 3.5 & -110.4 $\pm$ 8.1 & -130.1 $\pm$ 8.3 \\ \hline
        74 & NONE & MULLERMANDEL & 0.2 & MACLEOD\_LINEAR & 1 & 10 & FIXED & 0.2 & -125.5 $\pm$ 4.8 & -152 $\pm$ 3.7 & -132.9 $\pm$ 7.5 & -155.7 $\pm$ 7.7 \\ \hline
        75 & NONE & MULLERMANDEL & 0.2 & MACLEOD\_LINEAR & 1 & 10 & FIXED & 0.5 & -113 $\pm$ 4.7 & -137.5 $\pm$ 3.6 & -116.6 $\pm$ 6.8 & -138.5 $\pm$ 7.1 \\ \hline
        76 & NONE & MULLERMANDEL & 0.2 & MACLEOD\_LINEAR & 1 & 10 & FIXED & 1 & -106 $\pm$ 3.5 & -130.2 $\pm$ 2.5 & -131.5 $\pm$ 7.3 & -148.6 $\pm$ 7.7 \\ \hline
        77 & NONE & MULLERMANDEL & 0.2 & MACLEOD\_LINEAR & 1 & 100 & FIXED & 0.1 & -117 $\pm$ 6.1 & -143.6 $\pm$ 4.8 & -118.5 $\pm$ 8.2 & -143.3 $\pm$ 8.2 \\ \hline
        78 & NONE & MULLERMANDEL & 0.2 & MACLEOD\_LINEAR & 1 & 100 & FIXED & 0.2 & -120.4 $\pm$ 6.3 & -146.7 $\pm$ 5 & -120.3 $\pm$ 8.2 & -145.6 $\pm$ 8.1 \\ \hline
        79 & NONE & MULLERMANDEL & 0.2 & MACLEOD\_LINEAR & 1 & 100 & FIXED & 0.5 & -159.5 $\pm$ 5.6 & -192.6 $\pm$ 4.4 & -162.6 $\pm$ 7.9 & -193.3 $\pm$ 7.9 \\ \hline
        80 & NONE & MULLERMANDEL & 0.2 & MACLEOD\_LINEAR & 1 & 100 & FIXED & 1 & -114.8 $\pm$ 5.9 & -140.2 $\pm$ 4.7 & -119.6 $\pm$ 8.6 & -142 $\pm$ 8.6 \\ \hline
        81 & PESSIMISTIC & MULLERMANDEL & 0.2 & MACLEOD\_LINEAR & 0 & 0.1 & FIXED & 0.1 & -55.8 $\pm$ 2.8 & -68.2 $\pm$ 1.9 & -454.4 $\pm$ 29.6 & -445 $\pm$ 29.8 \\ \hline
        82 & PESSIMISTIC & MULLERMANDEL & 0.2 & MACLEOD\_LINEAR & 0 & 0.1 & FIXED & 0.2 & -46.2 $\pm$ 2.6 & -55.2 $\pm$ 1.8 & -430.3 $\pm$ 27.3 & -417.8 $\pm$ 27.4 \\ \hline
        83 & PESSIMISTIC & MULLERMANDEL & 0.2 & MACLEOD\_LINEAR & 0 & 0.1 & FIXED & 0.5 & -37.3 $\pm$ 2.1 & -44.3 $\pm$ 1.3 & -610.1 $\pm$ 30.3 & -592.9 $\pm$ 30.4 \\ \hline
        84 & PESSIMISTIC & MULLERMANDEL & 0.2 & MACLEOD\_LINEAR & 0 & 0.1 & FIXED & 1 & -28.3 $\pm$ 1.7 & -35.1 $\pm$ 1 & -797.6 $\pm$ 32.8 & -778 $\pm$ 32.9 \\ \hline
        85 & PESSIMISTIC & MULLERMANDEL & 0.2 & MACLEOD\_LINEAR & 0 & 1 & FIXED & 0.1 & -29.7 $\pm$ 1.2 & -34.5 $\pm$ 0.8 & -522.7 $\pm$ 11.9 & -504.2 $\pm$ 12.4 \\ \hline
        86 & PESSIMISTIC & MULLERMANDEL & 0.2 & MACLEOD\_LINEAR & 0 & 1 & FIXED & 0.2 & -29.4 $\pm$ 1.2 & -34.4 $\pm$ 0.8 & -637.1 $\pm$ 14 & -617.4 $\pm$ 14.4 \\ \hline
        87 & PESSIMISTIC & MULLERMANDEL & 0.2 & MACLEOD\_LINEAR & 0 & 1 & FIXED & 0.5 & -38.8 $\pm$ 1.3 & -47.2 $\pm$ 0.7 & -811.6 $\pm$ 22.7 & -793.5 $\pm$ 22.9 \\ \hline
        88 & PESSIMISTIC & MULLERMANDEL & 0.2 & MACLEOD\_LINEAR & 0 & 1 & FIXED & 1 & -26.3 $\pm$ 1.7 & -31.4 $\pm$ 1 & -787.6 $\pm$ 31.2 & -766.3 $\pm$ 31.4 \\ \hline
        89 & PESSIMISTIC & MULLERMANDEL & 0.2 & MACLEOD\_LINEAR & 0 & 10 & FIXED & 0.1 & -48.6 $\pm$ 3.5 & -57.9 $\pm$ 2.5 & -446.1 $\pm$ 36.7 & -433.7 $\pm$ 36.8 \\ \hline
        90 & PESSIMISTIC & MULLERMANDEL & 0.2 & MACLEOD\_LINEAR & 0 & 10 & FIXED & 0.2 & -32.1 $\pm$ 3.1 & -39.2 $\pm$ 2.2 & -493.7 $\pm$ 37.3 & -478.1 $\pm$ 37.4 \\ \hline
        91 & PESSIMISTIC & MULLERMANDEL & 0.2 & MACLEOD\_LINEAR & 0 & 10 & FIXED & 0.5 & -39.7 $\pm$ 2.2 & -48.5 $\pm$ 1.4 & -717.1 $\pm$ 37.9 & -700.4 $\pm$ 38 \\ \hline
        92 & PESSIMISTIC & MULLERMANDEL & 0.2 & MACLEOD\_LINEAR & 0 & 10 & FIXED & 1 & -14.7 $\pm$ 1.7 & -18.5 $\pm$ 1 & -776 $\pm$ 31.9 & -753.4 $\pm$ 32 \\ \hline
        93 & PESSIMISTIC & MULLERMANDEL & 0.2 & MACLEOD\_LINEAR & 0 & 100 & FIXED & 0.1 & -60.8 $\pm$ 3.7 & -74.2 $\pm$ 2.7 & -456.7 $\pm$ 38.2 & -448.5 $\pm$ 38.3 \\ \hline
        94 & PESSIMISTIC & MULLERMANDEL & 0.2 & MACLEOD\_LINEAR & 0 & 100 & FIXED & 0.2 & -51.8 $\pm$ 3.4 & -63.8 $\pm$ 2.4 & -541.1 $\pm$ 42.4 & -529.9 $\pm$ 42.5 \\ \hline
        95 & PESSIMISTIC & MULLERMANDEL & 0.2 & MACLEOD\_LINEAR & 0 & 100 & FIXED & 0.5 & -32.8 $\pm$ 2.4 & -43 $\pm$ 1.6 & -638.8 $\pm$ 37.5 & -624.3 $\pm$ 37.7 \\ \hline
        96 & PESSIMISTIC & MULLERMANDEL & 0.2 & MACLEOD\_LINEAR & 0 & 100 & FIXED & 1 & -28.1 $\pm$ 2 & -34.7 $\pm$ 1.2 & -737.9 $\pm$ 36 & -718.7 $\pm$ 36.2 \\ \hline
        97 & PESSIMISTIC & MULLERMANDEL & 0.2 & MACLEOD\_LINEAR & 0.25 & 0.1 & FIXED & 0.1 & -18.1 $\pm$ 2 & -23.2 $\pm$ 1.2 & -617.6 $\pm$ 30.6 & -598.2 $\pm$ 30.8 \\ \hline
        98 & PESSIMISTIC & MULLERMANDEL & 0.2 & MACLEOD\_LINEAR & 0.25 & 0.1 & FIXED & 0.2 & -34.6 $\pm$ 1.6 & -41.2 $\pm$ 0.9 & -752.2 $\pm$ 27.1 & -732.9 $\pm$ 27.3 \\ \hline
        99 & PESSIMISTIC & MULLERMANDEL & 0.2 & MACLEOD\_LINEAR & 0.25 & 0.1 & FIXED & 0.5 & -27 $\pm$ 1.3 & -31.9 $\pm$ 0.7 & -1012 $\pm$ 27.1 & -988.7 $\pm$ 27.4 \\ \hline
        100 & PESSIMISTIC & MULLERMANDEL & 0.2 & MACLEOD\_LINEAR & 0.25 & 0.1 & FIXED & 1 & -24.5 $\pm$ 1.3 & -29.2 $\pm$ 0.8 & -951.5 $\pm$ 27.5 & -928.5 $\pm$ 27.7 \\ \hline
    \end{tabular}}
    \caption{Hyper-parameters for the \citet{10.1093/mnras/stad1757} models and new Models 1-100 and the corresponding Bayes factor relative to the maximum likelihoods, for which the model is in bold. In green, we highlight the highest likelihood models, where the Bayes factor is $>-10$ for either the marginalised or unmarginalised likelihoods.}
    \label{tab:all_models}
\end{table*}

\begin{table*}
    \tiny
    \centering
     \begin{tabular}{|c|c|c|c|c|c|c|c|c|c|c|c|c|}
    \hline
        \multirow{2}{*}{Model} & \multirow{2}{*}{CHE} & Remnant mass & \multirow{2}{*}{$f_{\rm WR}$} & \multirow{2}{*}{$\gamma$ prescription} & \multirow{2}{*}{$f_\gamma$} & \multirow{2}{*}{$\alpha_{\rm CE}$} & $\dot{M}_a$ & \multirow{2}{*}{$f_{\rm MT}$} & $\log_{10} \mathcal{B}(\lambda|\mathcal{H})$ & $ \log_{10} \mathcal{B}(\lambda|\mathcal{H})$ & $\log_{10} \mathcal{B}(\lambda, N_\lambda|\mathcal{H})$ & $ \log_{10} \mathcal{B}(\lambda, N_\lambda|\mathcal{H})$ \\ 
        & & prescription & & & & & prescription & & & ($\rm SNR \geq 8$) & & ($\rm SNR \geq 8$) \\\hline
        101 & PESSIMISTIC & MULLERMANDEL & 0.2 & MACLEOD\_LINEAR & 0.25 & 1 & FIXED & 0.1 & -23.9 $\pm$ 1.2 & -28 $\pm$ 0.7 & -671.1 $\pm$ 16.5 & -650.1 $\pm$ 16.9 \\ \hline
        102 & PESSIMISTIC & MULLERMANDEL & 0.2 & MACLEOD\_LINEAR & 0.25 & 1 & FIXED & 0.2 & -27.6 $\pm$ 1.2 & -32.5 $\pm$ 0.8 & -918.5 $\pm$ 17.6 & -895.9 $\pm$ 18 \\ \hline
        103 & PESSIMISTIC & MULLERMANDEL & 0.2 & MACLEOD\_LINEAR & 0.25 & 1 & FIXED & 0.5 & -11.9 $\pm$ 1.2 & -14.7 $\pm$ 0.8 & -936.3 $\pm$ 17.7 & -911.4 $\pm$ 18 \\ \hline
        104 & PESSIMISTIC & MULLERMANDEL & 0.2 & MACLEOD\_LINEAR & 0.25 & 1 & FIXED & 1 & -11.4 $\pm$ 1.3 & -14.4 $\pm$ 0.7 & -941.1 $\pm$ 25.8 & -916.3 $\pm$ 26 \\ \hline
        105 & PESSIMISTIC & MULLERMANDEL & 0.2 & MACLEOD\_LINEAR & 0.25 & 10 & FIXED & 0.1 & -21.2 $\pm$ 2 & -26.2 $\pm$ 1.2 & -630.8 $\pm$ 31.4 & -611 $\pm$ 31.5 \\ \hline
        106 & PESSIMISTIC & MULLERMANDEL & 0.2 & MACLEOD\_LINEAR & 0.25 & 10 & FIXED & 0.2 & -22.6 $\pm$ 1.6 & -27.9 $\pm$ 0.9 & -824.5 $\pm$ 30.1 & -803.2 $\pm$ 30.3 \\ \hline
        107 & PESSIMISTIC & MULLERMANDEL & 0.2 & MACLEOD\_LINEAR & 0.25 & 10 & FIXED & 0.5 & -29.9 $\pm$ 1.4 & -35 $\pm$ 0.8 & -1013.1 $\pm$ 29.5 & -989.9 $\pm$ 29.7 \\ \hline
        108 & PESSIMISTIC & MULLERMANDEL & 0.2 & MACLEOD\_LINEAR & 0.25 & 10 & FIXED & 1 & -18.7 $\pm$ 1.3 & -21.3 $\pm$ 0.8 & -917.5 $\pm$ 26.3 & -892.5 $\pm$ 26.5 \\ \hline
        109 & PESSIMISTIC & MULLERMANDEL & 0.2 & MACLEOD\_LINEAR & 0.25 & 100 & FIXED & 0.1 & -32.6 $\pm$ 2.2 & -39.1 $\pm$ 1.4 & -577.3 $\pm$ 31.6 & -560 $\pm$ 31.7 \\ \hline
        110 & PESSIMISTIC & MULLERMANDEL & 0.2 & MACLEOD\_LINEAR & 0.25 & 100 & FIXED & 0.2 & -24.3 $\pm$ 1.7 & -30 $\pm$ 1 & -803.3 $\pm$ 31.8 & -782.6 $\pm$ 32 \\ \hline
        111 & PESSIMISTIC & MULLERMANDEL & 0.2 & MACLEOD\_LINEAR & 0.25 & 100 & FIXED & 0.5 & -30.6 $\pm$ 1.4 & -37 $\pm$ 0.8 & -1006.4 $\pm$ 30.8 & -984.6 $\pm$ 31 \\ \hline
        112 & PESSIMISTIC & MULLERMANDEL & 0.2 & MACLEOD\_LINEAR & 0.25 & 100 & FIXED & 1 & -27 $\pm$ 1.5 & -31.8 $\pm$ 0.9 & -876.4 $\pm$ 30.9 & -854.1 $\pm$ 31.1 \\ \hline
        113 & PESSIMISTIC & MULLERMANDEL & 0.2 & MACLEOD\_LINEAR & 0.5 & 0.1 & FIXED & 0.1 & -12.8 $\pm$ 1.2 & -15 $\pm$ 0.8 & -944.8 $\pm$ 17.6 & -919.3 $\pm$ 17.9 \\ \hline
        114 & PESSIMISTIC & MULLERMANDEL & 0.2 & MACLEOD\_LINEAR & 0.5 & 0.1 & FIXED & 0.2 & -12.9 $\pm$ 1.2 & -15.7 $\pm$ 0.8 & -962.4 $\pm$ 17.3 & -937.2 $\pm$ 17.7 \\ \hline
\rowcolor{Green!30}
        115 & PESSIMISTIC & MULLERMANDEL & 0.2 & MACLEOD\_LINEAR & 0.5 & 0.1 & FIXED & 0.5 & -7.6 $\pm$ 1.2 & -7.8 $\pm$ 0.8 & -977.8 $\pm$ 17.5 & -949.9 $\pm$ 17.8 \\ \hline
        116 & PESSIMISTIC & MULLERMANDEL & 0.2 & MACLEOD\_LINEAR & 0.5 & 0.1 & FIXED & 1 & -24.2 $\pm$ 1.5 & -29.1 $\pm$ 0.8 & -752.7 $\pm$ 25.5 & -731.6 $\pm$ 25.8 \\ \hline
        117 & PESSIMISTIC & MULLERMANDEL & 0.2 & MACLEOD\_LINEAR & 0.5 & 1 & FIXED & 0.1 & -22 $\pm$ 1.2 & -25.4 $\pm$ 1 & -969.7 $\pm$ 11.3 & -945.2 $\pm$ 11.9 \\ \hline
        118 & PESSIMISTIC & MULLERMANDEL & 0.2 & MACLEOD\_LINEAR & 0.5 & 1 & FIXED & 0.2 & -21.7 $\pm$ 1.2 & -26 $\pm$ 0.9 & -1105 $\pm$ 13.8 & -1080.4 $\pm$ 14.3 \\ \hline
        119 & PESSIMISTIC & MULLERMANDEL & 0.2 & MACLEOD\_LINEAR & 0.5 & 1 & FIXED & 0.5 & -10.5 $\pm$ 1.2 & -12.1 $\pm$ 0.9 & -1058.6 $\pm$ 14.6 & -1031.6 $\pm$ 15 \\ \hline
\rowcolor{Green!30}
        120 & PESSIMISTIC & MULLERMANDEL & 0.2 & MACLEOD\_LINEAR & 0.5 & 1 & FIXED & 1 & -9.6 $\pm$ 1.3 & -10.9 $\pm$ 0.7 & -777 $\pm$ 21.7 & -751.9 $\pm$ 21.9 \\ \hline
        121 & PESSIMISTIC & MULLERMANDEL & 0.2 & MACLEOD\_LINEAR & 0.5 & 10 & FIXED & 0.1 & -16.4 $\pm$ 1.2 & -19.9 $\pm$ 0.9 & -936 $\pm$ 13.7 & -911.7 $\pm$ 14.1 \\ \hline
        122 & PESSIMISTIC & MULLERMANDEL & 0.2 & MACLEOD\_LINEAR & 0.5 & 10 & FIXED & 0.2 & -13.7 $\pm$ 1.2 & -16.3 $\pm$ 0.9 & -914.3 $\pm$ 14.5 & -889.3 $\pm$ 14.9 \\ \hline
        123 & PESSIMISTIC & MULLERMANDEL & 0.2 & MACLEOD\_LINEAR & 0.5 & 10 & FIXED & 0.5 & -11.1 $\pm$ 1.2 & -12.5 $\pm$ 0.9 & -998.7 $\pm$ 15.8 & -971.9 $\pm$ 16.1 \\ \hline
        124 & PESSIMISTIC & MULLERMANDEL & 0.2 & MACLEOD\_LINEAR & 0.5 & 10 & FIXED & 1 & -13.5 $\pm$ 1.4 & -15.7 $\pm$ 0.8 & -780.6 $\pm$ 24 & -756.4 $\pm$ 24.2 \\ \hline
        125 & PESSIMISTIC & MULLERMANDEL & 0.2 & MACLEOD\_LINEAR & 0.5 & 100 & FIXED & 0.1 & -14.1 $\pm$ 1.2 & -16.7 $\pm$ 0.8 & -840.1 $\pm$ 18.5 & -815.8 $\pm$ 18.8 \\ \hline
        126 & PESSIMISTIC & MULLERMANDEL & 0.2 & MACLEOD\_LINEAR & 0.5 & 100 & FIXED & 0.2 & -22.5 $\pm$ 1.2 & -27.1 $\pm$ 0.8 & -848.5 $\pm$ 18.6 & -826.1 $\pm$ 18.9 \\ \hline
\rowcolor{Green!30}
        127 & PESSIMISTIC & MULLERMANDEL & 0.2 & MACLEOD\_LINEAR & 0.5 & 100 & FIXED & 0.5 & -9.8 $\pm$ 1.2 & -10.2 $\pm$ 0.8 & -934.8 $\pm$ 21.3 & -907.5 $\pm$ 21.6 \\ \hline
        128 & PESSIMISTIC & MULLERMANDEL & 0.2 & MACLEOD\_LINEAR & 0.5 & 100 & FIXED & 1 & -20.8 $\pm$ 1.7 & -25.8 $\pm$ 1 & -797.8 $\pm$ 31.9 & -776.3 $\pm$ 32.1 \\ \hline
        129 & PESSIMISTIC & MULLERMANDEL & 0.2 & MACLEOD\_LINEAR & 0.75 & 0.1 & FIXED & 0.1 & -27.7 $\pm$ 2.4 & -32 $\pm$ 1.6 & -452 $\pm$ 27.4 & -434.1 $\pm$ 27.5 \\ \hline
        130 & PESSIMISTIC & MULLERMANDEL & 0.2 & MACLEOD\_LINEAR & 0.75 & 0.1 & FIXED & 0.2 & -29.6 $\pm$ 2.5 & -36 $\pm$ 1.6 & -427.1 $\pm$ 26.5 & -411.8 $\pm$ 26.7 \\ \hline
        131 & PESSIMISTIC & MULLERMANDEL & 0.2 & MACLEOD\_LINEAR & 0.75 & 0.1 & FIXED & 0.5 & -32.2 $\pm$ 2.5 & -39.8 $\pm$ 1.6 & -556.2 $\pm$ 33.3 & -540.2 $\pm$ 33.5 \\ \hline
        132 & PESSIMISTIC & MULLERMANDEL & 0.2 & MACLEOD\_LINEAR & 0.75 & 0.1 & FIXED & 1 & -13.9 $\pm$ 2.1 & -17.2 $\pm$ 1.3 & -554.2 $\pm$ 28.9 & -533.7 $\pm$ 29.1 \\ \hline
        133 & PESSIMISTIC & MULLERMANDEL & 0.2 & MACLEOD\_LINEAR & 0.75 & 1 & FIXED & 0.1 & -26.4 $\pm$ 1.9 & -30.6 $\pm$ 1.1 & -369.1 $\pm$ 18.3 & -352.6 $\pm$ 18.6 \\ \hline
        134 & PESSIMISTIC & MULLERMANDEL & 0.2 & MACLEOD\_LINEAR & 0.75 & 1 & FIXED & 0.2 & -26.6 $\pm$ 1.9 & -32.9 $\pm$ 1.1 & -435 $\pm$ 20.8 & -419.4 $\pm$ 21.1 \\ \hline
        135 & PESSIMISTIC & MULLERMANDEL & 0.2 & MACLEOD\_LINEAR & 0.75 & 1 & FIXED & 0.5 & -16.6 $\pm$ 1.9 & -19.6 $\pm$ 1.1 & -487.8 $\pm$ 23.2 & -468 $\pm$ 23.5 \\ \hline
        136 & PESSIMISTIC & MULLERMANDEL & 0.2 & MACLEOD\_LINEAR & 0.75 & 1 & FIXED & 1 & -22.8 $\pm$ 2 & -26.1 $\pm$ 1.3 & -517.7 $\pm$ 26.7 & -497.8 $\pm$ 26.9 \\ \hline
        137 & PESSIMISTIC & MULLERMANDEL & 0.2 & MACLEOD\_LINEAR & 0.75 & 10 & FIXED & 0.1 & -37.7 $\pm$ 2.8 & -45 $\pm$ 1.9 & -424.6 $\pm$ 29.1 & -410.4 $\pm$ 29.2 \\ \hline
        138 & PESSIMISTIC & MULLERMANDEL & 0.2 & MACLEOD\_LINEAR & 0.75 & 10 & FIXED & 0.2 & -37.7 $\pm$ 2.6 & -45.1 $\pm$ 1.8 & -425 $\pm$ 27.5 & -410.8 $\pm$ 27.6 \\ \hline
        139 & PESSIMISTIC & MULLERMANDEL & 0.2 & MACLEOD\_LINEAR & 0.75 & 10 & FIXED & 0.5 & -25.5 $\pm$ 2.5 & -31.1 $\pm$ 1.7 & -482.9 $\pm$ 30.3 & -465.9 $\pm$ 30.4 \\ \hline
        140 & PESSIMISTIC & MULLERMANDEL & 0.2 & MACLEOD\_LINEAR & 0.75 & 10 & FIXED & 1 & -21.1 $\pm$ 2.5 & -24.2 $\pm$ 1.7 & -410.6 $\pm$ 26.8 & -392.1 $\pm$ 26.9 \\ \hline
        141 & PESSIMISTIC & MULLERMANDEL & 0.2 & MACLEOD\_LINEAR & 0.75 & 100 & FIXED & 0.1 & -40.5 $\pm$ 3 & -47.6 $\pm$ 2.1 & -429.5 $\pm$ 31.6 & -415.1 $\pm$ 31.8 \\ \hline
        142 & PESSIMISTIC & MULLERMANDEL & 0.2 & MACLEOD\_LINEAR & 0.75 & 100 & FIXED & 0.2 & -49.1 $\pm$ 3.1 & -58 $\pm$ 2.2 & -499.1 $\pm$ 36.1 & -485.5 $\pm$ 36.2 \\ \hline
        143 & PESSIMISTIC & MULLERMANDEL & 0.2 & MACLEOD\_LINEAR & 0.75 & 100 & FIXED & 0.5 & -43.2 $\pm$ 3.2 & -52.7 $\pm$ 2.3 & -470.7 $\pm$ 35.8 & -458 $\pm$ 35.9 \\ \hline
        144 & PESSIMISTIC & MULLERMANDEL & 0.2 & MACLEOD\_LINEAR & 0.75 & 100 & FIXED & 1 & -39.7 $\pm$ 3.2 & -47.9 $\pm$ 2.3 & -503.2 $\pm$ 38.6 & -488.6 $\pm$ 38.7 \\ \hline
        145 & PESSIMISTIC & MULLERMANDEL & 0.2 & MACLEOD\_LINEAR & 1 & 0.1 & FIXED & 0.1 & -25.9 $\pm$ 2.6 & -33.9 $\pm$ 1.7 & -491.5 $\pm$ 31.6 & -476.7 $\pm$ 31.8 \\ \hline
        146 & PESSIMISTIC & MULLERMANDEL & 0.2 & MACLEOD\_LINEAR & 1 & 0.1 & FIXED & 0.2 & -25.7 $\pm$ 2.6 & -32.7 $\pm$ 1.7 & -497 $\pm$ 32.3 & -481.2 $\pm$ 32.4 \\ \hline
        147 & PESSIMISTIC & MULLERMANDEL & 0.2 & MACLEOD\_LINEAR & 1 & 0.1 & FIXED & 0.5 & -33.3 $\pm$ 2.3 & -41.6 $\pm$ 1.5 & -557 $\pm$ 31.7 & -541.7 $\pm$ 31.9 \\ \hline
        148 & PESSIMISTIC & MULLERMANDEL & 0.2 & MACLEOD\_LINEAR & 1 & 0.1 & FIXED & 1 & -18.2 $\pm$ 2.3 & -24.2 $\pm$ 1.5 & -510.1 $\pm$ 29.8 & -492.9 $\pm$ 30 \\ \hline
        149 & PESSIMISTIC & MULLERMANDEL & 0.2 & MACLEOD\_LINEAR & 1 & 1 & FIXED & 0.1 & -32.9 $\pm$ 2.1 & -40 $\pm$ 1.3 & -513.2 $\pm$ 26.8 & -497.3 $\pm$ 27 \\ \hline
        150 & PESSIMISTIC & MULLERMANDEL & 0.2 & MACLEOD\_LINEAR & 1 & 1 & FIXED & 0.2 & -31.5 $\pm$ 2.1 & -38.8 $\pm$ 1.3 & -504.2 $\pm$ 26.2 & -488.7 $\pm$ 26.4 \\ \hline
        151 & PESSIMISTIC & MULLERMANDEL & 0.2 & MACLEOD\_LINEAR & 1 & 1 & FIXED & 0.5 & -31.6 $\pm$ 2 & -37.4 $\pm$ 1.2 & -558.6 $\pm$ 27.6 & -540.7 $\pm$ 27.8 \\ \hline
        152 & PESSIMISTIC & MULLERMANDEL & 0.2 & MACLEOD\_LINEAR & 1 & 1 & FIXED & 1 & -29.9 $\pm$ 2 & -36.9 $\pm$ 1.2 & -525.7 $\pm$ 26.3 & -509.4 $\pm$ 26.5 \\ \hline
        153 & PESSIMISTIC & MULLERMANDEL & 0.2 & MACLEOD\_LINEAR & 1 & 10 & FIXED & 0.1 & -43.7 $\pm$ 3.1 & -53.1 $\pm$ 2.2 & -500.7 $\pm$ 37 & -487.5 $\pm$ 37.1 \\ \hline
        154 & PESSIMISTIC & MULLERMANDEL & 0.2 & MACLEOD\_LINEAR & 1 & 10 & FIXED & 0.2 & -36 $\pm$ 3.1 & -44.7 $\pm$ 2.2 & -462.4 $\pm$ 34.4 & -448.9 $\pm$ 34.5 \\ \hline
        155 & PESSIMISTIC & MULLERMANDEL & 0.2 & MACLEOD\_LINEAR & 1 & 10 & FIXED & 0.5 & -34.7 $\pm$ 3.1 & -43 $\pm$ 2.2 & -401.5 $\pm$ 31 & -388.6 $\pm$ 31.2 \\ \hline
        156 & PESSIMISTIC & MULLERMANDEL & 0.2 & MACLEOD\_LINEAR & 1 & 10 & FIXED & 1 & -38.6 $\pm$ 2.6 & -47.4 $\pm$ 1.7 & -532.7 $\pm$ 33.4 & -518.3 $\pm$ 33.5 \\ \hline
        157 & PESSIMISTIC & MULLERMANDEL & 0.2 & MACLEOD\_LINEAR & 1 & 100 & FIXED & 0.1 & -63.8 $\pm$ 3.3 & -84.5 $\pm$ 2.4 & -542.5 $\pm$ 40.5 & -540.2 $\pm$ 40.6 \\ \hline
        158 & PESSIMISTIC & MULLERMANDEL & 0.2 & MACLEOD\_LINEAR & 1 & 100 & FIXED & 0.2 & -60.3 $\pm$ 3.5 & -77.2 $\pm$ 2.5 & -443.1 $\pm$ 35.2 & -438.5 $\pm$ 35.3 \\ \hline
        159 & PESSIMISTIC & MULLERMANDEL & 0.2 & MACLEOD\_LINEAR & 1 & 100 & FIXED & 0.5 & -38.4 $\pm$ 3.3 & -50.2 $\pm$ 2.4 & -487.3 $\pm$ 38.4 & -476.6 $\pm$ 38.5 \\ \hline
        160 & PESSIMISTIC & MULLERMANDEL & 0.2 & MACLEOD\_LINEAR & 1 & 100 & FIXED & 1 & -52.7 $\pm$ 3.4 & -66.9 $\pm$ 2.5 & -549.6 $\pm$ 43.3 & -540.5 $\pm$ 43.4 \\ \hline
        161 & OPTIMISTIC & MULLERMANDEL & 0.2 & MACLEOD\_LINEAR & 0 & 0.1 & FIXED & 0.1 & -56.6 $\pm$ 2.4 & -67.9 $\pm$ 1.6 & -853.3 $\pm$ 47 & -837.9 $\pm$ 47.1 \\ \hline
        162 & OPTIMISTIC & MULLERMANDEL & 0.2 & MACLEOD\_LINEAR & 0 & 0.1 & FIXED & 0.2 & -52.2 $\pm$ 2.4 & -63 $\pm$ 1.6 & -990.4 $\pm$ 54.8 & -973.4 $\pm$ 54.9 \\ \hline
        163 & OPTIMISTIC & MULLERMANDEL & 0.2 & MACLEOD\_LINEAR & 0 & 0.1 & FIXED & 0.5 & -30.8 $\pm$ 2 & -37.8 $\pm$ 1.2 & -970.6 $\pm$ 45.7 & -949.8 $\pm$ 45.8 \\ \hline
        164 & OPTIMISTIC & MULLERMANDEL & 0.2 & MACLEOD\_LINEAR & 0 & 0.1 & FIXED & 1 & -32.8 $\pm$ 1.7 & -40.6 $\pm$ 1 & -1043.1 $\pm$ 40.2 & -1022.5 $\pm$ 40.3 \\ \hline
        165 & OPTIMISTIC & MULLERMANDEL & 0.2 & MACLEOD\_LINEAR & 0 & 1 & FIXED & 0.1 & -30.6 $\pm$ 1.2 & -35.4 $\pm$ 0.8 & -860.2 $\pm$ 17.3 & -838.1 $\pm$ 17.7 \\ \hline
        166 & OPTIMISTIC & MULLERMANDEL & 0.2 & MACLEOD\_LINEAR & 0 & 1 & FIXED & 0.2 & -38.7 $\pm$ 1.2 & -45.4 $\pm$ 0.8 & -906.4 $\pm$ 17.6 & -885.9 $\pm$ 18 \\ \hline
        167 & OPTIMISTIC & MULLERMANDEL & 0.2 & MACLEOD\_LINEAR & 0 & 1 & FIXED & 0.5 & -30.3 $\pm$ 1.3 & -38 $\pm$ 0.7 & -1111.1 $\pm$ 29.2 & -1089.8 $\pm$ 29.4 \\ \hline
        168 & OPTIMISTIC & MULLERMANDEL & 0.2 & MACLEOD\_LINEAR & 0 & 1 & FIXED & 1 & -39.2 $\pm$ 1.5 & -47.7 $\pm$ 0.8 & -1139.6 $\pm$ 38.5 & -1118.9 $\pm$ 38.6 \\ \hline
        169 & OPTIMISTIC & MULLERMANDEL & 0.2 & MACLEOD\_LINEAR & 0 & 10 & FIXED & 0.1 & -55.5 $\pm$ 3 & -65.8 $\pm$ 2.1 & -925.8 $\pm$ 62.2 & -908.8 $\pm$ 62.2 \\ \hline
        170 & OPTIMISTIC & MULLERMANDEL & 0.2 & MACLEOD\_LINEAR & 0 & 10 & FIXED & 0.2 & -50.8 $\pm$ 2.8 & -60.9 $\pm$ 1.9 & -999 $\pm$ 63 & -981.2 $\pm$ 63.1 \\ \hline
        171 & OPTIMISTIC & MULLERMANDEL & 0.2 & MACLEOD\_LINEAR & 0 & 10 & FIXED & 0.5 & -27.8 $\pm$ 2 & -34.7 $\pm$ 1.2 & -1146.8 $\pm$ 52.9 & -1124.4 $\pm$ 53 \\ \hline
        172 & OPTIMISTIC & MULLERMANDEL & 0.2 & MACLEOD\_LINEAR & 0 & 10 & FIXED & 1 & -29.2 $\pm$ 1.6 & -35.2 $\pm$ 0.9 & -1118.6 $\pm$ 40.8 & -1095.6 $\pm$ 40.9 \\ \hline
        173 & OPTIMISTIC & MULLERMANDEL & 0.2 & MACLEOD\_LINEAR & 0 & 100 & FIXED & 0.1 & -63.9 $\pm$ 3.3 & -76.6 $\pm$ 2.4 & -680.9 $\pm$ 50.8 & -668.7 $\pm$ 50.9 \\ \hline
        174 & OPTIMISTIC & MULLERMANDEL & 0.2 & MACLEOD\_LINEAR & 0 & 100 & FIXED & 0.2 & -65.9 $\pm$ 2.9 & -82 $\pm$ 2 & -999.6 $\pm$ 64.6 & -987.9 $\pm$ 64.7 \\ \hline
        175 & OPTIMISTIC & MULLERMANDEL & 0.2 & MACLEOD\_LINEAR & 0 & 100 & FIXED & 0.5 & -33.9 $\pm$ 2.4 & -43.2 $\pm$ 1.5 & -872 $\pm$ 48.4 & -854.3 $\pm$ 48.5 \\ \hline
        176 & OPTIMISTIC & MULLERMANDEL & 0.2 & MACLEOD\_LINEAR & 0 & 100 & FIXED & 1 & -42.3 $\pm$ 1.9 & -53.4 $\pm$ 1.2 & -1169.6 $\pm$ 52.1 & -1151.5 $\pm$ 52.2 \\ \hline
        177 & OPTIMISTIC & MULLERMANDEL & 0.2 & MACLEOD\_LINEAR & 0.25 & 0.1 & FIXED & 0.1 & -44.7 $\pm$ 1.7 & -53.9 $\pm$ 1 & -1033.7 $\pm$ 41.5 & -1014.6 $\pm$ 41.7 \\ \hline
        178 & OPTIMISTIC & MULLERMANDEL & 0.2 & MACLEOD\_LINEAR & 0.25 & 0.1 & FIXED & 0.2 & -39.6 $\pm$ 1.5 & -47.4 $\pm$ 0.8 & -1170.3 $\pm$ 37.9 & -1148.8 $\pm$ 38.1 \\ \hline
        179 & OPTIMISTIC & MULLERMANDEL & 0.2 & MACLEOD\_LINEAR & 0.25 & 0.1 & FIXED & 0.5 & -30.5 $\pm$ 1.2 & -36.7 $\pm$ 0.7 & -1471.8 $\pm$ 35.2 & -1446.9 $\pm$ 35.4 \\ \hline
        180 & OPTIMISTIC & MULLERMANDEL & 0.2 & MACLEOD\_LINEAR & 0.25 & 0.1 & FIXED & 1 & -37.6 $\pm$ 1.3 & -43.8 $\pm$ 0.7 & -1286 $\pm$ 32.7 & -1262.2 $\pm$ 32.9 \\ \hline
        181 & OPTIMISTIC & MULLERMANDEL & 0.2 & MACLEOD\_LINEAR & 0.25 & 1 & FIXED & 0.1 & -27.7 $\pm$ 1.2 & -32.5 $\pm$ 0.8 & -1055.2 $\pm$ 22.5 & -1031.5 $\pm$ 22.8 \\ \hline
        182 & OPTIMISTIC & MULLERMANDEL & 0.2 & MACLEOD\_LINEAR & 0.25 & 1 & FIXED & 0.2 & -26.3 $\pm$ 1.2 & -32.2 $\pm$ 0.9 & -1305.9 $\pm$ 20 & -1281.6 $\pm$ 20.3 \\ \hline
        183 & OPTIMISTIC & MULLERMANDEL & 0.2 & MACLEOD\_LINEAR & 0.25 & 1 & FIXED & 0.5 & -28.7 $\pm$ 1.2 & -33.7 $\pm$ 0.8 & -1413.4 $\pm$ 23.6 & -1387.6 $\pm$ 23.9 \\ \hline
        184 & OPTIMISTIC & MULLERMANDEL & 0.2 & MACLEOD\_LINEAR & 0.25 & 1 & FIXED & 1 & -13.3 $\pm$ 1.3 & -16.5 $\pm$ 0.7 & -1214.8 $\pm$ 31.7 & -1188.3 $\pm$ 31.9 \\ \hline
        185 & OPTIMISTIC & MULLERMANDEL & 0.2 & MACLEOD\_LINEAR & 0.25 & 10 & FIXED & 0.1 & -37 $\pm$ 1.9 & -44.4 $\pm$ 1.1 & -1106.3 $\pm$ 47.6 & -1084.9 $\pm$ 47.8 \\ \hline
        186 & OPTIMISTIC & MULLERMANDEL & 0.2 & MACLEOD\_LINEAR & 0.25 & 10 & FIXED & 0.2 & -34.2 $\pm$ 1.5 & -40.6 $\pm$ 0.8 & -1144.1 $\pm$ 37.7 & -1121.4 $\pm$ 37.8 \\ \hline
        187 & OPTIMISTIC & MULLERMANDEL & 0.2 & MACLEOD\_LINEAR & 0.25 & 10 & FIXED & 0.5 & -23.2 $\pm$ 1.3 & -28.2 $\pm$ 0.7 & -1450.4 $\pm$ 36 & -1424.4 $\pm$ 36.2 \\ \hline
        188 & OPTIMISTIC & MULLERMANDEL & 0.2 & MACLEOD\_LINEAR & 0.25 & 10 & FIXED & 1 & -19.6 $\pm$ 1.3 & -23.7 $\pm$ 0.7 & -1276.4 $\pm$ 33.5 & -1250.4 $\pm$ 33.7 \\ \hline
        189 & OPTIMISTIC & MULLERMANDEL & 0.2 & MACLEOD\_LINEAR & 0.25 & 100 & FIXED & 0.1 & -30.5 $\pm$ 2 & -37.3 $\pm$ 1.3 & -908.4 $\pm$ 43.7 & -887.8 $\pm$ 43.8 \\ \hline
        190 & OPTIMISTIC & MULLERMANDEL & 0.2 & MACLEOD\_LINEAR & 0.25 & 100 & FIXED & 0.2 & -24.2 $\pm$ 1.6 & -30.7 $\pm$ 0.9 & -1150.8 $\pm$ 42.3 & -1128.1 $\pm$ 42.5 \\ \hline
        191 & OPTIMISTIC & MULLERMANDEL & 0.2 & MACLEOD\_LINEAR & 0.25 & 100 & FIXED & 0.5 & -26.7 $\pm$ 1.4 & -32.8 $\pm$ 0.8 & -1400 $\pm$ 40.7 & -1375.4 $\pm$ 40.8 \\ \hline
        192 & OPTIMISTIC & MULLERMANDEL & 0.2 & MACLEOD\_LINEAR & 0.25 & 100 & FIXED & 1 & -23.2 $\pm$ 1.4 & -29.6 $\pm$ 0.8 & -1332.4 $\pm$ 42.4 & -1308.4 $\pm$ 42.5 \\ \hline
        193 & OPTIMISTIC & MULLERMANDEL & 0.2 & MACLEOD\_LINEAR & 0.5 & 0.1 & FIXED & 0.1 & -14.8 $\pm$ 1.2 & -17.7 $\pm$ 0.8 & -1453.1 $\pm$ 25.4 & -1425 $\pm$ 25.7 \\ \hline
        194 & OPTIMISTIC & MULLERMANDEL & 0.2 & MACLEOD\_LINEAR & 0.5 & 0.1 & FIXED & 0.2 & -13.4 $\pm$ 1.2 & -16.1 $\pm$ 0.8 & -1266.2 $\pm$ 21.9 & -1238.8 $\pm$ 22.2 \\ \hline
        195 & OPTIMISTIC & MULLERMANDEL & 0.2 & MACLEOD\_LINEAR & 0.5 & 0.1 & FIXED & 0.5 & -15.7 $\pm$ 1.2 & -19.1 $\pm$ 0.8 & -1404.6 $\pm$ 22.4 & -1377.2 $\pm$ 22.7 \\ \hline
        196 & OPTIMISTIC & MULLERMANDEL & 0.2 & MACLEOD\_LINEAR & 0.5 & 0.1 & FIXED & 1 & -28.1 $\pm$ 1.4 & -33.3 $\pm$ 0.8 & -1105 $\pm$ 34 & -1081.3 $\pm$ 34.1 \\ \hline
        197 & OPTIMISTIC & MULLERMANDEL & 0.2 & MACLEOD\_LINEAR & 0.5 & 1 & FIXED & 0.1 & -12.6 $\pm$ 1.2 & -14.9 $\pm$ 1 & -1320.5 $\pm$ 13.9 & -1292.4 $\pm$ 14.4 \\ \hline
        198 & OPTIMISTIC & MULLERMANDEL & 0.2 & MACLEOD\_LINEAR & 0.5 & 1 & FIXED & 0.2 & -13.6 $\pm$ 1.2 & -16 $\pm$ 1 & -1436.1 $\pm$ 14.8 & -1407.5 $\pm$ 15.2 \\ \hline
        199 & OPTIMISTIC & MULLERMANDEL & 0.2 & MACLEOD\_LINEAR & 0.5 & 1 & FIXED & 0.5 & -12.6 $\pm$ 1.2 & -14.6 $\pm$ 1 & -1527.9 $\pm$ 18.2 & -1498.4 $\pm$ 18.5 \\ \hline
        200 & OPTIMISTIC & MULLERMANDEL & 0.2 & MACLEOD\_LINEAR & 0.5 & 1 & FIXED & 1 & -15.7 $\pm$ 1.2 & -19 $\pm$ 0.7 & -1201.4 $\pm$ 29.5 & -1174.9 $\pm$ 29.7 \\ \hline
    \end{tabular}
    \caption{Hyper-parameters for Models 101-200 and the corresponding Bayes factor relative to the maximum likelihoods, for which the model is in bold. In green, we highlight the highest likelihood models, where the Bayes factor is $>-10$ for either the marginalised or unmarginalised likelihoods.}
    \label{tab:all_models_2}
\end{table*}

\begin{table*}
    \tiny
    \centering
     \begin{tabular}{|c|c|c|c|c|c|c|c|c|c|c|c|c|}
    \hline
        \multirow{2}{*}{Model} & \multirow{2}{*}{CHE} & Remnant mass & \multirow{2}{*}{$f_{\rm WR}$} & \multirow{2}{*}{$\gamma$ prescription} & \multirow{2}{*}{$f_\gamma$} & \multirow{2}{*}{$\alpha_{\rm CE}$} & $\dot{M}_a$ & \multirow{2}{*}{$f_{\rm MT}$} & $\log_{10} \mathcal{B}(\lambda|\mathcal{H})$ & $ \log_{10} \mathcal{B}(\lambda|\mathcal{H})$ & $\log_{10} \mathcal{B}(\lambda, N_\lambda|\mathcal{H})$ & $ \log_{10} \mathcal{B}(\lambda, N_\lambda|\mathcal{H})$ \\ 
        & & prescription & & & & & prescription & & & ($\rm SNR \geq 8$) & & ($\rm SNR \geq 8$) \\\hline
        201 & OPTIMISTIC & MULLERMANDEL & 0.2 & MACLEOD\_LINEAR & 0.5 & 10 & FIXED & 0.1 & -17.5 $\pm$ 1.2 & -20.4 $\pm$ 0.9 & -1226.5 $\pm$ 18.2 & -1199.6 $\pm$ 18.5 \\ \hline
        202 & OPTIMISTIC & MULLERMANDEL & 0.2 & MACLEOD\_LINEAR & 0.5 & 10 & FIXED & 0.2 & -16.4 $\pm$ 1.2 & -20.2 $\pm$ 0.9 & -1404.4 $\pm$ 18.6 & -1377.4 $\pm$ 18.9 \\ \hline
        203 & OPTIMISTIC & MULLERMANDEL & 0.2 & MACLEOD\_LINEAR & 0.5 & 10 & FIXED & 0.5 & -14.6 $\pm$ 1.2 & -17.7 $\pm$ 0.9 & -1495.4 $\pm$ 21.6 & -1467.2 $\pm$ 21.8 \\ \hline
        204 & OPTIMISTIC & MULLERMANDEL & 0.2 & MACLEOD\_LINEAR & 0.5 & 10 & FIXED & 1 & -18.7 $\pm$ 1.3 & -22.2 $\pm$ 0.8 & -1176.7 $\pm$ 32.9 & -1150.8 $\pm$ 33.1 \\ \hline
        205 & OPTIMISTIC & MULLERMANDEL & 0.2 & MACLEOD\_LINEAR & 0.5 & 100 & FIXED & 0.1 & -14.7 $\pm$ 1.2 & -17.6 $\pm$ 0.8 & -1269.5 $\pm$ 22 & -1242.4 $\pm$ 22.3 \\ \hline
        206 & OPTIMISTIC & MULLERMANDEL & 0.2 & MACLEOD\_LINEAR & 0.5 & 100 & FIXED & 0.2 & -15.2 $\pm$ 1.2 & -18.5 $\pm$ 0.8 & -1312.8 $\pm$ 24.5 & -1285.8 $\pm$ 24.8 \\ \hline
        207 & OPTIMISTIC & MULLERMANDEL & 0.2 & MACLEOD\_LINEAR & 0.5 & 100 & FIXED & 0.5 & -14.2 $\pm$ 1.2 & -16.9 $\pm$ 0.8 & -1333.9 $\pm$ 27.1 & -1306.2 $\pm$ 27.3 \\ \hline
        208 & OPTIMISTIC & MULLERMANDEL & 0.2 & MACLEOD\_LINEAR & 0.5 & 100 & FIXED & 1 & -23.2 $\pm$ 1.6 & -29.2 $\pm$ 0.9 & -1156.9 $\pm$ 41.5 & -1133.6 $\pm$ 41.6 \\ \hline
        209 & OPTIMISTIC & MULLERMANDEL & 0.2 & MACLEOD\_LINEAR & 0.75 & 0.1 & FIXED & 0.1 & -24.8 $\pm$ 2.2 & -31 $\pm$ 1.4 & -979.3 $\pm$ 50.1 & -957.5 $\pm$ 50.2 \\ \hline
        210 & OPTIMISTIC & MULLERMANDEL & 0.2 & MACLEOD\_LINEAR & 0.75 & 0.1 & FIXED & 0.2 & -33.3 $\pm$ 2.2 & -39.7 $\pm$ 1.4 & -846 $\pm$ 44 & -825.6 $\pm$ 44.1 \\ \hline
        211 & OPTIMISTIC & MULLERMANDEL & 0.2 & MACLEOD\_LINEAR & 0.75 & 0.1 & FIXED & 0.5 & -23.5 $\pm$ 2.3 & -28.8 $\pm$ 1.4 & -823.9 $\pm$ 44.5 & -802.5 $\pm$ 44.6 \\ \hline
        212 & OPTIMISTIC & MULLERMANDEL & 0.2 & MACLEOD\_LINEAR & 0.75 & 0.1 & FIXED & 1 & -20.1 $\pm$ 2.1 & -25.5 $\pm$ 1.3 & -875.1 $\pm$ 44.1 & -853.3 $\pm$ 44.2 \\ \hline
        213 & OPTIMISTIC & MULLERMANDEL & 0.2 & MACLEOD\_LINEAR & 0.75 & 1 & FIXED & 0.1 & -21 $\pm$ 1.8 & -25.1 $\pm$ 1.1 & -743.4 $\pm$ 32.7 & -721.5 $\pm$ 32.9 \\ \hline
        214 & OPTIMISTIC & MULLERMANDEL & 0.2 & MACLEOD\_LINEAR & 0.75 & 1 & FIXED & 0.2 & -31.3 $\pm$ 1.7 & -36.1 $\pm$ 1 & -1014.4 $\pm$ 39.4 & -991 $\pm$ 39.5 \\ \hline
        215 & OPTIMISTIC & MULLERMANDEL & 0.2 & MACLEOD\_LINEAR & 0.75 & 1 & FIXED & 0.5 & -23.3 $\pm$ 1.8 & -27.5 $\pm$ 1.1 & -956.8 $\pm$ 40.8 & -933.1 $\pm$ 40.9 \\ \hline
        216 & OPTIMISTIC & MULLERMANDEL & 0.2 & MACLEOD\_LINEAR & 0.75 & 1 & FIXED & 1 & -22.5 $\pm$ 1.9 & -27.1 $\pm$ 1.1 & -1022.9 $\pm$ 44.9 & -999.2 $\pm$ 45 \\ \hline
        217 & OPTIMISTIC & MULLERMANDEL & 0.2 & MACLEOD\_LINEAR & 0.75 & 10 & FIXED & 0.1 & -29.7 $\pm$ 2.4 & -35.7 $\pm$ 1.6 & -851.6 $\pm$ 49.1 & -830.7 $\pm$ 49.2 \\ \hline
        218 & OPTIMISTIC & MULLERMANDEL & 0.2 & MACLEOD\_LINEAR & 0.75 & 10 & FIXED & 0.2 & -28 $\pm$ 2.4 & -33.1 $\pm$ 1.5 & -771.2 $\pm$ 43.8 & -750.3 $\pm$ 44 \\ \hline
        219 & OPTIMISTIC & MULLERMANDEL & 0.2 & MACLEOD\_LINEAR & 0.75 & 10 & FIXED & 0.5 & -23.1 $\pm$ 2.2 & -26.6 $\pm$ 1.4 & -771.1 $\pm$ 41 & -748.4 $\pm$ 41.1 \\ \hline
        220 & OPTIMISTIC & MULLERMANDEL & 0.2 & MACLEOD\_LINEAR & 0.75 & 10 & FIXED & 1 & -26.9 $\pm$ 2.1 & -31.8 $\pm$ 1.3 & -910.7 $\pm$ 46.3 & -888.1 $\pm$ 46.4 \\ \hline
        221 & OPTIMISTIC & MULLERMANDEL & 0.2 & MACLEOD\_LINEAR & 0.75 & 100 & FIXED & 0.1 & -45.1 $\pm$ 2.9 & -68.1 $\pm$ 2 & -831.5 $\pm$ 55.4 & -827.9 $\pm$ 55.5 \\ \hline
        222 & OPTIMISTIC & MULLERMANDEL & 0.2 & MACLEOD\_LINEAR & 0.75 & 100 & FIXED & 0.2 & -46.6 $\pm$ 2.8 & -56.1 $\pm$ 1.9 & -750.8 $\pm$ 48.4 & -734.6 $\pm$ 48.5 \\ \hline
        223 & OPTIMISTIC & MULLERMANDEL & 0.2 & MACLEOD\_LINEAR & 0.75 & 100 & FIXED & 0.5 & -30.1 $\pm$ 2.7 & -36.6 $\pm$ 1.8 & -843.9 $\pm$ 53.6 & -823.7 $\pm$ 53.7 \\ \hline
        224 & OPTIMISTIC & MULLERMANDEL & 0.2 & MACLEOD\_LINEAR & 0.75 & 100 & FIXED & 1 & -61.6 $\pm$ 2.9 & -72.5 $\pm$ 2 & -883.8 $\pm$ 57.8 & -867.9 $\pm$ 57.9 \\ \hline
        225 & OPTIMISTIC & MULLERMANDEL & 0.2 & MACLEOD\_LINEAR & 1 & 0.1 & FIXED & 0.1 & -37.3 $\pm$ 2.3 & -45.3 $\pm$ 1.5 & -936.9 $\pm$ 50 & -917.4 $\pm$ 50 \\ \hline
        226 & OPTIMISTIC & MULLERMANDEL & 0.2 & MACLEOD\_LINEAR & 1 & 0.1 & FIXED & 0.2 & -26.9 $\pm$ 2.3 & -34.3 $\pm$ 1.5 & -756.1 $\pm$ 41.9 & -737.5 $\pm$ 42 \\ \hline
        227 & OPTIMISTIC & MULLERMANDEL & 0.2 & MACLEOD\_LINEAR & 1 & 0.1 & FIXED & 0.5 & -37.3 $\pm$ 2.3 & -46.9 $\pm$ 1.5 & -868.2 $\pm$ 47.7 & -850.8 $\pm$ 47.8 \\ \hline
        228 & OPTIMISTIC & MULLERMANDEL & 0.2 & MACLEOD\_LINEAR & 1 & 0.1 & FIXED & 1 & -19.5 $\pm$ 2 & -25.2 $\pm$ 1.2 & -865.2 $\pm$ 41.8 & -843.8 $\pm$ 41.9 \\ \hline
        229 & OPTIMISTIC & MULLERMANDEL & 0.2 & MACLEOD\_LINEAR & 1 & 1 & FIXED & 0.1 & -31.5 $\pm$ 1.9 & -37.8 $\pm$ 1.1 & -859 $\pm$ 38.5 & -838.4 $\pm$ 38.6 \\ \hline
        230 & OPTIMISTIC & MULLERMANDEL & 0.2 & MACLEOD\_LINEAR & 1 & 1 & FIXED & 0.2 & -35.3 $\pm$ 1.9 & -43.5 $\pm$ 1.1 & -1031.6 $\pm$ 45.4 & -1011.5 $\pm$ 45.5 \\ \hline
        231 & OPTIMISTIC & MULLERMANDEL & 0.2 & MACLEOD\_LINEAR & 1 & 1 & FIXED & 0.5 & -32.8 $\pm$ 1.9 & -40.9 $\pm$ 1.1 & -863 $\pm$ 38.2 & -844.1 $\pm$ 38.3 \\ \hline
        232 & OPTIMISTIC & MULLERMANDEL & 0.2 & MACLEOD\_LINEAR & 1 & 1 & FIXED & 1 & -34.5 $\pm$ 1.9 & -41.1 $\pm$ 1.1 & -974 $\pm$ 42.8 & -952.7 $\pm$ 42.9 \\ \hline
        233 & OPTIMISTIC & MULLERMANDEL & 0.2 & MACLEOD\_LINEAR & 1 & 10 & FIXED & 0.1 & -55.5 $\pm$ 2.9 & -68.2 $\pm$ 2 & -766.6 $\pm$ 50.5 & -753.5 $\pm$ 50.6 \\ \hline
        234 & OPTIMISTIC & MULLERMANDEL & 0.2 & MACLEOD\_LINEAR & 1 & 10 & FIXED & 0.2 & -56.7 $\pm$ 2.6 & -68.8 $\pm$ 1.7 & -967.7 $\pm$ 56.8 & -952.2 $\pm$ 56.9 \\ \hline
        235 & OPTIMISTIC & MULLERMANDEL & 0.2 & MACLEOD\_LINEAR & 1 & 10 & FIXED & 0.5 & -43.7 $\pm$ 2.7 & -53.9 $\pm$ 1.8 & -844.4 $\pm$ 52.2 & -827.9 $\pm$ 52.3 \\ \hline
        236 & OPTIMISTIC & MULLERMANDEL & 0.2 & MACLEOD\_LINEAR & 1 & 10 & FIXED & 1 & -25 $\pm$ 2.4 & -31 $\pm$ 1.5 & -1012.4 $\pm$ 56.2 & -990.2 $\pm$ 56.3 \\ \hline
        237 & OPTIMISTIC & MULLERMANDEL & 0.2 & MACLEOD\_LINEAR & 1 & 100 & FIXED & 0.1 & -62.3 $\pm$ 3 & -80 $\pm$ 2.1 & -886.3 $\pm$ 59.5 & -877.1 $\pm$ 59.5 \\ \hline
        238 & OPTIMISTIC & MULLERMANDEL & 0.2 & MACLEOD\_LINEAR & 1 & 100 & FIXED & 0.2 & -52.9 $\pm$ 3.1 & -65.6 $\pm$ 2.2 & -820.6 $\pm$ 57.8 & -807 $\pm$ 57.9 \\ \hline
        239 & OPTIMISTIC & MULLERMANDEL & 0.2 & MACLEOD\_LINEAR & 1 & 100 & FIXED & 0.5 & -59.5 $\pm$ 3.1 & -75.1 $\pm$ 2.1 & -844.3 $\pm$ 58 & -833.4 $\pm$ 58 \\ \hline
        240 & OPTIMISTIC & MULLERMANDEL & 0.2 & MACLEOD\_LINEAR & 1 & 100 & FIXED & 1 & -44.2 $\pm$ 3 & -56.1 $\pm$ 2.1 & -741.1 $\pm$ 50.8 & -727.4 $\pm$ 50.8 \\ \hline
        241 & NONE & MULLERMANDEL & 0.1 & ISOTROPIC & 0 & 1 & THERMAL & 0.5 & -27.6 $\pm$ 1.4 & -42.1 $\pm$ 0.8 & -347.4 $\pm$ 12.2 & -341.7 $\pm$ 12.6 \\ \hline
        242 & NONE & MULLERMANDEL & 0.5 & ISOTROPIC & 0 & 1 & THERMAL & 0.5 & -18 $\pm$ 1.5 & -31.7 $\pm$ 0.9 & -246.3 $\pm$ 10.4 & -242 $\pm$ 10.9 \\ \hline
        243 & NONE & MULLERMANDEL & 1 & ISOTROPIC & 0 & 1 & THERMAL & 0.5 & -11.7 $\pm$ 1.8 & -15.8 $\pm$ 1.1 & -143.3 $\pm$ 8.7 & -132.7 $\pm$ 9.3 \\ \hline
\rowcolor{Green!30}
        244 & NONE & MULLERMANDEL & 5 & ISOTROPIC & 0 & 1 & THERMAL & 0.5 & -6.1 $\pm$ 2.8 & -11.2 $\pm$ 1.9 & -10.4 $\pm$ 4.4 & -12.7 $\pm$ 5.3 \\ \hline
        245 & NONE & MULLERMANDEL & 10 & ISOTROPIC & 0 & 1 & THERMAL & 0.5 & -13.5 $\pm$ 3.7 & -15.3 $\pm$ 2.7 & -10.4 $\pm$ 4.4 & -15.5 $\pm$ 5.1 \\ \hline
        246 & NONE & SCHNEIDER2020 & 0.1 & ISOTROPIC & 0 & 1 & THERMAL & 0.5 & -18.3 $\pm$ 5.1 & -48.7 $\pm$ 3.9 & -46.5 $\pm$ 10.5 & -69.5 $\pm$ 10.6 \\ \hline
        247 & NONE & SCHNEIDER2020 & 0.5 & ISOTROPIC & 0 & 1 & THERMAL & 0.5 & -28.3 $\pm$ 5.5 & -61.8 $\pm$ 4.3 & -42 $\pm$ 9.3 & -70.5 $\pm$ 9.4 \\ \hline
        248 & NONE & SCHNEIDER2020 & 1 & ISOTROPIC & 0 & 1 & THERMAL & 0.5 & -12 $\pm$ 6.1 & -39.5 $\pm$ 4.8 & -11 $\pm$ 7.8 & -38.1 $\pm$ 7.7 \\ \hline
\rowcolor{Green!30}
        249 & NONE & SCHNEIDER2020 & 5 & ISOTROPIC & 0 & 1 & THERMAL & 0.5 & -6.6 $\pm$ 8.2 & -33 $\pm$ 6.7 & -15 $\pm$ 8.5 & -53.4 $\pm$ 7.8 \\ \hline
        250 & NONE & SCHNEIDER2020 & 10 & ISOTROPIC & 0 & 1 & THERMAL & 0.5 & -14.2 $\pm$ 9 & -41.6 $\pm$ 7.4 & -38.8 $\pm$ 9.2 & -84.8 $\pm$ 8.4 \\ \hline
        251 & PESSIMISTIC & MULLERMANDEL & 0.1 & ISOTROPIC & 0 & 1 & THERMAL & 0.5 & -29.2 $\pm$ 1.3 & -34.9 $\pm$ 0.7 & -793.7 $\pm$ 21.1 & -773.1 $\pm$ 21.4 \\ \hline
        252 & PESSIMISTIC & MULLERMANDEL & 0.5 & ISOTROPIC & 0 & 1 & THERMAL & 0.5 & -30.6 $\pm$ 1.4 & -34.9 $\pm$ 0.8 & -636.8 $\pm$ 20 & -616.4 $\pm$ 20.3 \\ \hline
        253 & PESSIMISTIC & MULLERMANDEL & 1 & ISOTROPIC & 0 & 1 & THERMAL & 0.5 & -35.9 $\pm$ 1.6 & -40.3 $\pm$ 0.9 & -501.7 $\pm$ 18.7 & -483.2 $\pm$ 19 \\ \hline
        254 & PESSIMISTIC & MULLERMANDEL & 5 & ISOTROPIC & 0 & 1 & THERMAL & 0.5 & -19.6 $\pm$ 2.6 & -23.7 $\pm$ 1.7 & -77.7 $\pm$ 7.6 & -71.3 $\pm$ 8.2 \\ \hline
        255 & PESSIMISTIC & MULLERMANDEL & 10 & ISOTROPIC & 0 & 1 & THERMAL & 0.5 & -19.4 $\pm$ 3.5 & -21.8 $\pm$ 2.5 & -27.6 $\pm$ 5.7 & -26.1 $\pm$ 6.3 \\ \hline
        256 & PESSIMISTIC & SCHNEIDER2020 & 0.1 & ISOTROPIC & 0 & 1 & THERMAL & 0.5 & -53.1 $\pm$ 3.2 & -78.9 $\pm$ 2.3 & -480.4 $\pm$ 36.1 & -484 $\pm$ 36.2 \\ \hline
        257 & PESSIMISTIC & SCHNEIDER2020 & 0.5 & ISOTROPIC & 0 & 1 & THERMAL & 0.5 & -59 $\pm$ 3.5 & -87.9 $\pm$ 2.5 & -455.2 $\pm$ 36.3 & -462.4 $\pm$ 36.4 \\ \hline
        258 & PESSIMISTIC & SCHNEIDER2020 & 1 & ISOTROPIC & 0 & 1 & THERMAL & 0.5 & -50.9 $\pm$ 3.9 & -71 $\pm$ 2.9 & -314.8 $\pm$ 28.9 & -315.9 $\pm$ 29 \\ \hline
        259 & PESSIMISTIC & SCHNEIDER2020 & 5 & ISOTROPIC & 0 & 1 & THERMAL & 0.5 & -53.2 $\pm$ 6 & -88.7 $\pm$ 4.8 & -68.4 $\pm$ 10.4 & -98.4 $\pm$ 10.4 \\ \hline
        260 & PESSIMISTIC & SCHNEIDER2020 & 10 & ISOTROPIC & 0 & 1 & THERMAL & 0.5 & -51.9 $\pm$ 7.2 & -87.3 $\pm$ 5.7 & -49.9 $\pm$ 8.8 & -85.7 $\pm$ 8.5 \\ \hline
        261 & OPTIMISTIC & MULLERMANDEL & 0.1 & ISOTROPIC & 0 & 1 & THERMAL & 0.5 & -33.9 $\pm$ 1.2 & -41.6 $\pm$ 0.7 & -1324.5 $\pm$ 30.8 & -1301.8 $\pm$ 31 \\ \hline
        262 & OPTIMISTIC & MULLERMANDEL & 0.5 & ISOTROPIC & 0 & 1 & THERMAL & 0.5 & -37.3 $\pm$ 1.4 & -42.5 $\pm$ 0.8 & -845 $\pm$ 25.4 & -823.5 $\pm$ 25.7 \\ \hline
        263 & OPTIMISTIC & MULLERMANDEL & 1 & ISOTROPIC & 0 & 1 & THERMAL & 0.5 & -35.6 $\pm$ 1.6 & -41.1 $\pm$ 0.9 & -577.5 $\pm$ 21.4 & -559.1 $\pm$ 21.6 \\ \hline
        264 & OPTIMISTIC & MULLERMANDEL & 5 & ISOTROPIC & 0 & 1 & THERMAL & 0.5 & -26.1 $\pm$ 2.6 & -29.5 $\pm$ 1.7 & -89.8 $\pm$ 8.1 & -82.3 $\pm$ 8.6 \\ \hline
        265 & OPTIMISTIC & MULLERMANDEL & 10 & ISOTROPIC & 0 & 1 & THERMAL & 0.5 & -34.1 $\pm$ 3.5 & -41.1 $\pm$ 2.5 & -44 $\pm$ 5.8 & -46.6 $\pm$ 6.4 \\ \hline
        266 & OPTIMISTIC & SCHNEIDER2020 & 0.1 & ISOTROPIC & 0 & 1 & THERMAL & 0.5 & -79.7 $\pm$ 2.7 & -120.1 $\pm$ 1.9 & -1191.7 $\pm$ 71.8 & -1202.9 $\pm$ 71.8 \\ \hline
        267 & OPTIMISTIC & SCHNEIDER2020 & 0.5 & ISOTROPIC & 0 & 1 & THERMAL & 0.5 & -59.9 $\pm$ 3.2 & -96.4 $\pm$ 2.3 & -553.9 $\pm$ 40.8 & -567.3 $\pm$ 40.9 \\ \hline
        268 & OPTIMISTIC & SCHNEIDER2020 & 1 & ISOTROPIC & 0 & 1 & THERMAL & 0.5 & -66.5 $\pm$ 3.6 & -98.8 $\pm$ 2.7 & -419.8 $\pm$ 34.5 & -431.3 $\pm$ 34.6 \\ \hline
        269 & OPTIMISTIC & SCHNEIDER2020 & 5 & ISOTROPIC & 0 & 1 & THERMAL & 0.5 & -62.8 $\pm$ 5.9 & -97.9 $\pm$ 4.6 & -81.2 $\pm$ 10.7 & -110.3 $\pm$ 10.6 \\ \hline
        270 & OPTIMISTIC & SCHNEIDER2020 & 10 & ISOTROPIC & 0 & 1 & THERMAL & 0.5 & -45.9 $\pm$ 7.2 & -81.4 $\pm$ 5.8 & -44.8 $\pm$ 9 & -79.9 $\pm$ 8.7 \\ \hline
        271 & NONE & FRYER2012 & 0.1 & ISOTROPIC & 0 & 1 & THERMAL & 0.5 & -14.2 $\pm$ 2.2 & -19.4 $\pm$ 1.4 & -209.6 $\pm$ 13.5 & -197.7 $\pm$ 13.8 \\ \hline
        272 & NONE & FRYER2012 & 0.5 & ISOTROPIC & 0 & 1 & THERMAL & 0.5 & -11.7 $\pm$ 2.3 & -13.5 $\pm$ 1.5 & -150.5 $\pm$ 11.3 & -137.3 $\pm$ 11.8 \\ \hline
\rowcolor{Green!30}
        273 & NONE & FRYER2012 & 1 & ISOTROPIC & 0 & 1 & THERMAL & 0.5 & -4.4 $\pm$ 2.7 & -6.7 $\pm$ 1.8 & -86.4 $\pm$ 9.4 & -76.5 $\pm$ 9.9 \\ \hline
\rowcolor{Green!30}
        \textbf{274} & \textbf{NONE} & \textbf{FRYER2012} & \textbf{5} & \textbf{ISOTROPIC} & \textbf{0} & \textbf{1} & \textbf{THERMAL} & \textbf{0.5} & \textbf{-1.9 $\pm$ 3.8} & \textbf{-1.5 $\pm$ 2.8} & \textbf{-0.5 $\pm$ 5} & \textbf{0 $\pm$ 5.6} \\ \hline
\rowcolor{Green!30}
        275 & NONE & FRYER2012 & 10 & ISOTROPIC & 0 & 1 & THERMAL & 0.5 & -3.8 $\pm$ 4.7 & -2.5 $\pm$ 3.6 & -2.9 $\pm$ 5.2 & -8.1 $\pm$ 5.5 \\ \hline
        276 & NONE & FRYER2022 & 0.1 & ISOTROPIC & 0 & 1 & THERMAL & 0.5 & -15.9 $\pm$ 2.2 & -20.6 $\pm$ 1.4 & -210.8 $\pm$ 13.9 & -198.5 $\pm$ 14.2 \\ \hline
        277 & NONE & FRYER2022 & 0.5 & ISOTROPIC & 0 & 1 & THERMAL & 0.5 & -14 $\pm$ 2.5 & -19.5 $\pm$ 1.6 & -144.7 $\pm$ 11.7 & -135.5 $\pm$ 12.1 \\ \hline
\rowcolor{Green!30}
        278 & NONE & FRYER2022 & 1 & ISOTROPIC & 0 & 1 & THERMAL & 0.5 & -0.5 $\pm$ 2.7 & -1.4 $\pm$ 1.8 & -71 $\pm$ 8.8 & -60.6 $\pm$ 9.3 \\ \hline
\rowcolor{Green!30}
        279 & NONE & FRYER2022 & 5 & ISOTROPIC & 0 & 1 & THERMAL & 0.5 & -4.5 $\pm$ 3.8 & -3.8 $\pm$ 2.8 & -2.4 $\pm$ 5 & -2.2 $\pm$ 5.5 \\ \hline
\rowcolor{Green!30}
        280 & NONE & FRYER2022 & 10 & ISOTROPIC & 0 & 1 & THERMAL & 0.5 & -6.6 $\pm$ 4.7 & -7.5 $\pm$ 3.6 & -5.7 $\pm$ 5.2 & -13.2 $\pm$ 5.5 \\ \hline
        281 & PESSIMISTIC & FRYER2012 & 0.1 & ISOTROPIC & 0 & 1 & THERMAL & 0.5 & -21.8 $\pm$ 1.8 & -25.8 $\pm$ 1 & -607.3 $\pm$ 26.3 & -587 $\pm$ 26.5 \\ \hline
        282 & PESSIMISTIC & FRYER2012 & 0.5 & ISOTROPIC & 0 & 1 & THERMAL & 0.5 & -19.8 $\pm$ 1.9 & -22.4 $\pm$ 1.2 & -607.8 $\pm$ 28.9 & -585.9 $\pm$ 29.1 \\ \hline
        283 & PESSIMISTIC & FRYER2012 & 1 & ISOTROPIC & 0 & 1 & THERMAL & 0.5 & -28.1 $\pm$ 2.1 & -32.4 $\pm$ 1.3 & -459 $\pm$ 24.6 & -441.1 $\pm$ 24.8 \\ \hline
        284 & PESSIMISTIC & FRYER2012 & 5 & ISOTROPIC & 0 & 1 & THERMAL & 0.5 & -17.3 $\pm$ 3.6 & -19 $\pm$ 2.6 & -56.8 $\pm$ 8.7 & -49.7 $\pm$ 9.1 \\ \hline
        285 & PESSIMISTIC & FRYER2012 & 10 & ISOTROPIC & 0 & 1 & THERMAL & 0.5 & -27.7 $\pm$ 4.4 & -29.8 $\pm$ 3.3 & -33.9 $\pm$ 6.7 & -32.6 $\pm$ 7.1 \\ \hline
        286 & PESSIMISTIC & FRYER2022 & 0.1 & ISOTROPIC & 0 & 1 & THERMAL & 0.5 & -28.9 $\pm$ 1.8 & -33.7 $\pm$ 1 & -731.4 $\pm$ 31.4 & -710.4 $\pm$ 31.5 \\ \hline
        287 & PESSIMISTIC & FRYER2022 & 0.5 & ISOTROPIC & 0 & 1 & THERMAL & 0.5 & -23.8 $\pm$ 1.9 & -28.2 $\pm$ 1.2 & -628.8 $\pm$ 29.7 & -608.5 $\pm$ 29.9 \\ \hline
        288 & PESSIMISTIC & FRYER2022 & 1 & ISOTROPIC & 0 & 1 & THERMAL & 0.5 & -29.1 $\pm$ 2.2 & -33 $\pm$ 1.4 & -414 $\pm$ 23.5 & -396.5 $\pm$ 23.7 \\ \hline
        289 & PESSIMISTIC & FRYER2022 & 5 & ISOTROPIC & 0 & 1 & THERMAL & 0.5 & -18.5 $\pm$ 3.4 & -20.3 $\pm$ 2.5 & -60.5 $\pm$ 8.6 & -53.3 $\pm$ 9 \\ \hline
        290 & PESSIMISTIC & FRYER2022 & 10 & ISOTROPIC & 0 & 1 & THERMAL & 0.5 & -16.6 $\pm$ 4.4 & -16.6 $\pm$ 3.4 & -20.6 $\pm$ 6.5 & -17.9 $\pm$ 6.8 \\ \hline
        291 & OPTIMISTIC & FRYER2012 & 0.1 & ISOTROPIC & 0 & 1 & THERMAL & 0.5 & -26.3 $\pm$ 1.7 & -31.3 $\pm$ 0.9 & -1108.1 $\pm$ 42.3 & -1084.2 $\pm$ 42.4 \\ \hline
        292 & OPTIMISTIC & FRYER2012 & 0.5 & ISOTROPIC & 0 & 1 & THERMAL & 0.5 & -29.5 $\pm$ 1.9 & -34.3 $\pm$ 1.1 & -686 $\pm$ 31 & -665.5 $\pm$ 31.1 \\ \hline
        293 & OPTIMISTIC & FRYER2012 & 1 & ISOTROPIC & 0 & 1 & THERMAL & 0.5 & -27.7 $\pm$ 2.2 & -31.6 $\pm$ 1.4 & -469.3 $\pm$ 25.7 & -450.7 $\pm$ 25.9 \\ \hline
        294 & OPTIMISTIC & FRYER2012 & 5 & ISOTROPIC & 0 & 1 & THERMAL & 0.5 & -29.3 $\pm$ 3.5 & -32.5 $\pm$ 2.5 & -80.3 $\pm$ 9.5 & -73.7 $\pm$ 9.9 \\ \hline
        295 & OPTIMISTIC & FRYER2012 & 10 & ISOTROPIC & 0 & 1 & THERMAL & 0.5 & -30.9 $\pm$ 4.4 & -34.7 $\pm$ 3.3 & -39.4 $\pm$ 7 & -39.2 $\pm$ 7.3 \\ \hline
        296 & OPTIMISTIC & FRYER2022 & 0.1 & ISOTROPIC & 0 & 1 & THERMAL & 0.5 & -31.9 $\pm$ 1.6 & -38 $\pm$ 0.9 & -1167.4 $\pm$ 43.1 & -1144.3 $\pm$ 43.3 \\ \hline
        297 & OPTIMISTIC & FRYER2022 & 0.5 & ISOTROPIC & 0 & 1 & THERMAL & 0.5 & -31.5 $\pm$ 1.9 & -35.4 $\pm$ 1.2 & -703.5 $\pm$ 32.8 & -682.1 $\pm$ 33 \\ \hline
        298 & OPTIMISTIC & FRYER2022 & 1 & ISOTROPIC & 0 & 1 & THERMAL & 0.5 & -22.8 $\pm$ 2.2 & -25.4 $\pm$ 1.4 & -436.6 $\pm$ 24.3 & -417.2 $\pm$ 24.5 \\ \hline
        299 & OPTIMISTIC & FRYER2022 & 5 & ISOTROPIC & 0 & 1 & THERMAL & 0.5 & -26.7 $\pm$ 3.6 & -29.2 $\pm$ 2.6 & -74.4 $\pm$ 9.4 & -67.4 $\pm$ 9.8 \\ \hline
        300 & OPTIMISTIC & FRYER2022 & 10 & ISOTROPIC & 0 & 1 & THERMAL & 0.5 & -28.8 $\pm$ 4.4 & -31.3 $\pm$ 3.3 & -35.5 $\pm$ 6.8 & -34.5 $\pm$ 7.1 \\ \hline
    \end{tabular}
    \caption{Hyper-parameters for Models 201-300 and the corresponding Bayes factor relative to the maximum likelihoods, for which the model is in bold. In green, we highlight the highest likelihood models, where the Bayes factor is $>-10$ for either the marginalised or unmarginalised likelihoods.}
    \label{tab:all_models_3}
\end{table*}




\bsp	
\label{lastpage}
\end{document}